\documentclass{aa}  

\usepackage{graphicx}
\usepackage[colorlinks=true,citecolor=blue]{hyperref}

\usepackage{txfonts}
\usepackage{multirow}
\begin{document}

   \title{Mapping radial abundance gradients with {\it Gaia}-ESO open clusters:}

   \subtitle{Evidence of recent gas accretion in the Milky Way disk}

   \author{M. Palla
          \inst{1,2}
          \and
          L. Magrini
          \inst{3}
          \and
          E. Spitoni
          \inst{4}
          \and
          F. Matteucci
          \inst{4,5,6}
          \and
          C. Viscasillas Vázquez
          \inst{7}
          \and
          M. Franchini
          \inst{4}
          \and 
          M. Molero
          \inst{4,8}
          \and S. Randich
          \inst{3}
          }

   \institute{Dipartimento di Fisica e Astronomia “Augusto Righi”, Alma Mater Studiorum,        Università di Bologna, Via Gobetti 93/2, 40129 Bologna, Italy
         \and
             INAF – Osservatorio di Astrofisica e Scienza dello Spazio di Bologna, Via Gobetti 93/3, 40129 Bologna, Italy\\
             \email{marco.palla@inaf.it}
         \and
             INAF – Osservatorio Astrofisico di Arcetri, Largo E. Fermi 5, 50125 Firenze, Italy
         \and 
             INAF - Osservatorio Astronomico di Trieste, Via Tiepolo 11, I-34131 Trieste, Italy
         \and
             Dipartimento di Fisica, Sezione di Astronomia, Università degli studi di Trieste, Via G.B. Tiepolo 11, I-34143 Trieste, Italy
         \and 
             INFN - Sezione di Trieste, via A. Valerio 2, I-34100, Trieste, Italy
         \and
             Institute of Theoretical Physics and Astronomy, Vilnius University, Sauletekio av. 3, 10257 Vilnius, Lithuania
         \and 
             Institut für Kernphysik, Technische Universität Darmstadt, Schlossgartenstr. 2, Darmstadt 64289, Germany
             }

   \date{Received XXX; accepted XXX}

 
  \abstract
   {Recent evidences from spectroscopic surveys point towards the presence of a metal-poor, young stellar population in the low-$\alpha$/chemical thin disk. In this context, the investigation of the spatial distribution and time evolution of precise, unbiased abundances is fundamental to disentangle the scenarios of formation and evolution of the Galaxy.}
   {We study the evolution of abundance gradients in the Milky Way by taking advantage of a large sample of open star clusters, which are among the best tracers for this purpose. In particular, we use data from the last release of the {\it Gaia}-ESO survey.} 
   {We perform careful selection of open cluster member stars excluding those members that may be affected by biases in spectral analysis. The cleaned open clusters sample is compared with detailed chemical evolution models for the Milky Way, using well tested stellar yields and prescription for radial migration. Different scenarios of Galaxy evolution are tested to explain the data, i.e. the two-infall and the three-infall frameworks, suggesting that the chemical thin disk is formed by one or two subsequent gas accretion episodes, respectively.}
   {With the performed selection in cluster member stars, we still find a metallicity decrease between intermediate age ($1<$ Age/Gyr $<3$) and young (Age $<1$ Gyr) open clusters. 
   This decrease cannot be explained in the context of the two-infall scenario, even by accounting for the effect of migration and yield prescriptions. The three-infall framework, with its late gas accretion in the last 3 Gyr, can explain the low metallic content in young clusters. 
   However, we invoke a milder metal dilution for this gas infall episode relative to previous findings.}
   {To explain the observed low metallic content in young clusters, we propose that a late gas accretion episode triggering a metal dilution should have taken place, extending the framework of the three-infall model for the first time to the entire Galactic disk.}


   \maketitle

\section{Introduction}

A fundamental constraint to study the formation and chemical evolution of the Galaxy are abundance gradients along the Galactic disk.

Different stellar and nebular Galactic tracers that correspond to different epochs in the evolution of our Galaxy have been used to probe radial abundance gradients.
These are Open Clusters (OCs, e.g. \citealt{Randich03,Randich22,Magrini10,Yong12}), HII regions (e.g. \citealt{Balser11,Esteban17,Mendez22}), young massive O and B stars (e.g. \citealt{Daflon04,Braganca19}), Classical Cepheids (CCs, e.g. \citealt{Lemasle07,Lemasle08,Luck11,Genovali15,Kovtyukh22}), planetary nebulae (PNe, e.g. \citealt{Maciel04,Henry10,Stanghellini10,Stanghellini18}), and also field stars with precise stellar ages (e.g. \citealt{Anders17,Santos21}).

The plethora of information from all these tracers has been carefully interpreted by means of models, allowing us to understand fundamental properties for the Milky Way (MW) disk formation.
Among these, the inside-out mechanism (e.g. \citealt{Francois89,Chiappini01,Schonrich17}), variable star formation efficiency (SFE), i.e. higher in the inner regions than in the outer ones (e.g. \citealt{Colavitti09,Grisoni18,Palla20}) and radial gas flows (e.g. \citealt{Portinari00,Spitoni11,Bilitewski12,Cavichia14}).
Moreover, studies on tracers of the "old" stellar gradients have given important information on the impact of the process of stellar radial migration (e.g. \citealt{Minchev18,Willett23}).\\

On the other side, the analysis of spectroscopic data from ground based surveys, such as the APOGEE (e.g. \citealt{Hayden15,Queiroz20}), the {\it Gaia}-ESO (GES, e.g. \citealt{Recio14,Kordopatis15,Rojas16}), 
and the GALAH (\citealt{Buder19,Buder21}) ones, the accurate asteroseismic stellar ages (e.g. \citealt{Pinsonneault14,Pinsonneault18,Miglio21}) and the kinematics and dynamical properties provided by the Gaia mission (\citealt{Gaia16,Gaia18,Gaia21,Gaia23GENERAL}) have pointed towards the existence of two sequences of stars in the [$\alpha$/Fe]\footnote{[X/Y] $=$ $\log$(X/Y) $-$ $\log$(X$_\odot$/Y$_\odot$), where X, Y are the abundances of the object studied and X$_\odot$, Y$_\odot$ are solar abundances.} versus [Fe/H] abundance pattern in the MW disk: the so-called high-$\alpha$ and low-$\alpha$ sequences.

To explain the wealth of these data, \citet{Palla20,Palla22} (see also \citealt{Spitoni19,Spitoni21}) suggested the that presence of this feature (also known as $\alpha$-bimodality) may be connected to to a delayed ($\gtrsim 3$ Gyr) accretion of gas. The latter forges the low-$\alpha$ sequence of stars, while high-$\alpha$ stars are formed promptly in a gas infall episode occurring in the first phases of Galactic formation.
This scenario is confirmed by several models and simulations of the evolution of galactic disks (e.g. \citealt{Noguchi18,Grand18,Mackereth18,Buck20}) which suggested that the bimodality may be strictly connected to a delayed accretion of gas of primordial/metal-poor chemical composition.

Recently, to reproduce the chemical abundances from Gaia DR3 Radial Velocity Spectrometer (RVS) spectra in the solar vicinity (\citealt{Gaia23,RecioBlanco23}) and in particular 
a population of massive stars with evidences of a recent chemical impoverishment, \citet{Spitoni23} suggested a novel scenario of chemical evolution in which the low-$\alpha$ population of stars is generated by two distinct gas accretion episodes, with the latter infall happening at very recent times ($<3$ Gyr of age).
This scenario is constrained by the star formation histories for disk stars inferred from Gaia DR1 and DR2 color-magnitude diagrams (CMDs, \citealt{Bernard17,Ruiz20}), which show evidence for short episodes of enhanced star formation (hereafter, SF) in recent times. 
In this model, the enhanced SF is triggered by the gas infall, which in turn causes the visible chemical impoverishment in the young stellar population as observed by Gaia RVS.\\

In the light of these recent development for the MW disk formation and evolution, we exploit the OCs from the last data release of the {\it Gaia}-ESO survey (\citealt{Randich22}) to investigate the late evolution of radial chemical gradients in the Galaxy.
OCs are in fact considered excellent tracers of the chemical properties of the disk stellar populations of our Galaxy, including the spatial distribution of elemental abundances, especially when observed with high-resolution spectroscopy (\citealt{Spina22}).
To this regard, the {\it Gaia}-ESO survey (\citealt{Gilmore12,Gilmore22,Randich13,Randich22}), is the only survey performed on a $8$ m-class telescope, which put specific focus on the population of Galactic OCs. {\it Gaia}-ESO targeted OCs over a wide range of ages, distances, masses, and metallicities, observing large unbiased samples of cluster candidates, with a well-defined selection function (\citealt{Bragaglia22,Randich22}).

Of this sample, \citet{Magrini23} selected 62 OCs, with extended radial (up to R$>$15 kpc) and age (up to 7 Gyr) ranges, analysing the shape of radial gradients in chemical elements spanning different nucleosynthetic origin (from Oxygen to Europium) and their time evolution. 
They found that the gradients of most of chemical elements, including the metallicity [Fe/H], can be better approximated with a two-slope shape, steeper in the inner regions and rather flat in the outer ones. Shallower gradient slopes in outer regions were also observed in other studies using different tracers (OCs, e.g. \citealt{Carbajo24}; CCs, e.g. \citealt{DaSilva23}), suggesting a flat SFE law for large Galactocentric distances at variance with previously theorised (e.g. \citealt{Grisoni18, Palla20}). 
In addition, \citet{Magrini23} found that the youngest clusters in the sample (age $< 1$ Gyr) have lower metallicity than their older counterpart, even though the effect could be mitigated by avoid considering stars with low surface gravity.

In this work, we compare the above mentioned data sample with detailed chemical evolution models, which also account for the effect of stellar radial migration. 
We start from well tested models under the revised two-infall scenario (\citealt{Palla20,DaSilva23}), which successfully reproduce data from high-resolution surveys in different Galactocentric regions (see also \citealt{Spitoni21}) and were already adopted in the context of radial abundance gradients (\citealt{Palla20,DaSilva23}). 
To better investigate the observed behaviour in {\it Gaia}-ESO OCs, we then extend the comparison to the newly proposed scenario of the three-infall model (\citealt{Spitoni23}), expanding this framework for the first time to the whole MW disk and discussing its feasibility in the context of the adopted dataset.\\

The paper is organised as follows. 
In Section \ref{s:data} we describe the {\it Gaia}-ESO OCs sample and the additional dataset adopted in this work. In Section \ref{s:models}, we present the model framework used, i.e. from the model scenarios to the nucleosynthesis and radial migration prescriptions. In Section \ref{s:results} we present the comparison between the different model predictions and the observations, also discussing the implications of the obtained results. Finally, in Section \ref{s:conclusion} we draw some conclusions.

\defcitealias{Magrini23}{Ma23}
\defcitealias{DaSilva23}{DS23}

\section{Observational data}
\label{s:data}

\subsection{{\it Gaia}-ESO open clusters}
\label{ss:GES_OCs}

{\em Gaia}-ESO is a large public spectroscopic survey that observed the major components of our Galaxy with FLAMES@VLT from 2011-2018 \citep[see][for a full description of the survey]{Randich22, Gilmore22}. 
The final  release {\sc dr5.1}  is public and available at the ESO website from June 2023. {\em Gaia}-ESO made use of FLAMES with both the high-resolution spectrograph UVES (operating at a resolving power, R=47\,000) and the medium-resolution spectrograph GIRAFFE (R$\sim$20\,000). It observed open star clusters for about 30\% of its 340 nights. The observed clusters cover a wide range in age, distance, mass and metallicity \citep[see][]{Randich22}, with an unbiased selection of cluster candidates. 
Each cluster was observed with both UVES and GIRAFFE. In particular, UVES spectra cover a wide spectral range from 480.0~nm to 680.0~nm (U580) or from 420.0~nm to 620.0~nm (U520). The large spectral interval, combined with the high signal-to-noise ratios (S/N) and with the high-resolution, have enabled an unprecedented characterisation of a large sample of open clusters, observed and analysed in a homogeneous way \citep{Bragaglia22,  Hourihane2023A&A...676A.129H}. 
For the about 80 observed clusters, it was possible to obtain precise stellar parameters and abundances of more than 30 different ions, including those of elements belonging to all the main nucleosynthesis channels, from the lightest ones, such as Li to the heaviest one, such as Eu. 
The detailed chemistry of the open cluster sample in {\em Gaia}-ESO, combined with uniform ages and precise distances from {\em Gaia} \citep[e.g.][]{Cantat2020A&A...640A...1C} has been used, e.g., to calibrated age-sensitive abundance ratios, the so-called chemical clocks \citep{Casali2019A&A...629A..62C, Casali2020A&A...639A.127C, Viscasillas2022A&A...660A.135V}, to investigate the nucleosynthesis of neutron-capture elements \citep{Magrini2018A&A...617A.106M, Magrini2022Univ....8...64M, Van2023A&A...670A.129V, Molero2023MNRAS.523.2974M}, to study the evolution of Li abundance \citep{Randich2020A&A...640L...1R, Romano2021A&A...653A..72R, Magrini2021A&A...646L...2M, Magrini2021A&A...655A..23M}, and to study the shape and the evolution of the radial abundance gradients \citep{Jacobson2016A&A...591A..37J, Overbeek2017A&A...598A..68O, Magrini17, Spina2017A&A...601A..70S, Spina22, Magrini23, DaSilva23}.

In the present work, we make us of the sample of open clusters used in \citeauthor{Magrini23} (\citeyear{Magrini23}, hereafter \citetalias{Magrini23}), 
in which abundances for 25 chemical elements were provided for the 62 {\it Gaia}-ESO OCs older than 100~Myr.
The adopted cluster ages are the same as in \citet{Viscasillas2022A&A...660A.135V} and are derived from the homogeneous analysis of {\it Gaia} photometric and astrometric data by \citet{Cantat2020A&A...640A...1C} by means of an artificial neural network trained on a set of objects with well-determined parameters in the literature. 
Here, we consider the guiding radius (R$_{guide}$) as tracer of the OCs Galactocentic distance. These are computed as defined by \citet{Halle15,Halle18}, i.e. as the average between the minimum and maximum radius of the orbits. For the latter, we calculate them by using the \textsc{Galpy} code with the axisymmetric potential \textsc{MWPotential2014} (\citealt{Bovy15}).
For further information on the sample cluster parameters, as well as on their distribution in different quantities, we refer to \citet[]{Viscasillas2022A&A...660A.135V}.
For each cluster, the membership analysis was done using both radial velocities from {\em Gaia}-ESO and proper motions and parallaxes from {\em Gaia} {\sc edr3} \citep{Gaia21}, as described in \citet{Magrini2021A&A...646L...2M}, \citet{Viscasillas2022A&A...660A.135V} and \citet{Jackson2022MNRAS.509.1664J}.

\subsubsection{A restricted sample to avoid observational biases}
\label{ss:restricted_sample}

From a purely observational point of view, the cluster sample in  \citetalias{Magrini23} showed an unexpected behaviour regarding the evolution with time of the metallicity gradient:  the youngest clusters (age<1~Gyr) in the inner disk have lower metallicity than their older counterparts and  they outline a flatter gradient below that of the older population

To distinguish the real evolution of the gradient from  possible spectral analysis effects, \citetalias{Magrini23} restricted the sample of member stars per cluster. In fact, an investigation of the internal abundances of each cluster showed that there are trends of [Fe/H] versus stellar parameters, in particular gravity and microturbulence. 
These trends are not specific to {\em Gaia}-ESO but are present in all the considered spectroscopic surveys  (see Fig. 10 and Appendix in \citetalias{Magrini23}). 
This might be due to two  effects: the former is related to problems in modelling the atmospheres of giant stars, which affect the  spectral analysis of low-gravity giant stars, as noticed already in \citet{Casali2020A&A...643A..12C} and \citet{Spina22}, and the latter to  the effects of magnetic activity in young massive giant stars. Hence, the spectral  analysis of  giants ($\log g$ < 2.5) likely underestimates their [Fe/H] of about 0.1-0.2~dex due to the combination of these effects.

Therefore, in a conservative approach, in order to preserve the actual mean abundance of clusters, we  consider a restricted sample of member stars, where only stars with $\log g$  > 2.5 and $\xi$ < 1.8 km~$^{-1}$ are taken into account to compute the mean cluster abundances. 
The average abundances used in the present work, restricted in stellar parameters,  are reported in Tab.~\ref{tab_clusters}. In the Table we also report the different cluster parameters (i.e. age, distances, orbital parameters), which are the same as the ones introduced earlier in \ref{ss:GES_OCs} for the original {\it Gaia}-ESO OC sample.

\subsection{Additional data sets}

\subsubsection{{\em Gaia}-ESO field stars}

We select a sample of about 3800 field stars in {\em Gaia}-ESO {\sc dr5.1} following the criteria of \citet{Viscasillas2022A&A...660A.135V}, to which we refer for a complete description. 
Here are the basic steps of our selection and the computation of their ages. 
Our set of field stars is composed by both field stars (GES\_FLD keywords GES\_MW for general MW fields, GES\_MW\_BL for fields in the direction of the Galactic bulge, GES\_K2 for stars observed in  Kepler2 (K2) fields, GES\_CR for stars observed in CoRoT fields) and non members of the 62 open clusters considered in this work 
 (age$\,>\,$100~Myr).
 We applied some further quality selections: SNR > 20; $e T_{\rm eff} < 150$~K, $e{\rm log}\,g < 0.25$, $e {\rm [Fe/ H]}< 0.20$ and $e \xi <$ 0.20 $km~ s^ {-1}$. We also apply a further cut in abundance errors, considering only those values that have an $e A(El) < 0.1$.

The  selection function adopted in the {\em Gaia}-ESO survey for UVES observations \citep[see][]{Stonkut2016MNRAS.460.1131S} favours the main sequence turn-off (MSTO),  which constitute the majority of the sample. 
By construction, the sample of field stars has a limited extent in Galactocentric distances, and is, therefore,  used in our analysis for comparison with the model in the [X/Fe] vs [Fe/H] plane.

\subsubsection{Classical Cepheids in Da Silva et al. (2023)}

\citeauthor{DaSilva23} (\citeyear{DaSilva23}, hereafter \citetalias{DaSilva23}) provided the largest (1118 spectra, 356 objects) and most homogeneous spectroscopic sample for Galactic CCs with measured metallicity from optical high-resolution, high-S/N spectra.
The sample is distributed across the four Galactic quadrants and it ranges from the inner (${\rm R}\sim 5$ kpc) to the outer (${\rm R}\sim 25$ kpc) disk. For the distances, measurements were based either on trigonometric parallaxes from Gaia DR3 or on near-infrared period-luminosity relations (\citealt{Ripepi22}).
Due to the steadily variation in target's physical properties due to their natural radial oscillations, special care was dedicated to the estimate of the different atmospheric parameters ($T_{eff}$, $\log g$, $\xi$), which were verified using different approaches and/or diagnostics (see also \citealt{DaSilva22}). 

In this work, we take advantage of the CCs sample from \citetalias{DaSilva23} to have an additional observational probe for present-day metallicity gradients, in addition to the young OCs from \citetalias{Magrini23}.

\section{Chemical evolution of the Milky Way disk}
\label{s:models}

\begin{table*}
    \centering
    \caption{Summary of the main parameters of the two-infall model (2INF) adopted in this study.}
    \begin{tabular}{c c c c c c c c c c}
        \hline\\[-1.95ex]
         &$t_{max,1}$ & $\tau_1$ & $\tau_2$ & $\nu_1$ & $\nu_2$ & $\Sigma_1$ & $\Sigma_2$ & v$_{flows}$\\
         &(Gyr) & (Gyr) & (Gyr) & (Gyr$^{-1}$) & (Gyr$^{-1}$) &  &  & (km s$^{-1}$) \\[0.1cm]
         \hline\\[-1.95ex]
         2INF & 3.25  & 1  & $1.033\times$R(kpc)$-1.26$ & 2 & 5 (4 kpc) - 0.4 ($>$12 kpc) & $\propto e^{-{\rm R}/2.3}$  &  $\propto e^{-{\rm R}/3.5}$   &  -1 \\[0.1cm]
         \hline
    \end{tabular}\\[0.2cm]
    {\bf Notes.} All the above parameters are the same to the ones adopted in \citetalias{DaSilva23}. The negative sign on the radial flow speed indicates inward flows.
    \label{tab:2inf}
\end{table*}

In this Section, we present the main assumptions and features of the multi-zone chemical evolution models adopted in this work. 
In particular, in \ref{ss:2inf} we provide the details of the revised two-infall model proposed by \citet{Palla20} (see also \citetalias{DaSilva23}), whereas in \ref{ss:3inf} we describe the details of the three-infall framework from \citet{Spitoni23}, which we expand throughout this work.\\

For both of the chemical evolution models listed above, the basic equations that describe the chemical evolution of a given element $i$ are:
\begin{equation}
    \Dot{G}_i ({\rm R}, t) = -\psi({\rm R}, t)\, X_i({\rm R}, t)\, + \,R_i({\rm R}, t)\, +\, \Dot{G}_{i,inf}({\rm R}, t)\, +\, \Dot{G}_{i,{\rm R_f}},
    \label{eq:basic_chemevo}
\end{equation}
where $G_i({\rm R}, t) = X_i({\rm R}, t)\,G({\rm R}, t)$ is the fraction of the gas mass in the form of an element $i$ and $G({\rm R}, t)$ is the fractional gas mass.
$X_i({\rm R}, t)$ represents the abundance fraction in mass of a given element $i$, with the summation over all elements in the gas mixture being equal to unity.

The first term on the right-hand side of Eq. \eqref{eq:basic_chemevo} corresponds to the rate at which an element $i$ is removed from the ISM due to star formation. The star formation rate (hereafter, SFR) is parametrised according to the Schmidt-Kennicutt law (\citealt{Kennicutt98}):
\begin{equation}
    \psi({\rm R}, t) = \nu \, \Sigma_{gas}({\rm R}, t)^k,
    \label{eq:SFR}
\end{equation}
where $\Sigma_{gas}$ is the surface gas density, $k = 1.5$ is the law index and $\nu$ is the star formation efficiency (SFE). 

$R_i({\rm R, t})$ (see \citealt{Palla20b} for the complete expression) takes into account the nucleosynthesis from low-intermediate mass stars (LIMS, $m < 8 {\rm M_\odot}$), core collapse (CC) SNe (Type II and Ib/c, $m > 8 {\rm M_\odot}$) and Type Ia SNe. 
For these latter, we assume the single-degenerate scenario and in particular the delay-time-distribution (DTD) by \citet{MatteucciRecchi01}. This choice can be considered a good compromise to describe the delayed pollution from the entire Type Ia SN population as it enables us to obtain abundance patterns that are similar to those obtained with other DTDs (see \citealt{Palla21} for details).
$R_i({\rm R}, t)$ output is also weighted by the initial mass function (IMF). Here, we adopt the IMF by \citet{Kroupa93}, which is preferred to reproduce the characteristics of the MW disk \citep{Romano05}.

The last term of Eq. \eqref{eq:basic_chemevo} refers to radial inflows of gas, which here are implemented following  \citet{Portinari00} (see also \citealt{Palla20} for a detailed description of the implementation).
In our models, we use a constant speed pattern, i.e. with $v_{flow}= 1\, {\rm km s^{-1}}$ across all radii, as suggested by \citet{Palla20}. 
Low speeds, i.e. small radial inflow motion, are also suggested by previous chemical evolution studies (e.g. \citealt{Bilitewski12,Mott13,Vincenzo20}) as well as observations of external galaxies (e.g. \citealt{Wong04,DiTeodoro21}).

In general, we ignore the effect of Galactic winds on chemical evolution of the MW disk. In fact, by studying the Galactic fountains originated by the explosions of Type II SNe in OB associations, \citet{Meioli08,Meioli09} and \citet{Spitoni09} found that metals fall back to approximately the same Galactocentric region  from where they were ejected. 
Moreover, \citet{Spitoni09} computed the typical timescale of the fallback of this material, finding a value of 0.1 Gyr. 
These results were also later supported by cosmological simulation of galactic discs of virial mass $> 10^{11}$ M$_\odot$, i.e. encompassing the MW, which showed that the majority of the mass ejected by the disc is reaccreted on short timescales and close to the ejection location (e.g. 
\citealt{Hopkins23} and references therein).
Therefore, galactic winds are likely to produce a negligible effect on the global chemical evolution of the Galaxy.

\subsection{The two-infall model}
\label{ss:2inf}

In the two-infall model formalism, the model assumes two consecutive gas accretion episodes feeding the MW disk, forming the so-called high-$\alpha$ and low-$\alpha$ sequences observed in the Galactic disk. Therefore, the third term on the right-hand side of Eq. \eqref{eq:basic_chemevo} can be expressed in this way:
\begin{multline}
    \Dot{G}_{i,inf}({\rm R},t)=A({\rm R})\,X_{i,1inf}({\rm R})\,e^{-\frac{t}{\tau_1}} +\\ +\theta(t-t_{max,1})\, B({\rm R}) \, X_{i,2inf}({\rm R})\, e^{-\frac{t-t_{max,1}}{\tau_2}},
    \label{eq:infall_2inf}    
\end{multline}
where $G_{i,inf}({\rm R},t)$ is the infalling material in the form of element $i$ and $X_{i,Jinf}$ is the abundance of the same element for the $J$-th infall.  $\tau_1$ and $\tau_2$ are the timescales of the two infall  episodes,  while $t_{max,1}$ indicates  the  time  of  maximum infall,  which  is  also  the  delay  between  the  first  and  the  second infall episodes.  $\theta$ is the Heavyside step function, while $A({\rm R})$ and $B({\rm R})$ coefficients are set to reproduce the present-day total surface mass density of the high-$\alpha$ and low-$\alpha$ disks at different Galactocentric radii.
These latter are assumed to have exponential profiles, in this way:
\begin{equation}
     \frac{\Sigma_J({\rm R})}{{\rm M_\odot\, pc^{-2}}} \, = \, \Sigma_{0,J} \,  e^{-{\rm R}/{\rm R}_{d,J}},
    \label{eq:thin_surf}
\end{equation}
where the disk scale length ${\rm R}_{d,J}$ is 3.5 kpc for the low-$\alpha$ disk ($J=2$) and 2.3 kpc for the high-$\alpha$ one ($J=1$, see \citealt{Palla20} and references therein). The quantities $\Sigma_{0,J}$ are tuned to reproduce the total surface mass density in the solar neighbourhood as provided by \citet{McKee15} of $47.1 \pm 3.4$ M$_\odot$ pc$^{-2}$.\\

In this work, we take advantage of the model prescriptions adopted in \citet{Palla20} and later revised in \citetalias{DaSilva23}.
The model is thus a revised version of the two-infall paradigm (see also \citealt{Spitoni19}) in which two consecutive gas accretion episodes are separated by a delay of $t_{max,1}=3.25$ Gyr. 
The fairly large delay relative to the "classical" two-infall paradigm (of 1 Gyr, see \citealt{Chiappini97,Romano10}), allow us to reproduce [$\alpha$/Fe] vs. [Fe/H] abundance diagrams (e.g. \citealt{Palla20,Spitoni21}) throughout the MW disk as well as stellar ages trends (e.g. \citealt{Spitoni19,Nissen20}) in the solar vicinity. 

Going into more detail, in the first infall (forging the high-$\alpha$ sequence) the timescale of gas accretion is fixed at $\tau_1=$1 Gyr at all radii, with also a fixed SFE of $\nu=2$ Gyr$^{-1}$.
For the second gas-infall episode (forming the low-$\alpha$ sequence), the timescale for gas accretion $\tau_2$ increases linearly with radius according to the inside-out scenario (following \citealt{Romano00,Chiappini01} law).
In order to reproduce the slope observed in radial abundance gradients of CCs in \citetalias{DaSilva23}, as well as the  gradients in other physical quantities such as SFR and gas density (e.g. \citealt{Staler05,Nakanishi03,Nakanishi06}), the SFE of the second infall episode is variable depending on the galactocentric radius, with values between $\nu = 5$ Gyr$^{-1}$ (at R$=4$ kpc) and $\nu = 0.4$ Gyr$^{-1}$ (from R$>12$ kpc). In addition, inward radial gas flows with a constant velocity of $v_{flow}=1\, {\rm km \,s^{-1}}$ are also needed to reproduce the gradients, as discussed above.

A summary of all the main parameters adopted in this work for the two-infall model are listed in Table \ref{tab:2inf}.

\subsection{The three-infall model}
\label{ss:3inf}

\defcitealias{Spitoni23}{Sp23}

In their recent work, \citeauthor{Spitoni23} (\citeyear{Spitoni23}, hereafter \citetalias{Spitoni23}) proposed a new chemical evolution framework for the Galactic disk components, constrained by the star formation histories inferred from CMD analyses in Gaia DR1 and DR2 (\citealt{Bernard17,Ruiz20}). 
These works revealed enhanced SF activity within the last 2-3 Gyrs. This is mimicked in the chemical evolution model by including a recent (age $<3$ Gyr) gas infall episode, which triggers this enhanced SF at late times. 
In this way, \citetalias{Spitoni23} were also able to reproduce the new abundance ratios provided by the General Stellar Parametriser-spectroscopy module for the Gaia DR3 (\citealt{Gaia23,RecioBlanco23}), which show a chemical impoverishment in the young population of stars in the solar neighbourhood.

In this work, we extend the approach presented in \citetalias{Spitoni23} to the whole MW disk, in order to investigate abundance gradients.\\

In this model, the functional form of the gas infall rate is:
\begin{multline}
    \Dot{G}_{i,inf}({\rm R},t)=
    A({\rm R})\,X_{i,1inf}({\rm R})\,e^{-\frac{t}{\tau_1}}\, +\\ 
    \hspace{0.9cm} +\, \theta(t-t_{max,1})\, B({\rm R}) \, X_{i,2inf}({\rm R})\, e^{-\frac{t-t_{max,1}}{\tau_2}} \, +\\ 
    +\,  \theta(t-t_{max,2})\, C({\rm R}) \, X_{i,3inf}({\rm R})\, e^{-\frac{t-t_{max,2}}{\tau_3}},
    \label{eq:infall_3inf}  
\end{multline}
where $\tau_3$ is the timescales of the third gas accretion episode, $t_{max,2}$ is the Galactic time associated to the start of the third infall, and $C({\rm R})$ is the coefficient to reproduce the present-day total surface density of the third accretion episode $\Sigma_3$. Here, the sum between the latter and the total surface density of the second infall ($\Sigma_2$) is equal to the density profile as described in \ref{ss:2inf} for the low-$\alpha$ disk in the two-infall model.
All the other variables are as in Eq. \eqref{eq:infall_2inf}.
In fact, the new model uses the framework presented in \ref{ss:2inf}, but split the low-$\alpha$ sequence into two distinct gas accretion episodes in order to mimic the recent enhanced SF activity. 
Therefore, we will leave unchanged all the parameters adopted for the first two infall episodes, such as the infall timescales and the SFEs at different Galactocentric radii, as well as the IMF (from \citealt{Kroupa93}).

For what concern the third additional gas accretion, instead, we set its starting time at $t_{max,2}=11$ Gyr (as in \citetalias{Spitoni23}) for all the Galactocentric radii. For the other parameters, i.e. the SFE $\nu_3$, the timescale of infall $\tau_3$ and the total surface mass density accreted by the third gas infall, 
we test different parametrisations, which are listed in Tab. \ref{tab:3inf_param}:
\begin{enumerate}
    \item First, we adopt a setup very similar to the one proposed by \citetalias{Spitoni23} (3INF-1), which is shown in Tab. \ref{tab:3inf_param} upper row. Here, the parameters of the third infall, i.e. $t_{max,2}$, $\tau_3$, $\Sigma_2$/$\Sigma_3$ are the same of the latter paper. Concerning the SFE $\nu_3$, rather than fixing the value adopted in \citetalias{Spitoni23}, we fix the proportion between the SFE in the second and third infall episode to be similar to the one used in that paper;
    \item to test further the viability of the three-infall scenario in the context of radial gradients, we allow to vary the third infall parameters relative to the values adopted in the model 3INF-1. 
    While leaving constant the starting time of the third infall (to be consistent with SF peaks as show by, e.g., \citealt{Ruiz20}) we act on all the other physical parameters, i.e. the infall timescale $\tau_3$, the SFE $\nu_3$ and for the ratio between the baryonic mass accreted by the second and third infall $\Sigma_2$/$\Sigma_3$. The parameter for this setup (3INF-2) are shown in Tab. \ref{tab:3inf_param} bottom row. 
\end{enumerate}

\begin{table}[]
    \centering
    \caption{3rd infall parameters adopted in the models in this paper.}
    \begin{tabular}{c c c c c}
        \hline\\[-1.95ex]
         &$t_{max,2}$ & $\tau_3$ & $\nu_3$ & $\Sigma_2$/$\Sigma_3$ \\
         &(Gyr) & (Gyr) & & \\[0.1cm]
         \hline\\[-1.95ex]
         3INF-1 & 11    &   0.15    &       (1/3)$\times \nu_2$    &       2.33\\[0.115cm]
         3INF-2 & 11    &   1    &       (2/3)$\times \nu_2$     &     10(4 kpc)-3.5(>12 kpc)\\
         [0.1cm]
         \hline
    \end{tabular}\\[0.2cm]
    {\bf Notes. } For 3INF-1, $\tau_3$ and $\Sigma_2$/$\Sigma_3$ parameters are from \citet{Spitoni23}. For $\nu_3$, we consider a similar proportion to the one used by \citet{Spitoni23} for the solar vicinity, i.e. R$=8$ kpc.
    \label{tab:3inf_param}
\end{table}

\subsection{Nucleosynthesis prescriptions}
\label{ss:nucleo}

The nucleosynthesis prescriptions and the implementation of the stellar yields are fundamental ingredients for chemical evolution models. 
LIMS, massive stars and Type Ia SNe play a fundamental role in shaping the [X/Fe] vs. [Fe/H] abundance patterns as well as radial abundance gradients of the elements of the Periodic Table. 

In this work, we mainly adopt the prescriptions listed below:
\begin{itemize}
    \item for LIMS, we use the yield set from \citet{Ventura13,Ventura18,Ventura20}, also comprising the domains of super-AGB stars ($6<m/$M$_\odot<8-9$) and supersolar metallicities.
    
    \item for massive stars we use the stellar yields from \citet{Limongi18}. In particular, we adopt the "mixed $v_{rot}$\footnote{in \citet{Limongi18}, different yield grids considering different stellar rotation velocities $v_{rot}$ were built.} set", i.e. the one used in the best model (MWG-12) in \citet{Romano19} for the MW disk. 
    
    \item for Type Ia SNe we adopt the stellar yields from \citet{Iwa99} (W7 model) which are extensively used in chemical evolution literature (e.g. \citealt{Romano10,Prantzos18,Palla20} among others).
\end{itemize}

However, underlying uncertainties in stellar evolution and nucleosynthesis theory may limit our chances of getting a firm grasp on the evolutionary scenario for the Galaxy. 

For this reason, we also test other stellar yields for massive stars and Type Ia SNe. 
In particular, we run additional models taking advantage of the stellar yields from \citet{Koba06,Koba11} for massive stars. Concerning Type Ia SNe, instead, we also adopt either yields from \citet{Leung18} (benchmark model) or \citet{Leung20} (bubble detonation pattern model), which represent some of the most recent Type Ia SN models for different progenitor classes, i.e. near-Chandrasekhar mass white dwarfs (near-$M_{Ch}$, \citealt{Leung18}) and sub-Chandrasekhar mass white dwarfs (sub-$M_{Ch}$, \citealt{Leung20}, see \citealt{Koba20,Palla21} for more details).

\subsection{Accounting for migration and observational uncertainties}
\label{ss:migr_err}

The overall picture on MW shows also evident signatures of stellar migration, both on the theoretical and observational sides (e.g. \citealt{Schonrich09,Minchev11,Minchev18,Kordopatis15}), even when specifically focusing on OCs (e.g. \citealt{Anders17,Spina21,Myers22}).
Therefore, we include in the models stellar radial migration prescriptions from the literature to account for this phenomenon. In particular, we implement migration in the chemical evolution model by adopting the approach already tested in \citet{Palla22} for MW disk stars.
It is worth noting that OCs are more massive than single stars, and therefore the effect of the interactions with perturbing structures should be in principle less pronounced than for field stars (e.g. \citealt{Zhang21}; \citetalias{Magrini23}), especially for young-intermediate ages (see \citealt{Viscasillas23}). However, a model for radial migration valid for field stars is appropriate within our work, as it gives at least a robust upper limit on the effect of radial migration on OCs, which is a necessary mechanism to properly explain their radii, abundances and ages trends in the Galactic disk.

In the following, we provide some details on the implementation, which is extensively described in \citet{Frankel18}.
Here, migration is seen as a result of a diffusion process, which is the effect of repeated and transient torques on stars by features such as spiral arms or a bar, and it is treated in a parametrical way.
Following \citet{Sanders15} and adapting their parameterization to a Galactocentric radius coordinate, the probability for a star to be currently at a Galactocentric radius ${\rm R}_f$, given that it was born at a radius ${\rm R}_0$ and at a certain age can be written as:
\begin{equation}
    \ln p({\rm R}_f\, |\, {\rm R}_0, Age) = \ln(c_3) \, -\frac{({\rm R}_f- {\rm R}_0)^2}{2 \, \sigma_{RM} \, (Age/10\, {\rm Gyr})} ,
\end{equation}
where $\sigma_{RM}$ is the radial migration strength and $c_3$ a normalization constant ensuring that stars do not migrate to negative radii.
For $\sigma_{RM}$ we adopt a value of 3, as found by \citet{Frankel20} as a result of their Bayesian fitting procedure of APOGEE red clump stars in the low-$\alpha$ disk.
The value above mentioned refers to the churning\footnote{for churning (also known as cold torquing), we mean the change in orbit angular momentum that change orbit's radius. For blurring, we intend the increase the radial oscillation amplitudes around the guiding radius of the orbit.} strength, as we already account for blurring effect by considering OCs guiding radius (R$_{guide}$) rather than their present-day Galactocentric radius R$_{GC}$.

It is worth noting that the migration framework adopted is in 1D, i.e. it is not considering azimuthal variations in migration strength. In this way, it can be fully integrated within the chemical evolution models described previously in this Section. The 1D assumption is robust, as various works showed rather small azimuthal abundance variations in galaxies ISM (e.g. \citealt{Kreckel19}).
Another possible limitation of the adopted migration model is the assumption of no radial or temporal dependence on the migration strength. However, including these dependencies means adding further and uncertain assumptions on the inventory of speeds and strengths of spiral and bar patterns during Galactic evolution (see \citealt{Frankel18}). In any case, the calibration of the model parameters on a sample of low-$\alpha$ disc stars extending up to radii $\simeq 15$ kpc (\citealt{Frankel20}) allows for a radial and temporally averaged estimate of migration in a radial range that comprises all the OCs described in Section \ref{s:data}, except one (Br29).\\

We also account for the effect of observational uncertainties on the chemical abundances of the predicted stellar populations in our model.  

In particular, we add at each Galactic time $t$ a random error to the abundances of the stars formed at $t$ (see also \citealt{Spitoni19,Palla22}). 
In this way, we have for each chemical element a "new abundance", defined as:
\begin{equation}
    [{\rm X/H}]_{new}(t)=[{\rm X/H}](t)+\mathcal{N}([{\rm X/H}],\sigma_{[{\rm X/H]}}),
    \label{eq:new_abund}
\end{equation}
where $\mathcal{N}$ is a random function with Normal distribution.
In order to have fair comparison with the \citetalias{Magrini23} data set adopted in this study, the standard deviation $\sigma_{[{\rm X/H]}}$ corresponds to average spread observed within the OCs in the age intervals of interest, i.e. $0.1<Age$/Gyr$<1$, $1<Age$/Gyr$<3$, $3<Age$/Gyr$<7$.

\section{Results}
\label{s:results}

In this Section, we show the results of the comparison between the models presented in Section \ref{s:models} and the \citetalias{Magrini23} OC sample.
In particular, in \ref{ss:results_2inf} we look at the prediction of the two-infall model. 
In \ref{ss:results_3inf} we instead show the gradients obtained by means of the three-infall scenario in the light of the data at disposal, discussing such a scenario in \ref{ss:discuss}.

\subsection{Comparing observed gradients with the standard two-infall scenario}
\label{ss:results_2inf}

\begin{figure}
    \centering
    \includegraphics[width=0.95\columnwidth]{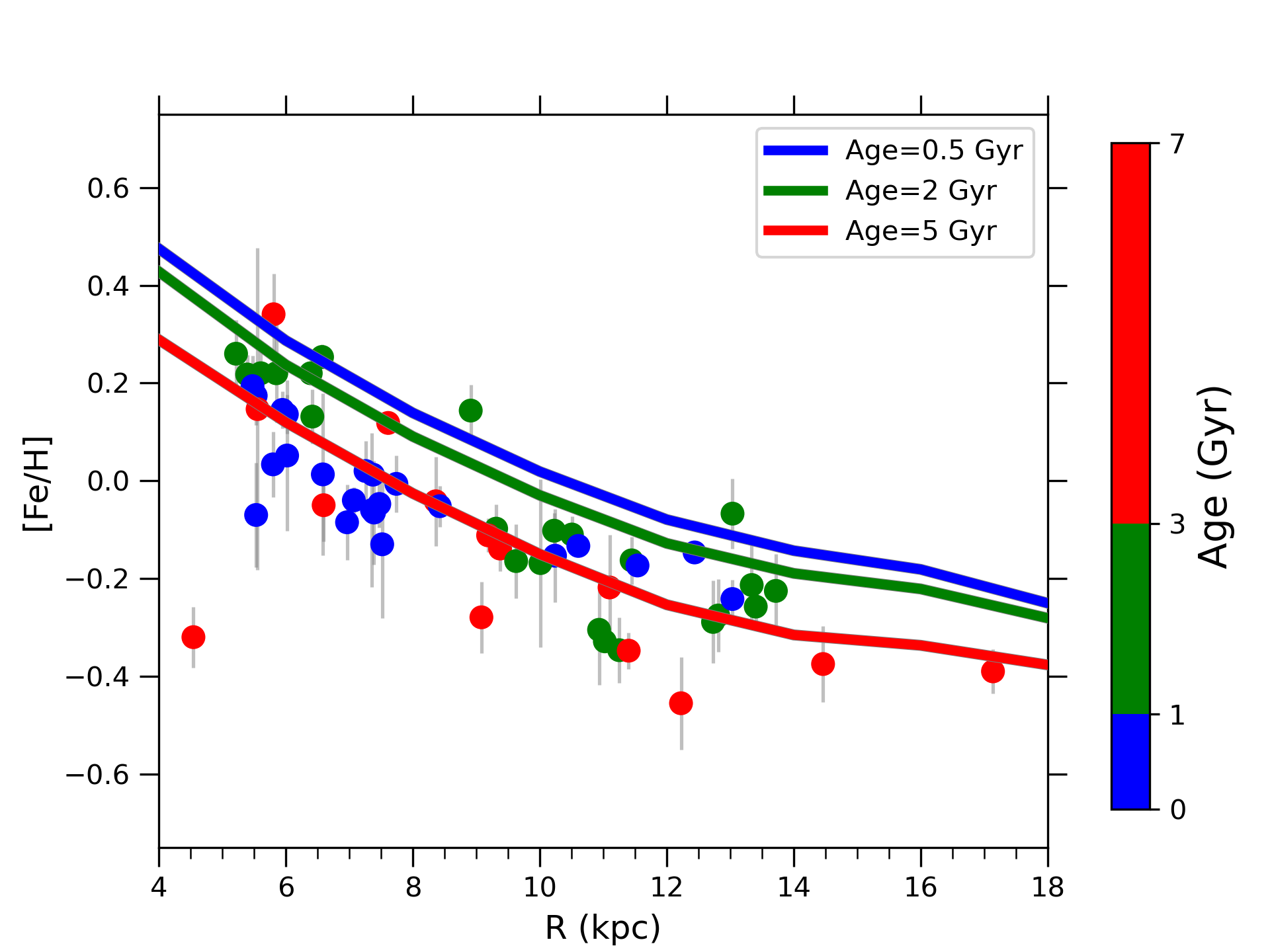}
    \caption{Time evolution of the radial [Fe/H] gradient as predicted by the two-infall model. Filled circles with errorbars are the OC sample by \citet{Magrini23}, which are divided in three age bins: young ($Age< 1$ Gyr, blue points), intermediate ($1 < Age$/Gyr$ < 3$, green points) and old ($Age > 3$ Gyr, red points). Solid lines are the results for the [Fe/H] gradient as predicted by the two-infall model at 0.5 Gyr (blue lines), 2 Gyr (green lines) and 5 Gyr (red lines). In this plot, and in the following Figures, we use the guiding radius of the orbit, computed as the average between the Apogalacticon and Perigalacticon radii (see \ref{ss:GES_OCs}), as an indication of the location of each cluster in the disk.}
    \label{fig:gradient_2inf}
\end{figure}

In Fig. \ref{fig:gradient_2inf}, we show the time evolution of the [Fe/H] as predicted by the two-infall model and compared to the OC data presented in \citetalias{Magrini23}. 
From this sample, we remove the Blanco 1 cluster due to its extremely high internal spread observed in metallicity ($\sigma>0.5$ dex), which may hide problems in abundance derivation of their members.
Coming back to Fig. \ref{fig:gradient_2inf}, we see that the model captures the general trend of the data: the gradient slope clearly decreases going towards larger radii, in agreement with the trend shown by OCs. This result confirms the conclusion by \citetalias{DaSilva23} of a flattening of the gradient at R$\gtrsim 12$ kpc, which requires a flat behavior of the SFE at large Galactocentric radii.

However, if we focus on different age bins, we note that OCs with $Age<1$ Gyr (blue points) are clearly below the prediction from the model. Moreover, we note that the metallicity of these OCs is lower than that of older clusters, i.e. the ones with $1<Age$/Gyr$<3$ (green points). Therefore, such a decrease in metallicity with decreasing age cannot be reproduced by genuine chemical evolution model predictions in a scenario of continuous SF, as the one of the two-infall model in the age ranges investigated in this study: subsequent stellar generations are progressively enriching the ISM in metals, as demonstrated by model lines in Fig. \ref{fig:gradient_2inf}.\\ 

\begin{figure*}
    \centering
    \includegraphics[width=0.33\textwidth]{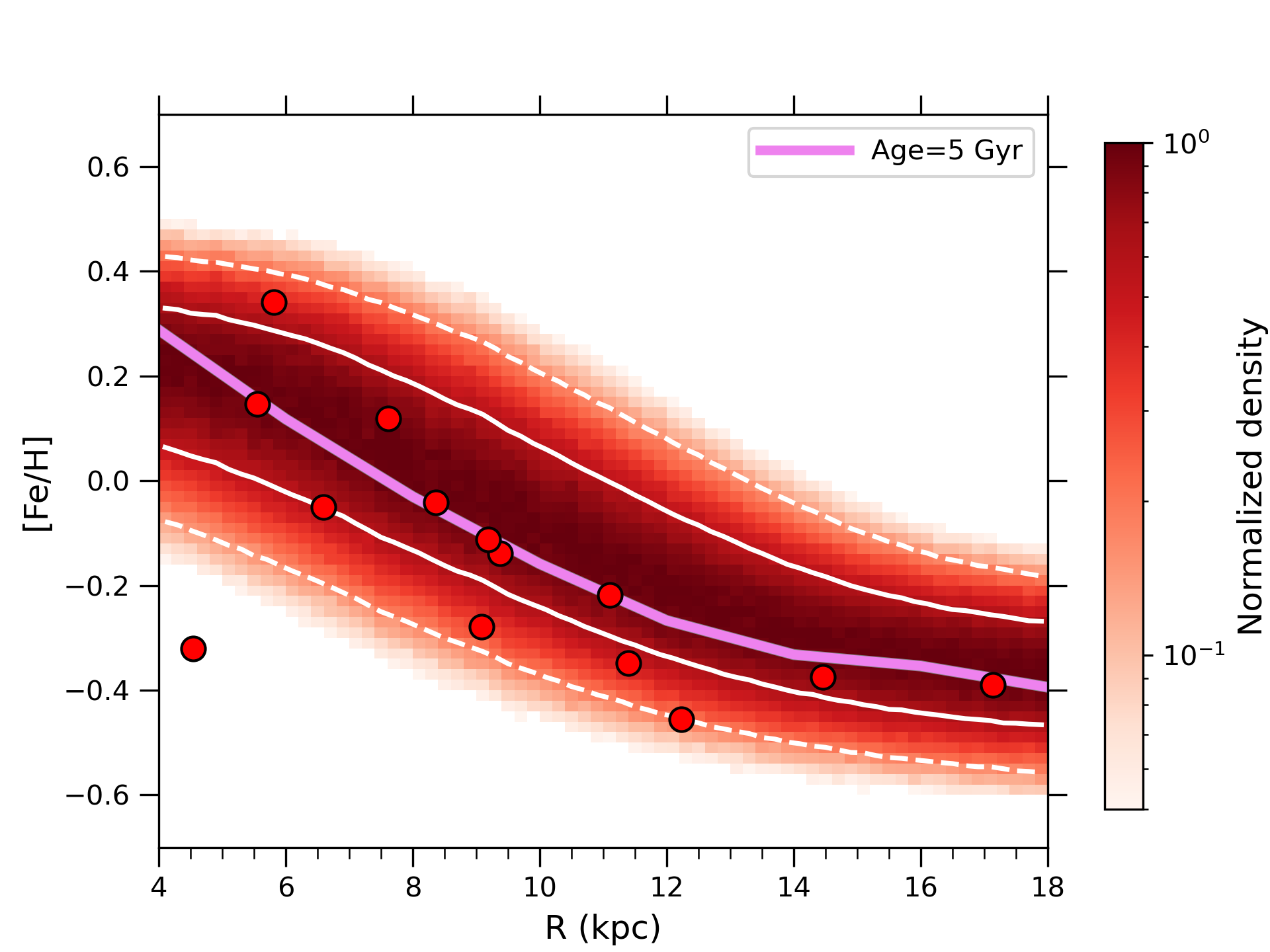}
    \includegraphics[width=0.33\textwidth]{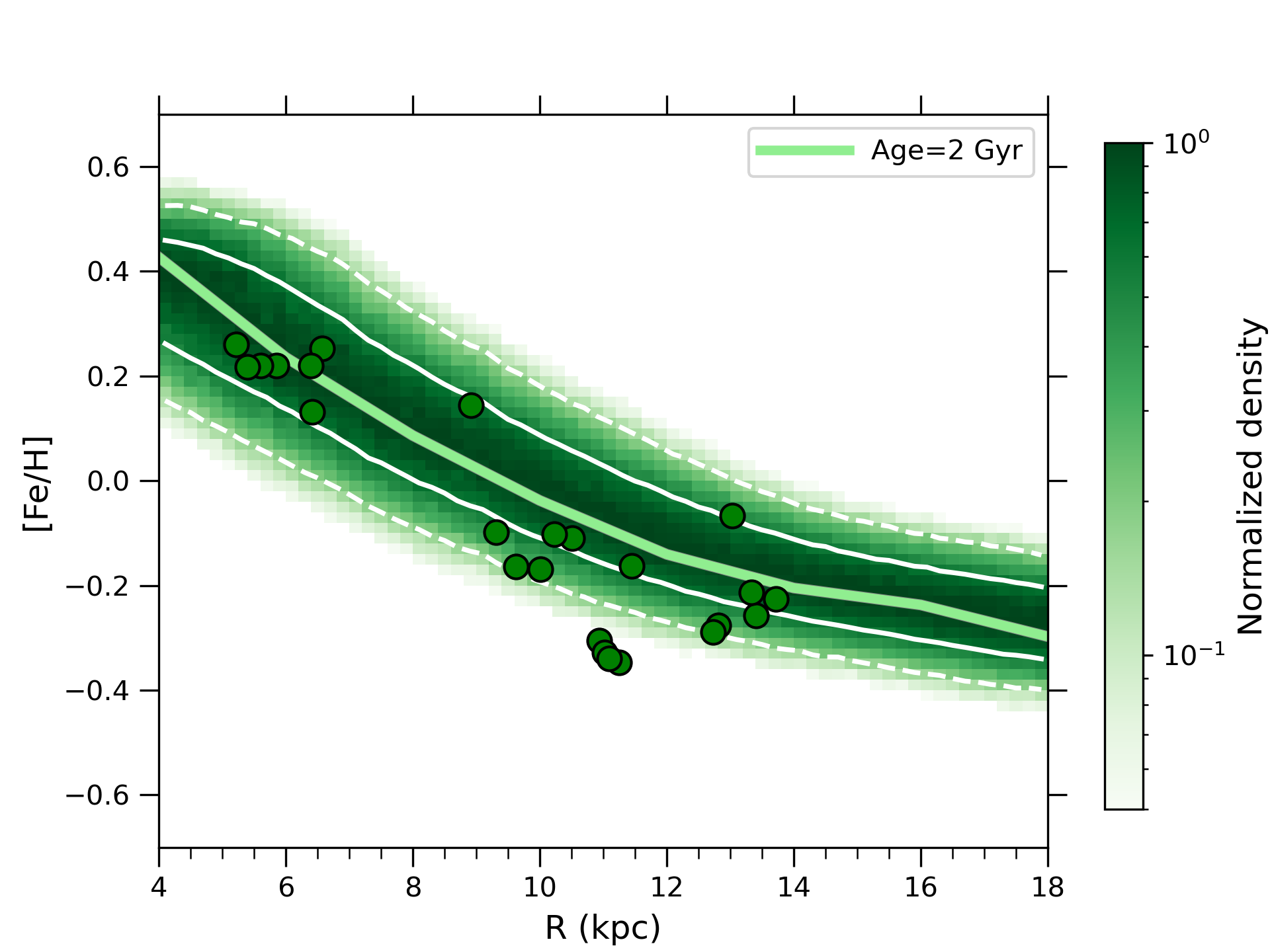}
    \includegraphics[width=0.33\textwidth]{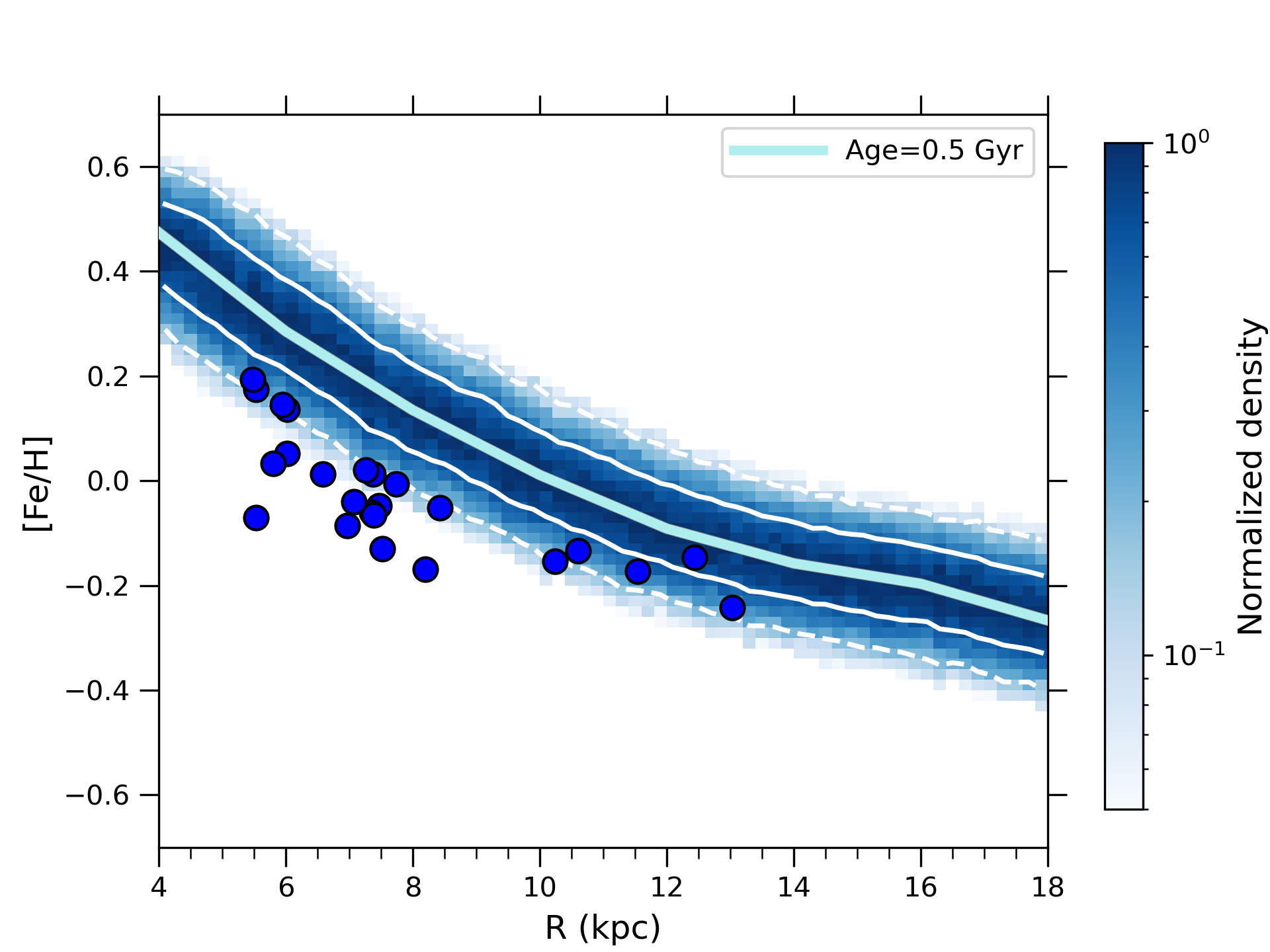}
    \caption{Time evolution of the radial [Fe/H] gradient as predicted by the two-infall model at $Age>3$ Gyr (left panel), $1<Age$/Gyr$<3$ (central panel), and $Age<1$ Gyr (right panel) for the two-infall model, including stellar migration and abundance uncertainties (see \ref{ss:migr_err}).
    Density plots show the normalised density (in $\log$ scale) of stars as predicted by the model in given Galactocentric bin of 0.2 kpc width. 
    White contour lines show the limits within are contained the 68\% (solid) and the 95\% (dashed) of the predicted stellar distribution in a given radial bin.
    Solid lines show the results for the [Fe/H] gradient as predicted by the genuine chemical tracks, i.e. the ones shown in Fig. \ref{fig:gradient_2inf}, at Ages=0.5 Gyr (cyan line), 2 Gyr (green line) and 5 Gyr (magenta line). 
    Data are the same as in Fig. \ref{fig:gradient_2inf}}
    \label{fig:gradient_2inf_evo}
\end{figure*}

For this reason, we try to explore if the inclusion of effect of stellar migration and abundance uncertainties/spread within OCs (see \ref{ss:migr_err} for details) may reconcile predictions and observations. The results are shown in Fig. \ref{fig:gradient_2inf_evo}. 
Here, the density plots represent the probability of finding a star with an abundance [X/H] at a Galactocentric radius R in a certain age bin, i.e. $Age>3$ Gyr (left panel) $1<Age$/Gyr$<3$ (central panel) and $Age<1$ Gyr (right panel), according to the predictions by the two-infall model. 
In order to highlight the spread caused by the inclusion of the different effects, the density plot in Fig. \ref{fig:gradient_2inf_evo} (as well as subsequent Figures) is represented in $\log$ scale.
By looking at the left and central panels of the Figure, we can see that the model predictions generally well capture the observed slope and spread within \citetalias{Magrini23} OC sample, with just the exception of few objects with $10\lesssim$R/kpc$\lesssim12$ in the intermediate age bin. On the other hand, the right panel clearly shows that young clusters are clearly overestimated by model predictions at different radii.
This is also demonstrated by the plotted solid and dashed contour lines, showing the limits in which are contained the 68 and 95\% of the predicted stars, respectively. Almost all the observed clusters are in fact outside the 2$\sigma$ of the distribution.\\

However, as noted by \citetalias{Magrini23}, abundances in stars with lower surface gravity $\log g$ and higher mictroturbolence parameter $\xi$ are susceptible to artifacts in stellar spectral analysis (see \ref{ss:restricted_sample}). In turn, this may affect the reliability of the obtained OC abundances especially for young clusters, which are more prone to contain such stars.
Therefore, we decide to remove stars with $\log g < 2.5$ and $\xi > 1.8$ km s$^{-1}$, as suggested by \citetalias{Magrini23}, to minimise the possible bias introduced in stellar abundance determination within clusters. Despite such a choice costs a not negligible number of OCs for the data-model comparison, this is the only way we can assure robust estimations for different chemical abundances.

\begin{figure*}
    \centering
    \includegraphics[width=0.33\textwidth]{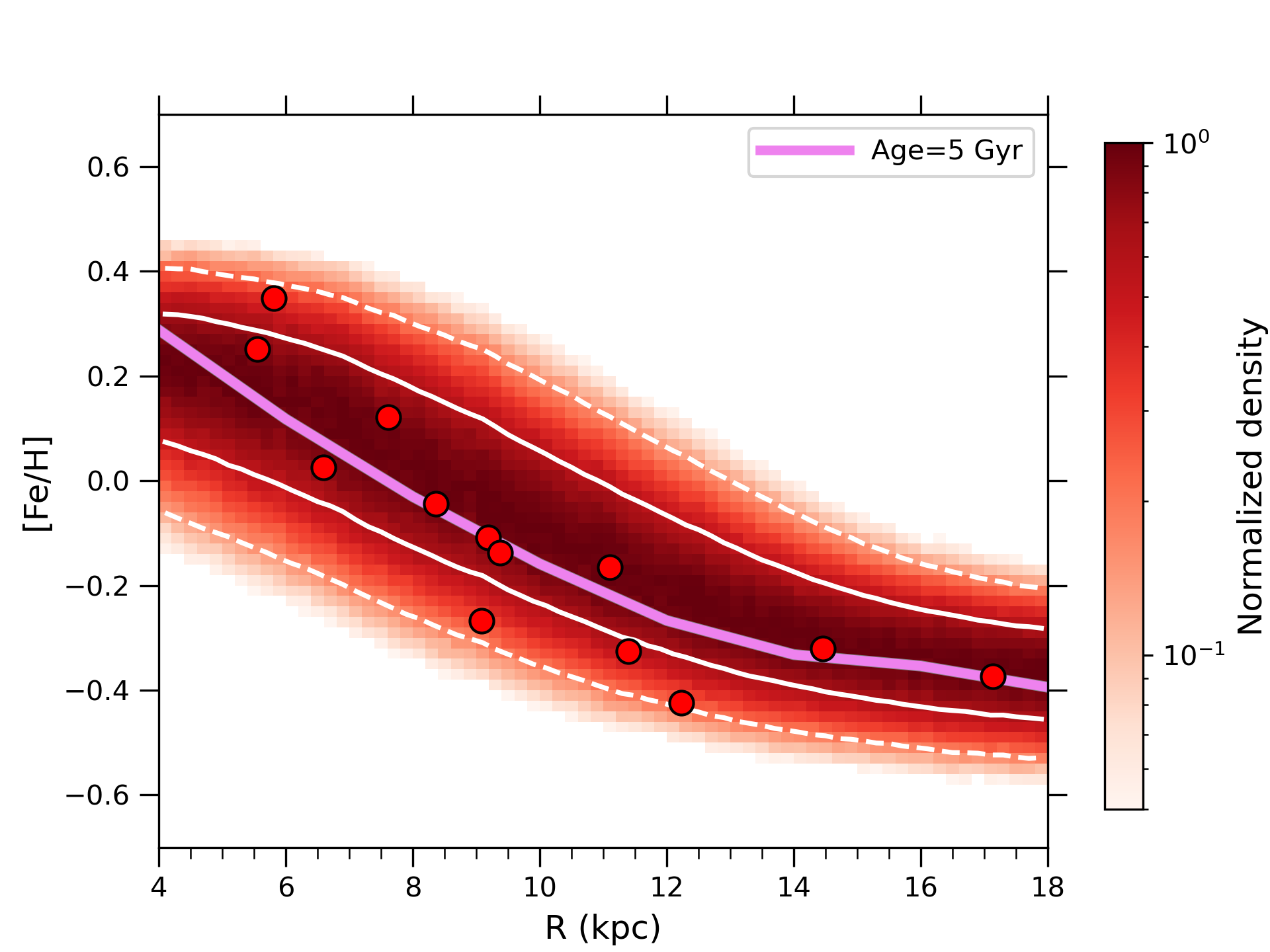}
    \includegraphics[width=0.33\textwidth]{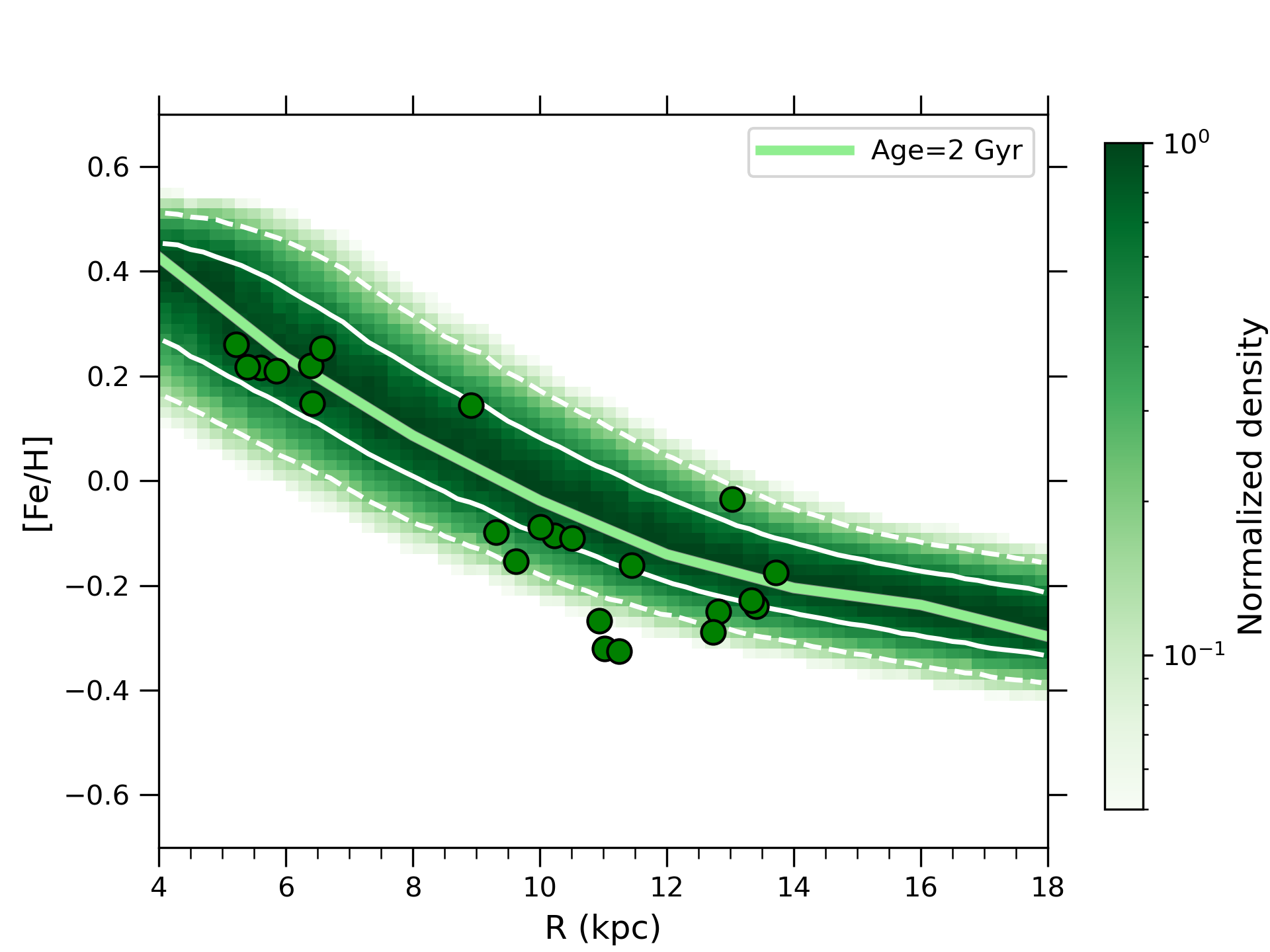}
    \includegraphics[width=0.33\textwidth]{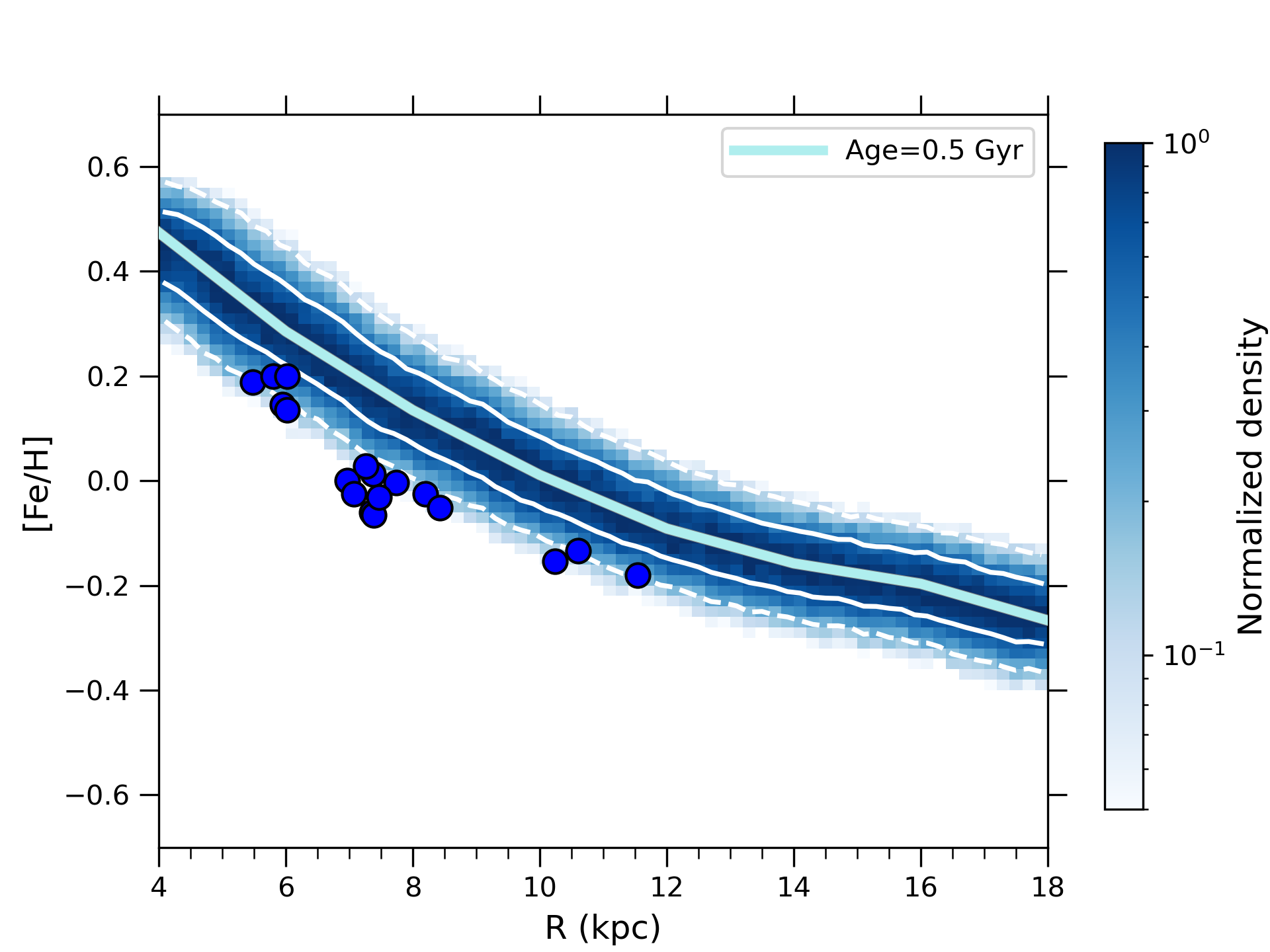}
    \caption{Same as Fig. \ref{fig:gradient_2inf_evo},  but with data restricted to OCs members with $\log g>$2.5 and $\xi < 1.8$ km s$^{-1}$ (restricted sample).}
    \label{fig:gradient_2inf_restricted}
\end{figure*}

In Fig. \ref{fig:gradient_2inf_restricted}, we show the comparison between the models already shown in Fig. \ref{fig:gradient_2inf_evo} and the \citetalias{Magrini23} OC sample with imposed cuts on individual stellar parameters (hereafter, restricted sample).
As mentioned above, the old and intermediate age bins are barely affected by the cut in stellar parameters and therefore the data-model comparison the left and central panel is very similar to what already shown in Fig. \ref{fig:gradient_2inf_evo}. For the right panel, we see instead that the restricted sample shows in general larger metallicities than those observed in the full OC sample. Nonetheless, the observed [Fe/H] are still clearly overestimated by the prediction of two-infall model, with most of the clusters still outside the 2$\sigma$ of the predicted distribution.

This is also highlighted in Tab. \ref{tab:comparison_fits}, where we show intercepts and slopes of gradient linear fits for the full and restricted OC samples and the two-infall model predictions in different age bins. The fits are obtained in the radial range below R $<12$ kpc: this choice is done to avoid the influence of the gradient flattening at large radii, which is found to start around this Galactocentric distance (see, e.g. \citealt{Carraro07,Magrini17,Donor20}; \citetalias{DaSilva23}).
\begin{table*}
    \centering
    \caption{Comparison between OC sample by \citet{Magrini23} and two-infall model [Fe/H] gradient slopes and intercepts for R$<12$ kpc at different ages. 
    Data fits are shown for both the full OC sample and the one restricted to OCs members with $\log g>$2.5 and $\xi < 1.8$ km s$^{-1}$.} 
    \begin{tabular}{c c | c  c  c}
         \hline\\[-1.95ex]
         &  & Full sample & Restricted sample &  Two-infall model\\[0.2cm]
         \hline\\[-1.95ex]
         Young Ages &  Slope (dex kpc$^{-1}$) & $-0.0497 \pm 0.0096$ & $-0.0627 \pm 0.0078$ &  $-0.0689 \pm 0.0051$  \\
         ($<1$ Gyr) &  Intercept (dex) & $0.3477 \pm 0.0723$ &  $0.4866 \pm 0.0610$ & $0.7190 \pm 0.0432$ \\[0.1cm]
         \hline\\[-1.95ex]
         Intermediate Ages & Slope (dex kpc$^{-1}$) & $-0.0927 \pm 0.0087$ & $-0.0862 \pm 0.0085$ & $-0.0690 \pm	0.0052$ \\
         ($1<$ Gyr $<3$) &  Intercept (dex) & $0.7725 \pm 0.0776$ & $0.7332 \pm 0.0743$  &  $0.6714	\pm 0.0440$	\\[0.1cm]
         \hline\\[-1.95ex]
         Old Ages &  Slope (dex kpc$^{-1}$) & $-0.0479 \pm 0.0276$ & $-0.0968 \pm 0.0167$ &  $-0.0677 \pm 0.0037$ \\
         ($>3$ Gyr) &  Intercept (dex) & $0.3042 \pm 0.2300$ & $0.7841 \pm 0.1437$ &  $0.5376 \pm 0.0315$ \\[0.1cm]
         \hline
    \end{tabular}\\[0.1cm]
    \label{tab:comparison_fits}
\end{table*}
For the young age bin, the restricted sample shows a significant increase in the fit intercept relative to the full sample, but still this is around 0.2 dex lower than what predicted by the two-infall model. Therefore, even by considering stellar migration and abundance uncertainties effects, these are not sufficient to explain the low [Fe/H] imprinted in the data sample.

We also checked whether uncertainties on cluster ages can be a source of bias for such a result. 
We consider either the cases of (i) "internal" uncertainties for the age derivation method adopted in this work and (ii) adoption of different isochrones  grids and/or different photometric datasets to derive ages.
For (i), we perform 1000 Monte Carlo resamplings of the age of the clusters according to their uncertainties. We consider an age uncertainty of $\log(Age) = 0.2$ dex, as found in \citet{Cantat2020A&A...640A...1C} for clusters older than $\log(Age/$yr$)>8.5$ in their validation sample\footnote{ \citet{Cantat2020A&A...640A...1C} actually found an uncertainty range between 0.1 and 0.2 dex in $\log(Age)$, but we leave 0.2 dex also to account for the age precision in the adopted training set, i.e. of $\log(Age) = 0.1$ dex.}. Even by randomly perturbing the cluster ages according to the uncertainties, we still find a decrease $>0.15$ dex (precisely, from 0.70 to 0.52 dex) between the intercepts in the intermediate and young age bins relative to the ones shown in Table \ref{tab:comparison_fits} for the restricted sample. 
For (ii), instead, we refer to \citet{Jeffries23}, who compared different sets of age determinations for OCs in {\it Gaia}-ESO, including a large fraction of the OCs used in this work. We find that the difference between the mean literature ages (from \citealt{Jeffries23}) and the ages adopted in this work is always lower than the uncertainty of $\log(Age)=0.2$ dex adopted for our resampling test: therefore, the distribution of gradient in different age bins will remain similar even when changing the ingredients in the age derivation.\\

\begin{figure}
    \centering
    \includegraphics[width=0.95\columnwidth]{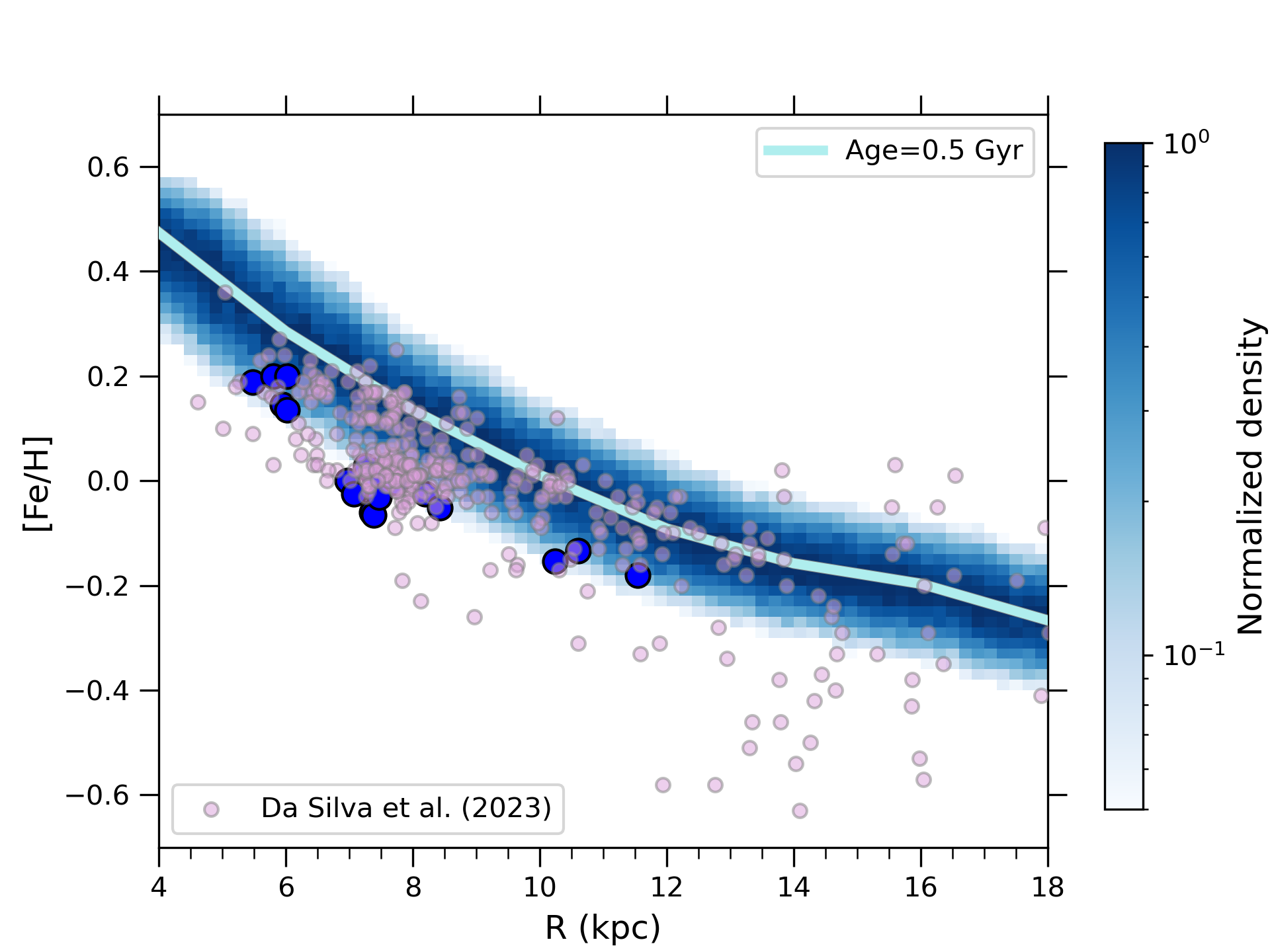}
    \caption{Same as Fig. \ref{fig:gradient_2inf_restricted}, but showing only the results for the young age bin ($< 1$ Gyr), with the addition of Classical Cepheids data sample from \citet{DaSilva23} (pink filled circles).}
    \label{fig:gradients_Cepheids_2inf}
\end{figure}

To further inspect the low [Fe/H] abundances by OCs at young ages, we also consider the metallicities derived for the CCs sample in \citetalias{DaSilva23} (see Section \ref{s:data}).  
It is worth reminding that the latter sample provided homogeneous derivation of Fe abundances relative to the OC sample adopted here, preventing any additional observational bias.
Both the datasets are shown in Fig. \ref{fig:gradients_Cepheids_2inf}, together with the prediction for the young age bin by the two-infall model. 
It is worth noting that here and in the following Figures, we avoid showing the contours of the predicted stellar distribution to avoid overcrowding the plots.
Despite the much larger abundance spread observed in \citetalias{DaSilva23} sample, the Figure  shows that the bulk of Cepheids observed in \citetalias{DaSilva23} up to a radius about $\sim 10$ kpc are also underabundant relative to the model predictions. For larger radii instead, the very large spread in [Fe/H] and the relevant abundance uncertainties for distant Cepheids (we do not show them here for sake of readability, but they arrive up to 0.25 dex), do not allow us to draw strong conclusions, despite the hint of a flattening of the radial metallicity gradient at large Galactocentric distances remains (see \citetalias{DaSilva23} and references therein).
In any case, the superposition between the two datasets, which we remind are obtained using different gradient tracers, strengthen the findings described in this Section.\\

As described in \ref{ss:nucleo}, we also run additional simulations using different yields for massive stars and Type Ia SNe. In this way, we test the dependence of the observed present-day gradient overestimation on nucleosynthetic calculations.

Results for these runs are shown in Appendix \ref{appendixA}. They highlight that the theoretical predictions with different stellar yields are very similar, with an overestimation of the present-day gradient in the two-infall model and a similar evolution of the predicted gradient through cosmic time.  
This denotes that the conclusions of the evolutionary scenarios on the [Fe/H] gradient are marginally affected by the nucleosynthesis prescriptions. Therefore, in the rest of paper, we will proceed by using the sets of yields adopted throughout this Section.

\subsection{A late time metal dilution: the three-infall model}
\label{ss:results_3inf}

To explain the unexpected decrease in the [Fe/H] gradient at late times, we explore the scenario proposed by \citetalias{Spitoni23} for the solar vicinity, where a late-time burst of SF (suggested by Gaia CMD analysis, e.g. \citealt{Ruiz20}) is fueled by gas accretion, which in turn causes a metal dilution in the ISM gas.
In particular, we extend such a scenario to the whole disk than the solar neighbourhood. \\

We start by adopting a setup for the third gas accretion very similar to the one by \citet{Spitoni23}, which is shown in Table \ref{tab:3inf_param} upper row. 
Here, the parameters of the third infall are the same of the latter paper, except for the SFE $\nu_3$, for which we fix the proportion between the SFE in the second and third infall episode to be similar to the one used in \citet{Spitoni23}. 
In this way, we are able to preserve the gradient slope observed in the data at different ages: if we apply a flat SFE with radius for the third infall, this will significantly decrease the gradient slope, which is not really seen in the observations (see Table \ref{tab:comparison_fits}).
Finally, at variance with \citetalias{Spitoni23}, we adopt a primordial chemical composition for the third gas infall. However, we also perform an additional run with an infall enrichment as in \citetalias{Spitoni23} (1/5 enriched with an abundance pattern as the one predicted for the high-$\alpha$ phase at [Fe/H]$=-0.75$ dex), finding negligible differences.

\begin{figure}
    \centering
    \includegraphics[width=0.95\columnwidth]{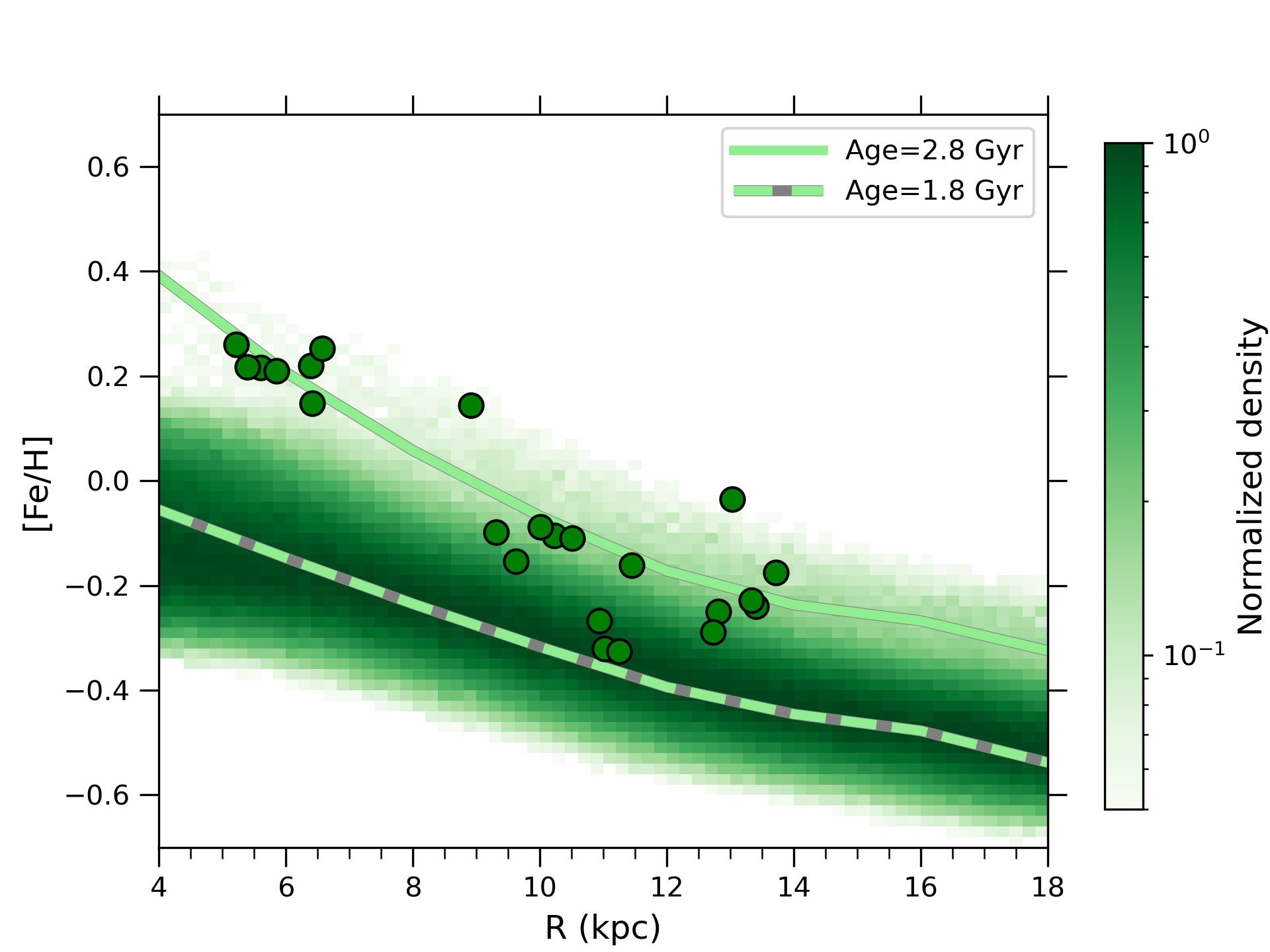}
    \includegraphics[width=0.95\columnwidth]{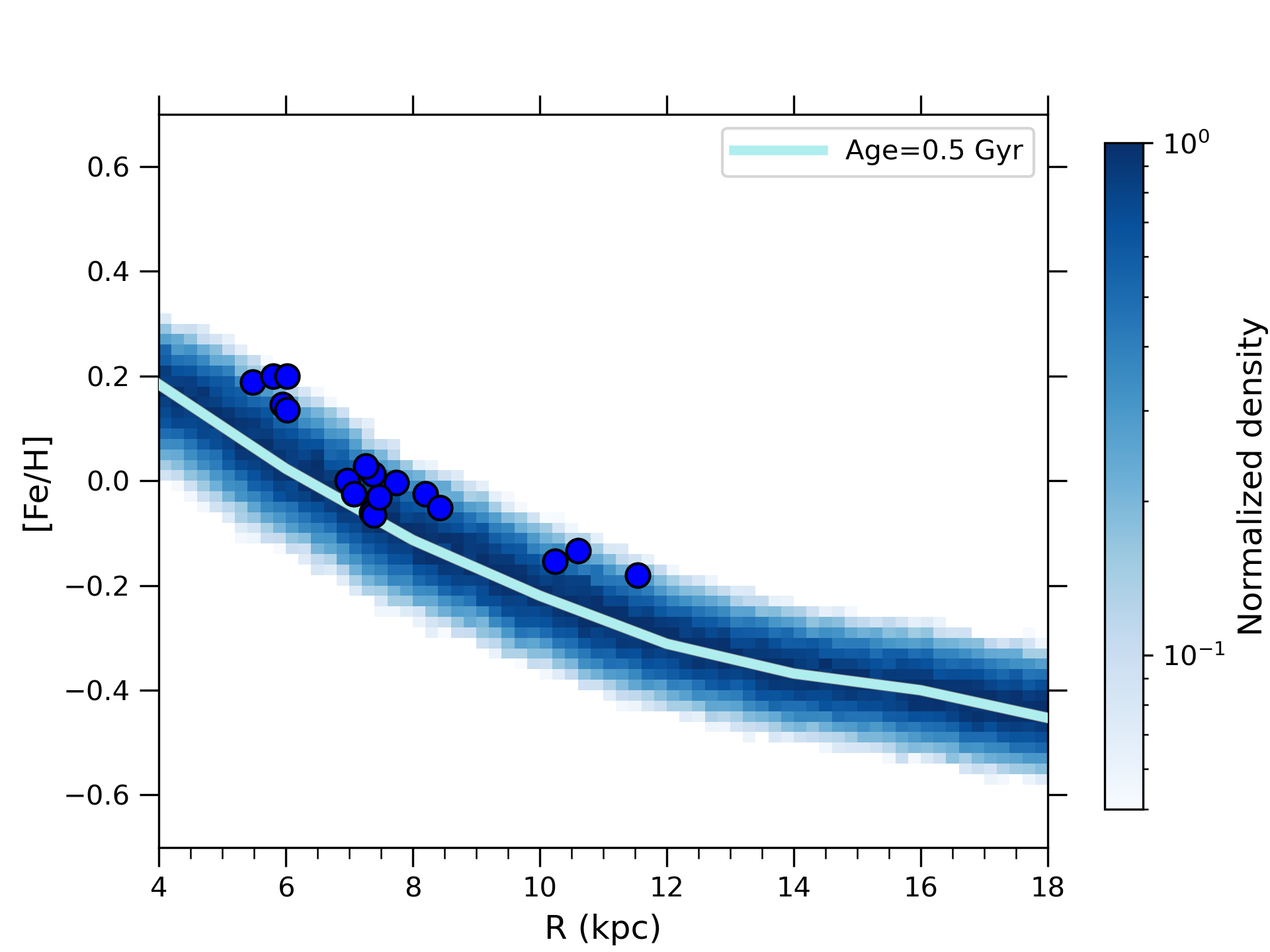}
    \caption{Time evolution of the radial [Fe/H] gradients at $1<Age$/Gyr$<3$ (upper panel) and $Age<1$ Gyr (lower panel) for the model 3INF-1, including stellar migration and average OC spread (see \ref{ss:migr_err}). Lines show the results for the [Fe/H] gradient as predicted by the model at Ages=0.5 Gyr (cyan line), 1.8 Gyr (green dashed line) and 2.8 Gyr (green solid line). 
    Data are the same as in Fig. \ref{fig:gradient_2inf_restricted}.}
    \label{fig:gradients_3inf_Sptioni}
\end{figure}

In Fig. \ref{fig:gradients_3inf_Sptioni}, we show the predicted evolution of the [Fe/H] gradient by the three-infall model with parameters as for the 3INF-1 setup. The results are shown for the last 3 Gyr, i.e. the ages at which this late gas accretion episode is actually acting and changing the gradient evolution.
Fig. \ref{fig:gradients_3inf_Sptioni} shows that the effect of metal dilution by the gas accretion is too strong. In particular, the upper panel highlights that we miss to reproduce the metal-rich stars in the intermediate age bin. This is due to the very strong gas dilution happening 2.7 Gyr ago, which leads the bulk of stellar production at subsolar metallicity, even at small Galactocentric radii. This is instead not seen in OC observations which, despite of the spread seen at individual radial distances, only marginally cover the region where the model expects most of the data. 

\begin{figure*}
    \centering
    \includegraphics[width=0.95\textwidth]{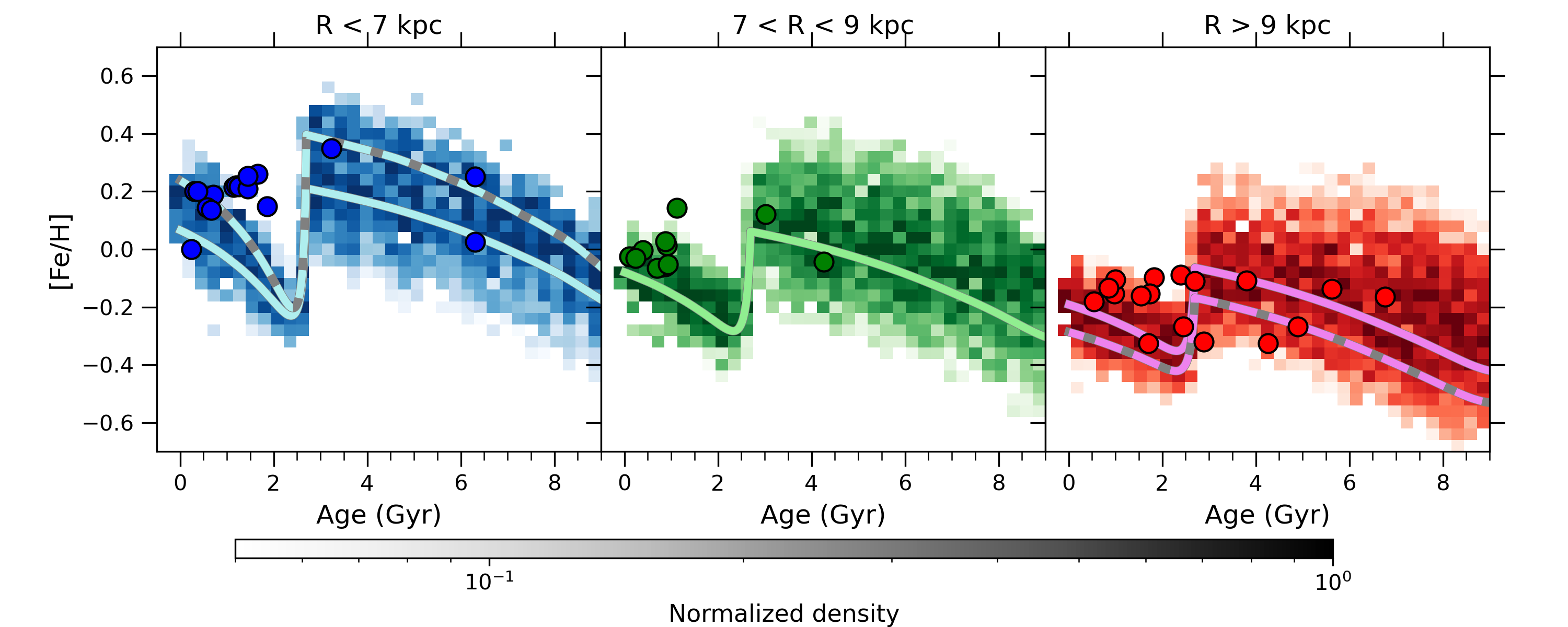}
    \caption{Age-[Fe/H] relation for the model 3INF-1 in different radial ranges, i.e. R$<7$ kpc (left panel), $7<$R/kpc$<9$ (central panel) and R$>9$ kpc (right panel), including stellar migration and OC spread (see \ref{ss:migr_err}). The density plot show the normalised density of stars (in $\log$ scale) as predicted by the model in a given age bin of 0.25 Gyr width.
    Lines are genuine chemical tracks at 6, 8 and 10 kpc (solid) and 4 and 12 kpc (dashed).
    Filled circles represent the restricted sample within OCs in \citet{Magrini23}.}
    \label{fig:age_FeH_Spitoni}
\end{figure*}

This is also reflected by looking at Fig. \ref{fig:age_FeH_Spitoni}, where we show the Age-[Fe/H] relation as predicted by the model 3INF-1 in different radial bins, i.e. inner (R $<7$ kpc, left panel), solar ($7<$R/kpc$<9$, central panel) and outer (R$>9$ kpc, right panel).
Even accounting for migration and abundance uncertainties effects (see density bins in the Figure), the prominent metal dilution prevents a good agreement with the data. This is especially evident at small Galactocentric radii (left panel), where the model fails to reproduce the age-metallicity trend within the last 3 Gyr. 
The situation is less dramatic at larger radii, where a large fraction of the OCs falls within the range of values allowed by the predictions. However, in the central and right panels we also observe that the genuine chemical evolution predictions by the model at 8 and 10 kpc (solid lines in central and left panels, respectively), slightly underestimate the metal content observed in young clusters.
The slight underestimation of the present-day gradient by the model 3INF-1 is in fact observed in Fig. \ref{fig:gradients_3inf_Sptioni} lower panel.\\

To test further the viability of the three-infall scenario in the context of radial gradients, we allow to vary the third infall parameters relative to the values proposed in \citetalias{Spitoni23}.
While leaving constant the starting time of the third infall (to be consistent with SF peaks as show by, e.g., \citealt{Ruiz20}) we act on all the other physical parameters, i.e. the infall timescale $\tau_3$, the SFE $\nu_3$ and for the ratio between the baryonic mass accreted by the second and third infall $\Sigma_2$/$\Sigma_3$, as already shown in Tab. \ref{tab:3inf_param} bottom row.

In addition, we allow a mild chemical enrichment for the infalling gas during the third gas accretion, with half of the gas enriched at a level of [Fe/H]=-0.75 dex with abundance pattern as the one predicted for the high-$\alpha$ phase at that metallicity. 
This assumption is justified in the light of two possible invoked physical mechanisms behind the observed recent peak in SF (e.g. \citealt{Isern19,Mor19,Ruiz20}). On one hand, it is suggested a tight connection between the SF peak and the last pericentric passage of Sagittarius dSph (\citealt{Ruiz20,RocaFabrega21}): the gas retained by Sagittarius after its first encounter with the Galaxy may have been definitely stripped in its second pericentric passage, contributing to the gas accretion together with steady cooling flow of gas from the hot corona. On the other side, \citet{Nepal24} proposed that the peak in MW disk SF was triggered by intense MW bar activity, which was shown to trigger enhanced SF in galaxies both from a theoretical (e.g. \citealt{Baba20}) and observational (with Integral Field Spectroscopy on local galaxies, e.g. \citealt{Lin20}) points of view. 
Without discriminating between the two scenarios, it is therefore likely that a mild chemical enrichment have to be present in this late stage of Galactic evolution.

\begin{figure}
    \centering
    \includegraphics[width=0.95\columnwidth]{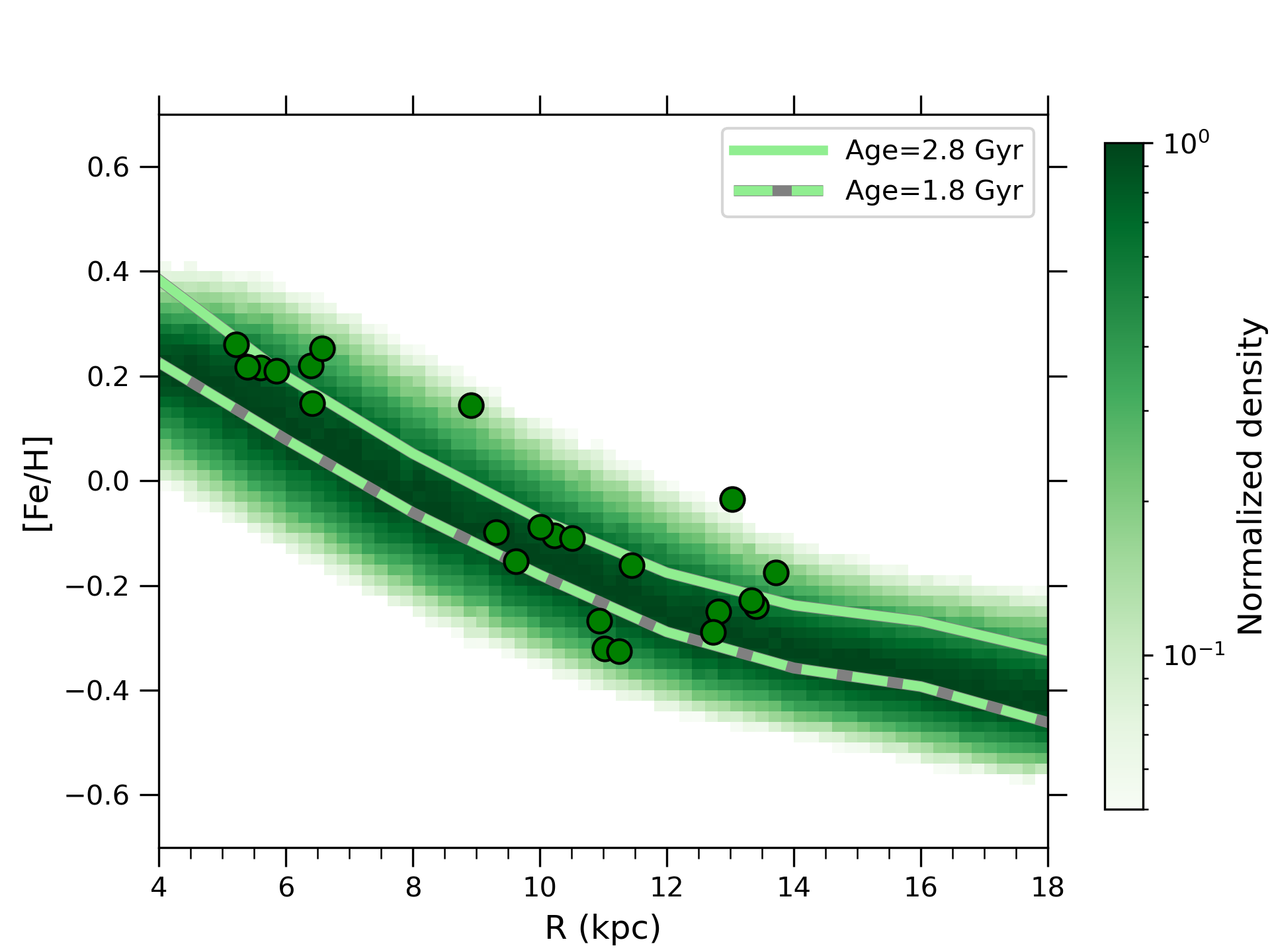}
    \includegraphics[width=0.95\columnwidth]{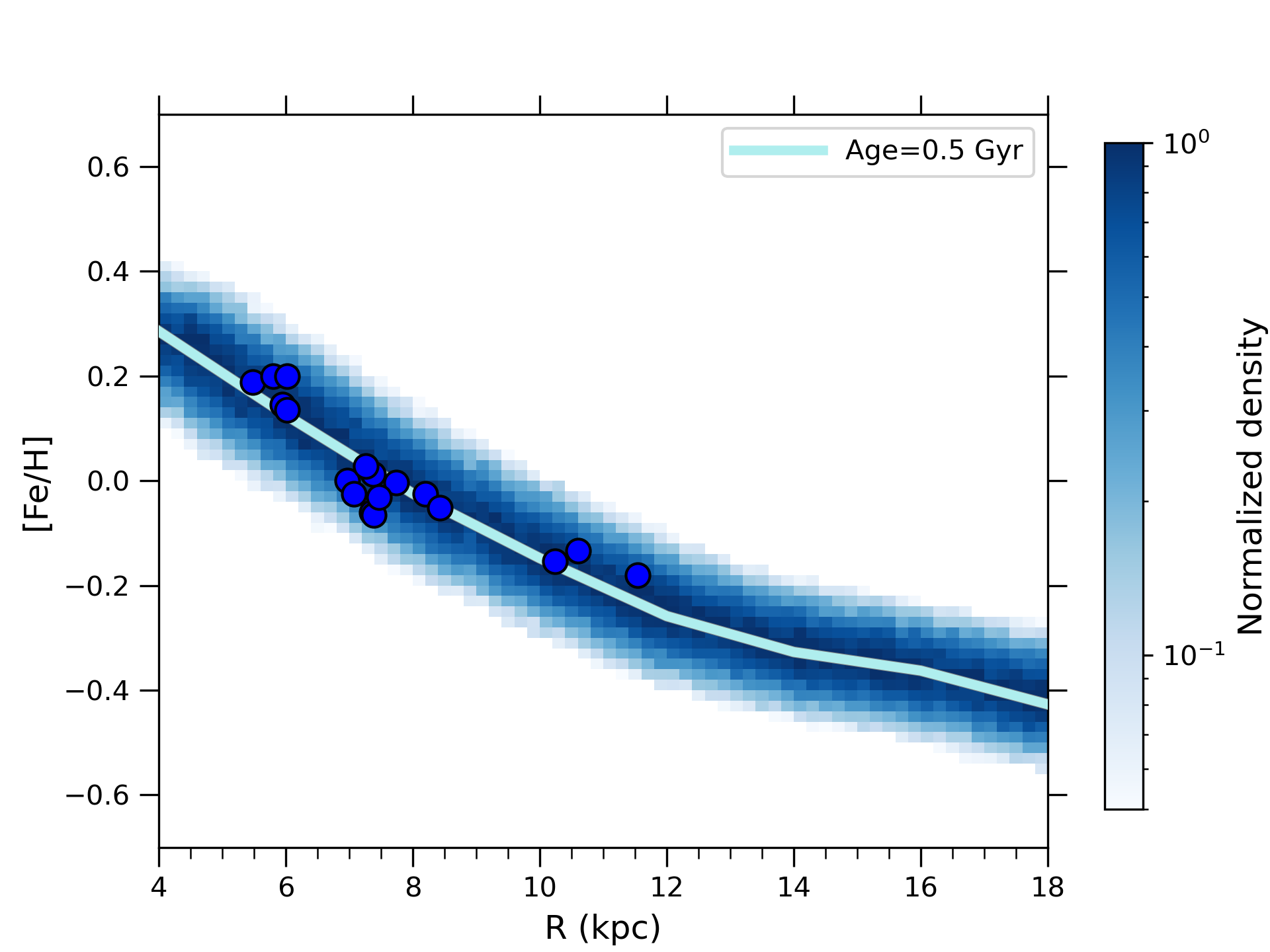}
    \caption{Same as Fig. \ref{fig:gradients_3inf_Sptioni}, but for the model 3INF-2.}
    \label{fig:gradients_3inf_Palla}
\end{figure}

In Fig. \ref{fig:gradients_3inf_Palla} we show the predicted evolution of the [Fe/H] gradient by the three-infall model with parameters as for the 3INF-2 setup.
In the upper panel, showing the metallicity gradients in the age bin 1-3 Gyr, we note that  the dilution effect is milder relative to what seen in Fig. \ref{fig:gradients_3inf_Sptioni}. This is  highlighted by the green solid and dashed lines, showing that the chemical evolution track before and almost 1 Gyr after the onset of the third infall, which have a metallicity difference of around  0.2 dex.  The smaller dilution allows us to capture the trend and spread observed in OCs in this age range: here, the location of data points broadly correspond to the regions with predicted higher probability density of stars, at variance with what happens in Fig. \ref{fig:gradients_3inf_Sptioni}.
The good agreement between OC data and the 3INF-2 model is also seen in Fig. \ref{fig:gradients_3inf_Palla} lower panel, with all the OCs in the youngest age bin falling within the range of values allowed by the model.

\begin{figure*}
    \centering
    \includegraphics[width=0.95\textwidth]{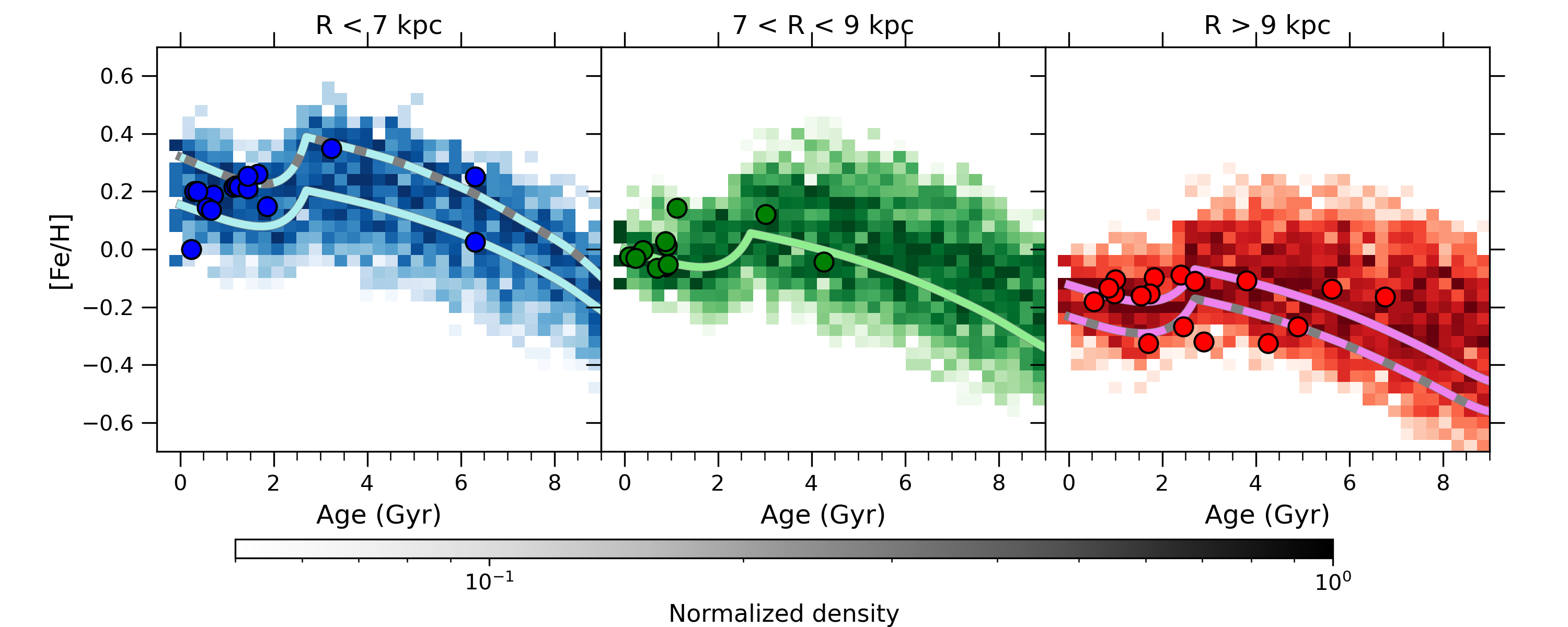}
    \caption{Same as Fig. \ref{fig:age_FeH_Spitoni}, but for the model 3INF-2.}
    \label{fig:age_FeH_Palla}
\end{figure*}

As done for model 3INF-1, we also compare the observed age-metallicity relations at different radii with the prediction of the model 3INF-2. These are shown in Fig. \ref{fig:age_FeH_Palla}. 
As already noted in Fig. \ref{fig:gradients_3inf_Palla}, here the dilution by the third gas accretion episode is much less prominent relative to the one seen in Fig. \ref{fig:age_FeH_Spitoni} for the model 3INF-1. This difference can be explained by the larger infall timescale for model 3INF-2: here, the slower gas accumulation allows dying stars to progressively pollute with metals the ISM, leaving a more prolonged but less marked decrease in metallicty. This behaviour better reproduces the observed age-metallicity trends, as no sharp dilution effects are seen in our restricted sample.
Moreover, the slightly larger SFE $\nu_3$ in model 3INF-2 allows a better agreement between the chemical evolution tracks at different radii (solid and dashed lines in Fig. \ref{fig:age_FeH_Palla}) and the metal content in young clusters, which is instead underestimated by model 3INF-1.\\

To probe even further the three-infall scenario, we looked also at other chemical elements than Fe. However, in doing this we are more prone to the intrinsic uncertainties in models related to stellar nucleosynthesis, which are less significant when probing the Fe gradient (see Appendix \ref{appendixA}).
To limit as much as possible this problem, we focus on two $\alpha$-elements whose [$\alpha$/Fe] vs. [Fe/H] abundance patterns are generally well reproduced throughout the whole MW metallicity range with the main sets of yield adopted in this work (see \ref{ss:nucleo}), i.e. O and Si (see \citealt{Prantzos18,Romano19}). 
Other $\alpha$-elements, such as Mg, are severely underestimated when adopting massive stars \citet{Limongi18} prescriptions (\citealt{Prantzos18,Palla22}), while most of Fe-peak elements also suffer of the additional uncertainty on the progenitor nature of Type Ia SNe, which severely alters the abundance trends (\citealt{Koba20,Palla21}).

\begin{figure*}
    \centering
    \includegraphics[width=0.475\textwidth]{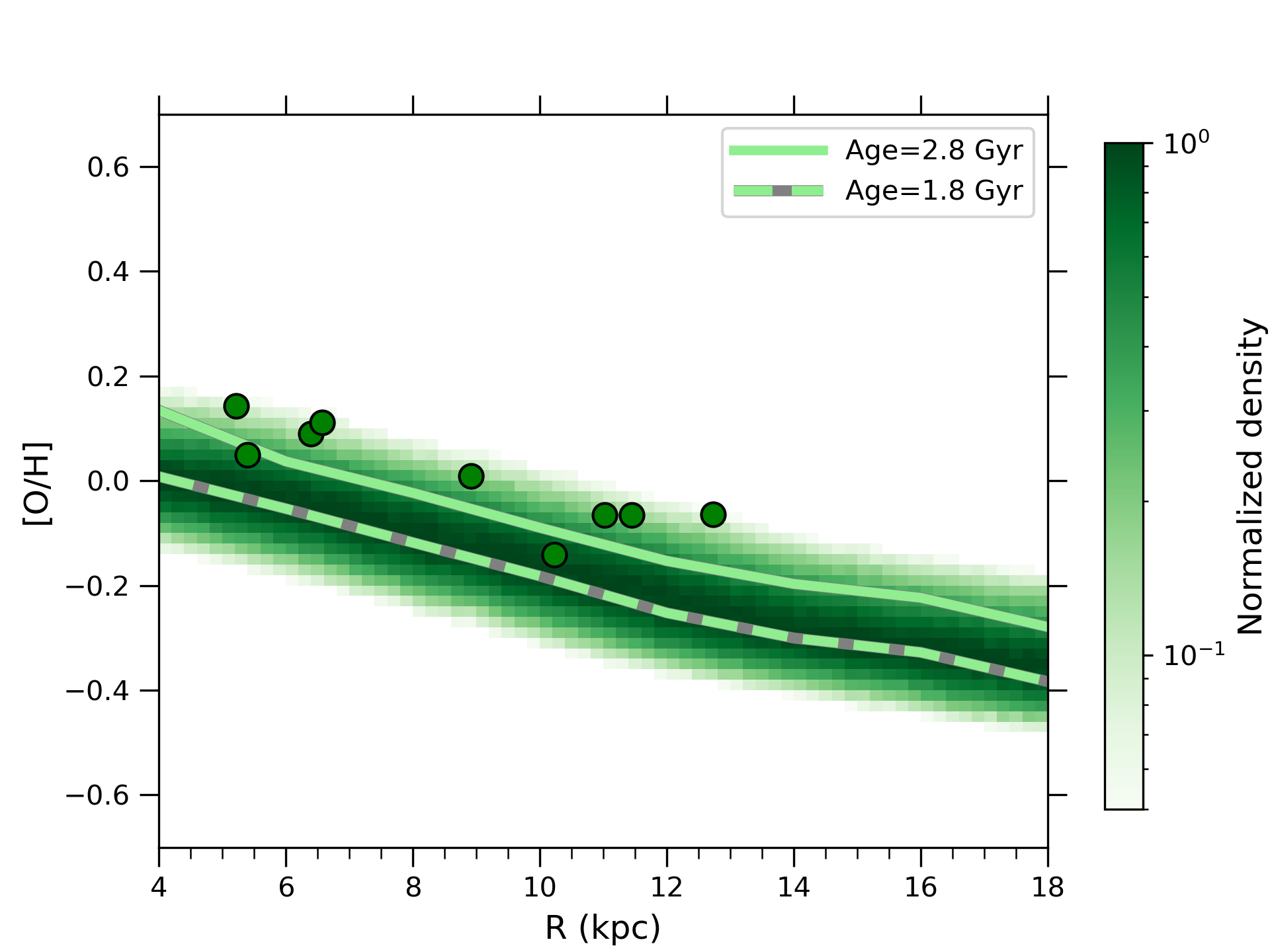}
    \includegraphics[width=0.475\textwidth]{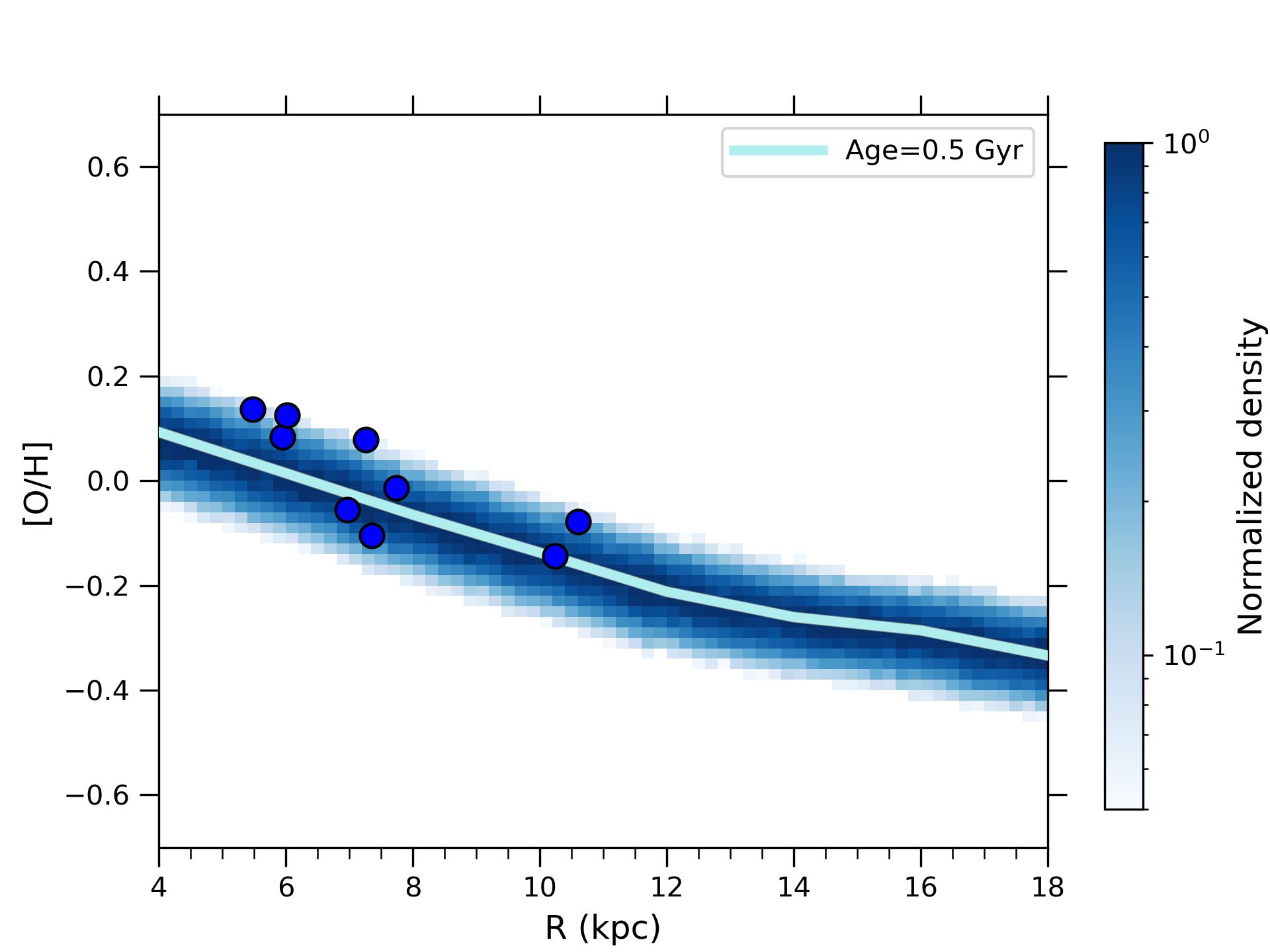}\\
    \includegraphics[width=0.475\textwidth]{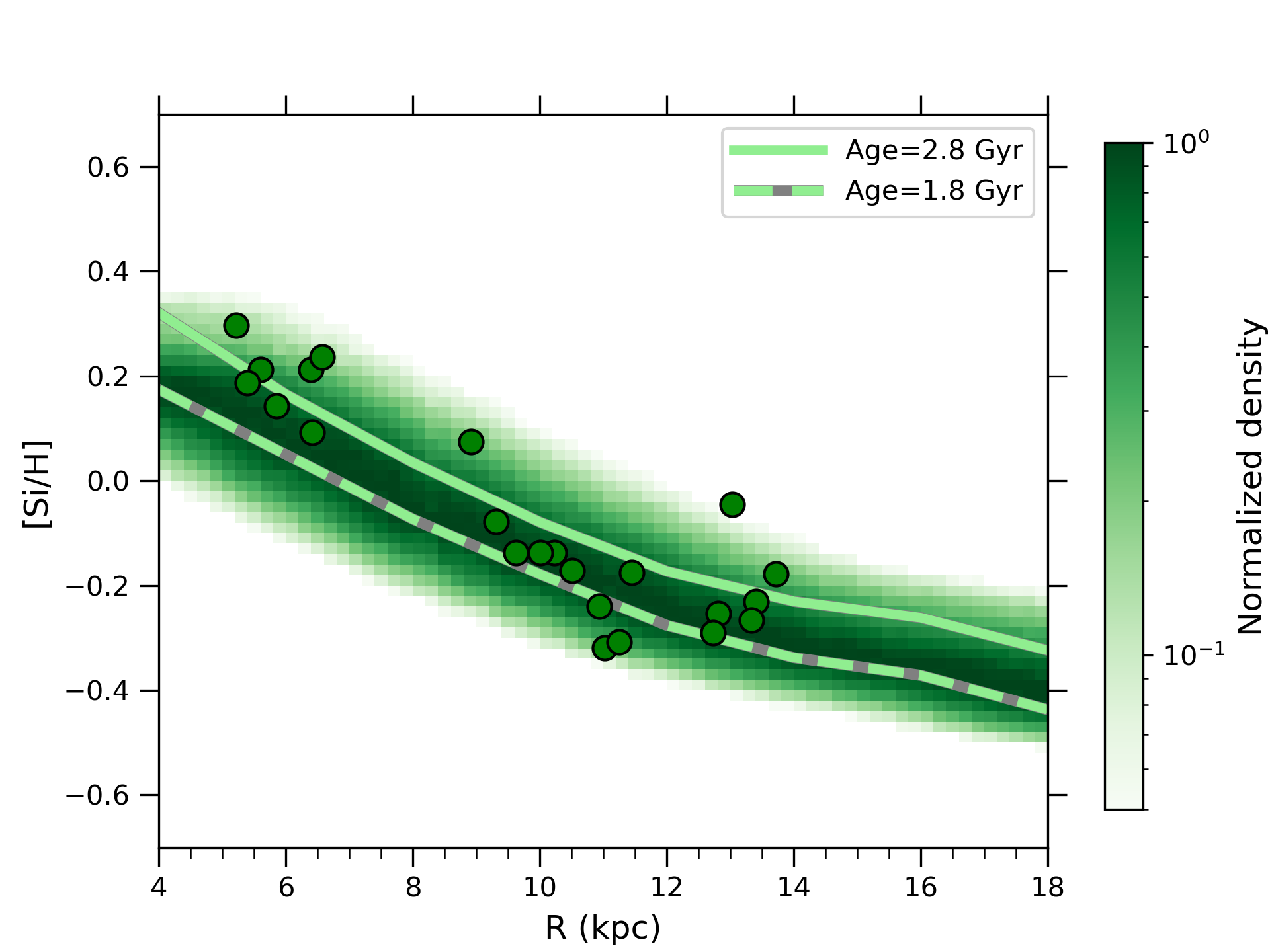}
    \includegraphics[width=0.475\textwidth]{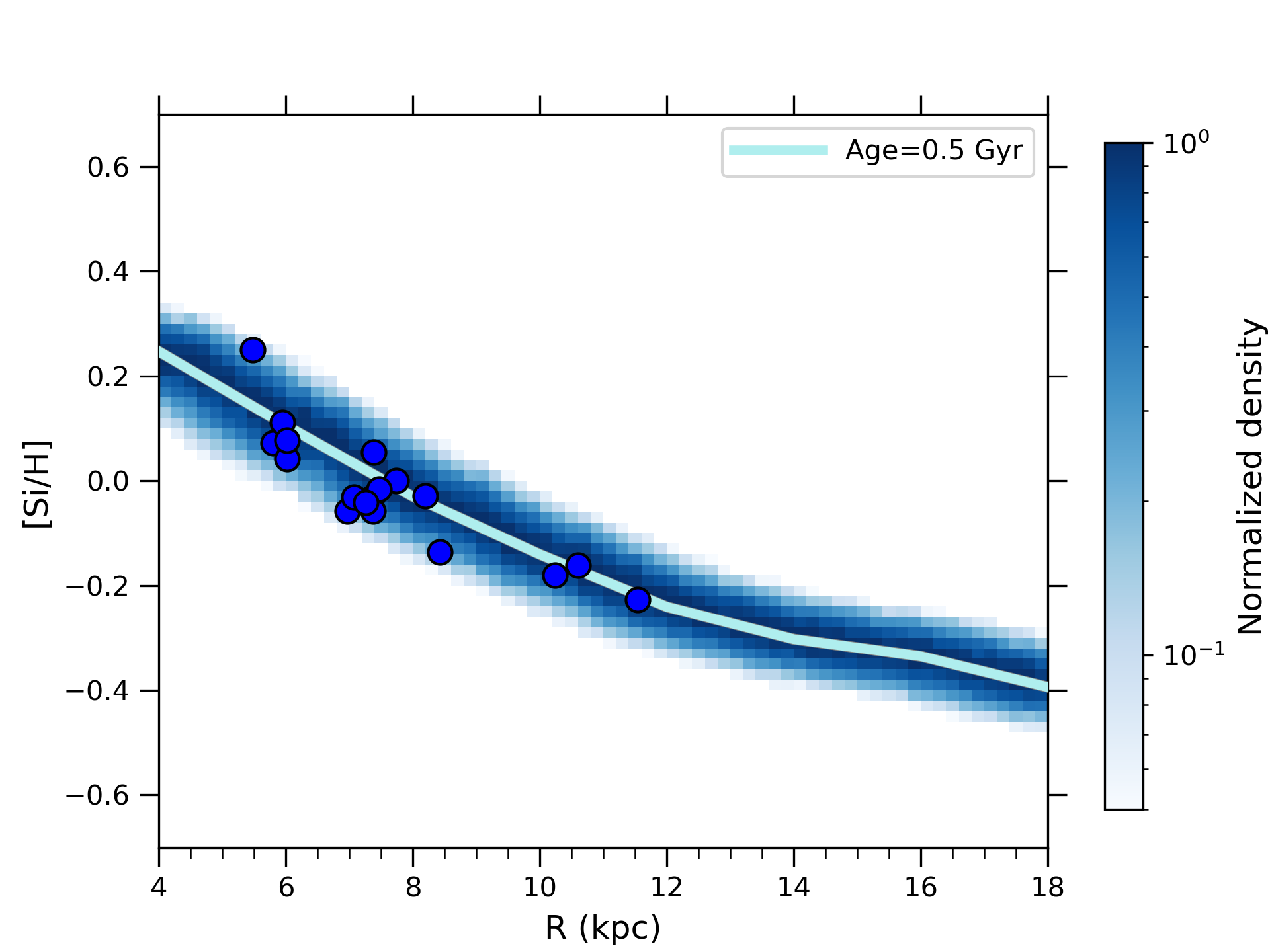}
    \caption{Radial and [O/H] (upper panels) and [Si/H] (lower panels) gradients at $1<Age$/Gyr$<3$ (left panels) and $Age<1$ Gyr (right panels) for the model 3INF-2, including stellar migration and average OC spread (see \ref{ss:migr_err}). Lines show the results for the [Fe/H] gradient as predicted by the model at Ages=0.5 Gyr (cyan lines), 1.8 Gyr (green dashed lines) and 2.8 Gyr (green solid lines). 
    Filled circles represent the restricted sample within OCs in \citet{Magrini23}.}
    \label{fig:gradient_alpha_3inf}
\end{figure*}

In Fig. \ref{fig:gradient_alpha_3inf} we show the evolution of [O/H] (upper panels) and [Si/H] (lower panels) as predicted by the model 3INF-2, compared with our restricted sample.
The Figure highlights a global agreement between the model scenario and the data. [Si/H] gradient resembles the trends already described in this Section for Fe, sharing similar gradients slopes with the [Fe/H] gradient in different age ranges. This reflects in a rather flat [Si/Fe] gradient throghout the last Gyr of galactic evolution.
On the other hand, the predicted and observed [O/H] gradient is much shallower, resulting in a positive [O/Fe] gradient. For the O gradient, we also note that in the 1-3 Gyr age bin (Fig. \ref{fig:gradient_alpha_3inf} upper left panel) all OC abundances lay in the upper end of the range of values allowed by the model, not showing any sign of particular spread as opposed to Fe and Si. However, it is worth noting that for O in general we have to rely on a much lower number of abundance data, which limit us in understanding the real motivation of this discrepancy. Nonetheless, the [O/H] gradient in the youngest age bin (Fig. \ref{fig:gradient_alpha_3inf} upper right panel) is reproduced by the model predictions, well in agreement with our proposed scenario.

\begin{figure*}
    \centering
    \includegraphics[width=0.95\textwidth]{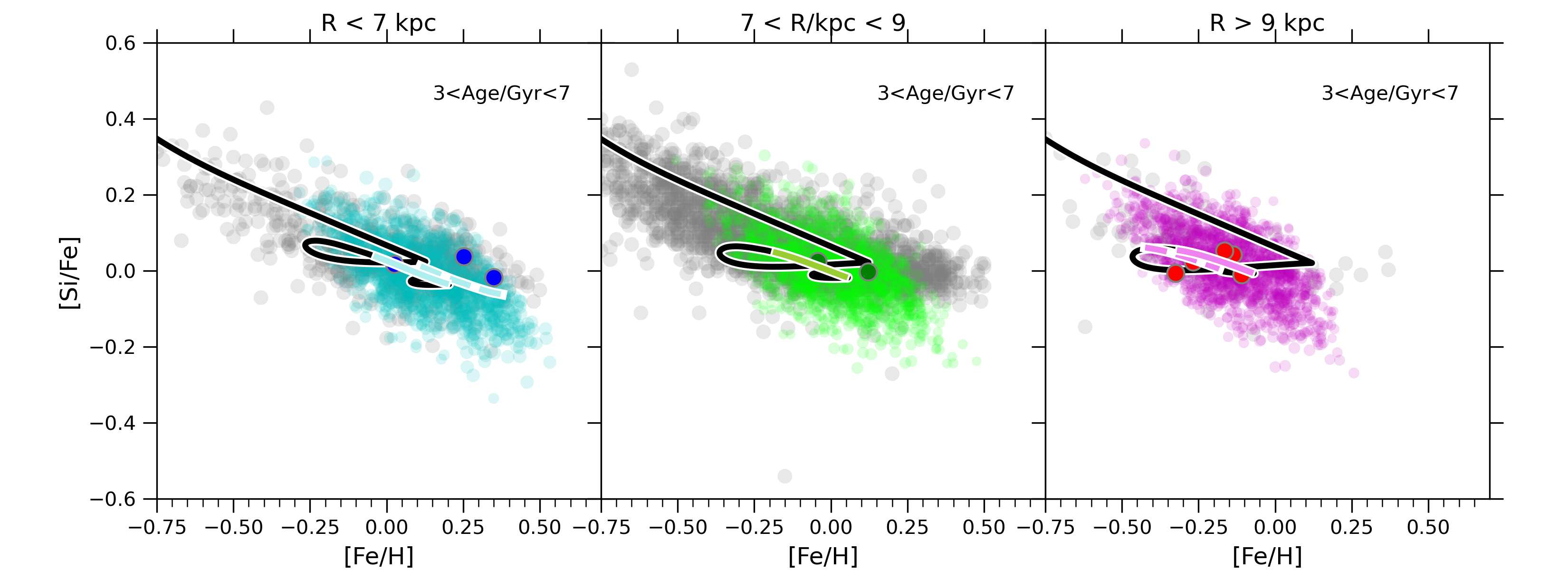}
    \includegraphics[width=0.95\textwidth]{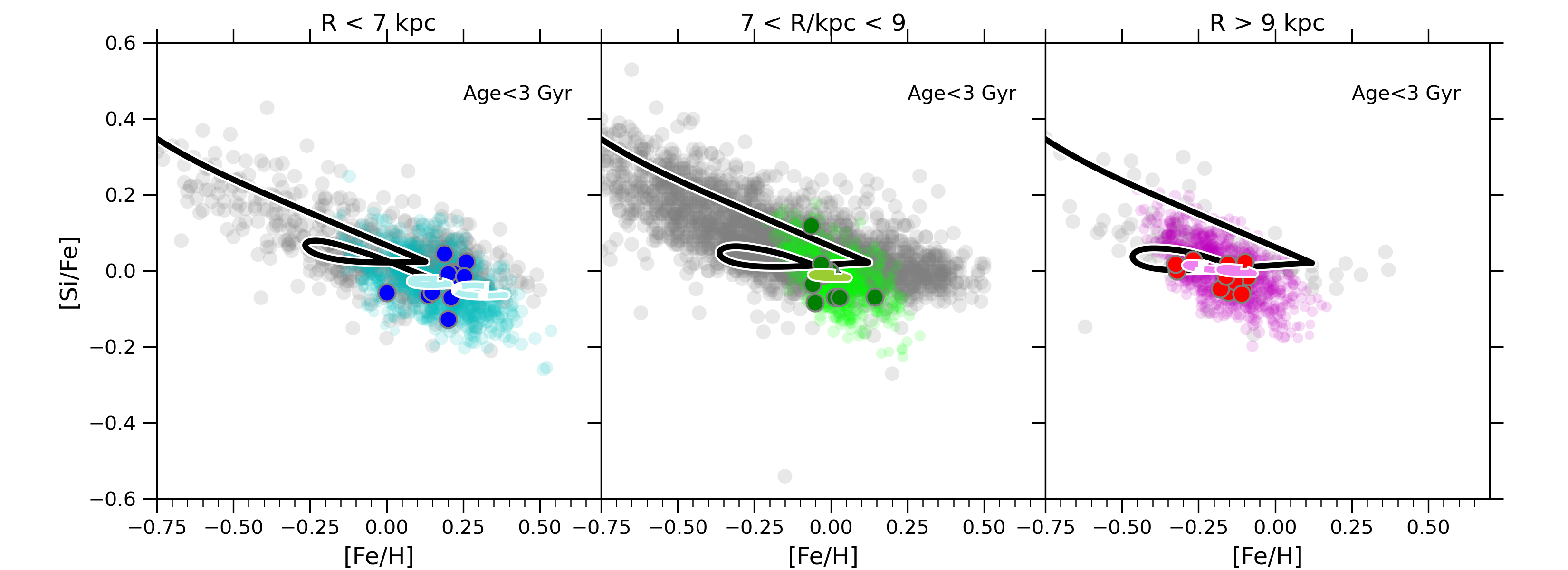}
    \caption{[Si/Fe] vs. [Fe/H] evolution for the model 3INF-2 in different radial ranges, i.e. R$< 7$ kpc (left panels), $7 <$R/kpc$< 9$ (central panels) and R$> 9$ kpc (right panels). The upper panels show results for ages$>3$ Gyr, while the lower panels for ages$<3$ Gyr. The shaded cyan, light green and magenta areas are the model prediction in a certain radial range taking into account the effect of stellar migration and OC spread (see \ref{ss:migr_err}). The solid lines represent genuine chemical evolution tracks at 6, 8 and 10 kpc and are colored in the age range considered in the respective panel. Colored dashed lines are the same as colored solid lines but for the radii of 4 kpc (left panels) and 12 kpc (right panel). 
    Colored filled circles represent the restricted sample within OCs in \citet{Magrini23}. Grey points are selected field stars from the {\it Gaia}-ESO survey.}
    \label{fig:SiFe_FeH}
\end{figure*}

In Fig. \ref{fig:SiFe_FeH} we also report the evolution of the OCs and the three-infall scenario in the [X/Fe] vs. [Fe/H] diagram in different radial regions at different ages. In particular we show the results for Si, as we can rely on a much larger sampling of OCs, for a more thorough comparison relative to O. 
In Fig. \ref{fig:SiFe_FeH} upper panels we show the comparison for ages larger than 3 Gyr, i.e. before the onset of the third infall. Here, the model predictions including the effects of migration and abundance uncertainties are shown in each radial region with colored areas. Moreover, to guide the eye, the genuine chemical tracks at 6, 8 and 10 kpc (in the left, central and right panel, respectively) are highlighted with color between 3 and 7 Gyr, i.e. the age range considered in these panels. Also, in the left and right panel tracks for 4 and 12 kpc in the age range considered are shown.
Despite of the small data sampling in this age range, we can say that data are well within the predicted [Si/Fe] vs. [Fe/H] ranges considering migration and abundance uncertainties effects.
In Fig. \ref{fig:SiFe_FeH} lower panel we show instead the predictions for ages below 3 Gyr, the ones interested by the late gas accretion. As for the upper panel, we highlight the parts of the tracks with ages $<$3 Gyr with color. Also in this case, the agreement between predictions and observations is remarkable, 
with all the clusters falling within the range of values allowed by model predictions.

\subsection{Discussion}
\label{ss:discuss}

Throughout this work we investigated the behaviour in radial abundance gradients as traced by  \citetalias{Magrini23} {\it Gaia}-ESO OCs, which show a decrease in their metal content towards the youngest population.
Within the framework of our models of chemical evolution, that also include the effects of stellar radial migration (following \citealt{Frankel18,Frankel20}), we observe that the observed metal impoverishment of young OCs should be mainly caused by a chemical dilution from a recent gas infall episode, which in turn is triggering a late enhanced SF activity. 
These results confirm the conclusion by \citetalias{Spitoni23}, who found the necessity of very late ($\lesssim 3$ Gyr of age) gas accretion episode to explain the abundance ratios in the solar neighbourhood observed in Gaia DR3 RVS spectra, and extend it to a much larger range of Galactocentric radii.

However, our preferred scenario generally requires a milder metal dilution than in \citetalias{Spitoni23} in this late gas infall. In particular, our best scenario results to be more similar to the "weak dilution" scenario in \citetalias{Spitoni23}, with smaller contribution of the late infall to the total mass budget and larger SFE relative to their best model.
The difference in the results can be explained by the different data set adopted in this work to calibrate our best model. In fact, in \citetalias{Spitoni23} the model was tuned to reproduce the abundances from young massive stars from Gaia RVS spectra (\citealt{Gaia23,RecioBlanco23}) and cross matched targets in APOGEE DR17 (\citealt{APOGEE17}, see their Fig. 14). \citetalias{Magrini23} showed that such a class of objects may suffer of biases in the determination of stellar abundances, with an evident decrease in their metallic content, and this in turn critically influence our view on [X/Fe] vs. [Fe/H] diagrams and abundance gradients.
This is also shown in this work, where we see how massive giants with low $\log g$ can influence our view on the gradients of young stellar populations (see \ref{ss:results_2inf}). 
As shown in \citet{Magrini23}, this problem  concerns spectral analysis in general and is present in various surveys (e.g. APOGEE, GALAH) regardless of the analysis method and spectral range. An effort is needed to improve the model atmospheres of low gravity giant stars, and to include the effect of the magnetic field to advance the analysis of these stars. Spectral analysis of these stars will have even more implications in the future because in the ELT era  the brightest giant stars  will give us detailed information about chemical abundances in distant galaxies \citep{Roederer2024ExA....57...17R}.

In the current situation, the choice of imposing motivated cuts in the stellar parameters of OC members allows us to be confident of being less prone to abundance systematics.
Moreover, we compare the results of our model with all the different diagnostics available, i.e. abundance gradients in different chemical abundances as well as age-metallicity relations and [X/Fe] vs. [Fe/H] abundance patterns in different Galactocentric regions, all of them showing a good agreement between data and models. We also verify that the proposed scenario agrees with the observed values at solar radius and gradients of different physical quantities, such as SFR and gas surface densities (see Appendix \ref{appendixB}).
In this way, we ensure that our proposed scenario of chemical evolution is robust in the context of the MW disk formation.

Nonetheless, the search for unbiased samples with abundances from high-resolution spectroscopy comes at expenses of the sample size. In fact, our results are still limited by the moderate sample size of OCs within our restricted sample ($\simeq 50$). 
More data are certainly needed to better probe the three-infall scenario of chemical evolution, imposing more stringent limits and maybe revising the values of the physical parameters that have been adopted in this work.
In particular, further sampling 
in age and in Galactocentric radii will be fundamental to pursue this goal. As for example, it is worth noting that for moderate to young ages we clearly lack of data at ${\rm R} \gtrsim 12$ kpc. This crucially limits our ability to draw firm conclusions on the evolution of Galactic outer regions, whose evolutionary trends are only suggested following a framework similar to that found for the innermost disk.
However, reliable tracers of gradient evolution as OCs are clearly missing from the outer galaxy \citep[see, e.g.][]{Cantat2020A&A...640A...1C} 
The CCs have started to be observed with high statistical significance also in the outermost regions of the Galaxy (e.g. \citealt{DaSilva23,Ripepi24}), but these objects only inform us on the present-day situation. Therefore, accurate abundances and ages of the stellar populations of the outermost Galaxy are needed, and planned (e.g. 4MOST, \citealt{4MOSTpaper}; WEAVE, \citealt{WEAVEpaper}; PLATO, \citealt{PLATOpaper}
) and proposed (e.g. WST, \citealt{WSTpaper}) facilities will undoubtedly help us in this search.

\section{Summary and conclusions}
\label{s:conclusion}

In this paper we studied the evolution of radial abundance gradients in the Milky Way (MW) by taking advantage of the sample of open clusters (OCs) from the last data release of the {\it Gaia}-ESO survey (\citealt{Randich22,Magrini23}). 
OCs, in fact, are among the best tracers of the shape and time evolution of the radial metallicity gradient due to their ages and distances, which can be properly measured by isochrone fitting of the complete sequence, and to high-resolution spectroscopic observations that provide precise abundances. 
{\em Gaia}-ESO dedicated about 30\% of its observing time to provide the largest sample of precise and homogeneous stellar parameters and abundances of member stars in Galactic OCs.

From a theoretical perspective, we started from the well tested revised two-infall model of chemical evolution (e.g. \citealt{Palla20}, see also \citealt{Spitoni19}), which successfully reproduces data from high-resolution surveys, such as APOGEE (\citealt{Ahumada19}), and extended the comparison to the newly proposed scenario of the three-infall model (\citealt{Spitoni23}), proposed in the light of constraints given by Gaia star formation history (e.g. \citealt{Ruiz20}) and abundance ratios (\citealt{Gaia23}).
For all the probed scenarios, our models take into account the effects of stellar radial migration, by including well tested prescription from the literature (\citealt{Frankel18,Frankel20}), allowing us to directly see its impact on the late evolution of radial abundance gradients.\\

Our main considerations and conclusions are thus summarised as follows:
\begin{enumerate}
    \item conservatively, we excluded stars whose spectral analysis can generate metallicity bias.  This choice reduces the trends between stellar parameters and metallicity within the same cluster.  This solution does not solve the problem of analysing massive and/or low-gravity giants, but allows us to circumvent it until progress is made on atmosphere models for these stars and including magnetic activity;
    
    \item despite of restricting the OC sample to stars which should not hide biases in spectroscopic analysis, we find a metallicity decrease between intermediate age (1-3 Gyr) and young ($<$1 Gyr) OCs.
    We show that the radial metallicity gradient as traced by young OCs is overestimated by the predictions of the two-infall scenario, even by accounting for the effect of stellar migration in the model. We also check whether the adoption of different nucleosynthetic yields for massive stars and Type Ia SNe, i.e. the main contributors to Fe enrichment, may affect this result. We find negligible differences between the different runs, confirming the conclusion;
    
    \item to explain the observed low metallic content in young clusters, we propose that a late gas accretion episode triggering a metal dilution should have taken place. This is in agreement with the proposed scenario of the three-infall model by \citet{Spitoni23} for the solar vicinity, which explains the recent star formation history and abundance ratios as derived by {\it Gaia} satellite, as a consequence of recent gas infall episode triggering enhanced SF at recent ages. 
    It is worth noting that in this work, for the first time we extend the three-infall scenario to the whole MW disk;
    
    \item at variance with the best model presented in \citet{Spitoni23}, we invoke a milder metal dilution for this late gas infall episode. In particular, our best scenario requires smaller contribution of the late infall to the mass budget forging the low-$\alpha$ disk (factor $\sim$1.5-4 lower), while larger infall timescale ($\tau_3\simeq1$ Gyr instead of $\tau_3\simeq 0.1$ Gyr) and star formation efficiency ($\nu_3\simeq(2/3)\nu_2$ instead of $\nu_3\simeq(1/2)\nu_2$).
    The difference in this results can be explained by the different data sample adopted in this work. In fact, our model are thought to reproduce a sample of OCs cleaned of stars subject to biases in chemical abundance determination (\citealt{Magrini23}, see also point 1 of this Section), whose class is instead considered in \citet{Spitoni23} (see \citealt{Gaia23,RecioBlanco23}).
\end{enumerate}

Further data are definitely needed to probe the new three-infall scenario of chemical evolution on the whole MW disk, imposing more stringent limits and maybe revising the values of the physical parameters that have been adopted in this work.

However, the constraints coming from high-resolution, unbiased chemical abundances, precise ages and star formation histories from multiple tracers (e.g. \citealt{Isern19,Mor19,Ruiz20}) allow us to consider the proposed model as a robust and viable scenario for the MW disk formation.

\begin{acknowledgements}
    The authors thank the referee for the careful reading of the manuscript and the useful comments improving the paper content.
      MP acknowledges financial support from the project "LEGO – Reconstructing the building blocks of the Galaxy by chemical tagging" granted by the Italian MUR through contract PRIN2022LLP8TK\_001.
       LM,  ES, MF, and SR thank INAF for the support (Large Grant EPOCH) and MP, LM, and CVV for the MiniGrant Checs. LM, MF, and SR acknowledge financial support under the National Recovery and Resilience Plan (NRRP), Mission 4, Component 2, Investment 1.1, Call for tender No. 104 published on 2.2.2022 by the Italian Ministry of University and Research (MUR), funded by the European Union – NextGenerationEU– Project ‘Cosmic POT’  Grant Assignment Decree No. 2022X4TM3H  by the Italian Ministry of Ministry of University and Research (MUR).
       MM thanks the Deutsche Forschungsgemeinschaft (DFG, German Research Foundation) – Project-ID 279384907 – SFB 1245, the State of Hessen within the Research Cluster ELEMENTS (Project ID 500/10.006) for financial support.
       This research was supported by the Munich Institute for Astro-, Particle and BioPhysics (MIAPbP) which is funded by the Deutsche Forschungsgemeinschaft (DFG, German Research Foundation) under Germany´s Excellence Strategy – EXC-2094 – 390783311.
       Based on data products from observations made with ESO
       Telescopes at the La Silla Paranal Observatory under programmes 188.B-3002, 193.B-0936, and 197.B-1074. 
       These data products have been processed by the Cambridge Astronomy Survey Unit (CASU) at the Institute of Astronomy, University of Cambridge, and by the FLAMES/UVES reduction team at INAF/Osservatorio Astrofisico di Arcetri. These data have been obtained from the {\it Gaia}-ESO Survey Data Archive, prepared and hosted by the Wide Field Astronomy Unit, Institute for Astronomy, University of Edinburgh, which is funded by the UK Science and Technology Facilities Council. 
       This work makes use of results from the European Space Agency (ESA) space mission {\it Gaia}. Gaia data are being processed by the Gaia Data Processing and Analysis Consortium (DPAC). Funding for the DPAC is provided by national institutions, in particular the institutions participating in the Gaia MultiLateral Agreement (MLA). The Gaia mission website is \hyperlink{https://www.cosmos.esa.int/gaia}{https://www.cosmos.esa.int/gaia}. The Gaia archive website is \hyperlink{https://archives.esac.esa.int/gaia}{https://archives.esac.esa.int/gaia}.

\end{acknowledgements}


   \bibliographystyle{aa} 
   \bibliography{3inf_gradient} 

\begin{thebibliography}{155}
\expandafter\ifx\csname natexlab\endcsname\relax\def\natexlab#1{#1}\fi

\bibitem[{{Abdurro'uf} {et~al.}(2022){Abdurro'uf}, {Accetta}, {Aerts}, {Silva
  Aguirre}, {Ahumada}, {Ajgaonkar}, {Filiz Ak}, {Alam}, {Allende Prieto},
  {Almeida}, {Anders}, {Anderson}, {Andrews}, {Anguiano}, {Aquino-Ort{\'\i}z},
  {Arag{\'o}n-Salamanca}, {Argudo-Fern{\'a}ndez}, {Ata}, {Aubert},
  {Avila-Reese}, {Badenes}, {Barb{\'a}}, {Barger}, {Barrera-Ballesteros},
  {Beaton}, {Beers}, {Belfiore}, {Bender}, {Bernardi}, {Bershady}, {Beutler},
  {Bidin}, {Bird}, {Bizyaev}, {Blanc}, {Blanton}, {Boardman}, {Bolton},
  {Boquien}, {Borissova}, {Bovy}, {Brandt}, {Brown}, {Brownstein}, {Brusa},
  {Buchner}, {Bundy}, {Burchett}, {Bureau}, {Burgasser}, {Cabang}, {Campbell},
  {Cappellari}, {Carlberg}, {Wanderley}, {Carrera}, {Cash}, {Chen}, {Chen},
  {Cherinka}, {Chiappini}, {Choi}, {Chojnowski}, {Chung}, {Clerc}, {Cohen},
  {Comerford}, {Comparat}, {da Costa}, {Covey}, {Crane}, {Cruz-Gonzalez},
  {Culhane}, {Cunha}, {Dai}, {Damke}, {Darling}, {Davidson}, {Davies},
  {Dawson}, {De Lee}, {Diamond-Stanic}, {Cano-D{\'\i}az}, {S{\'a}nchez},
  {Donor}, {Duckworth}, {Dwelly}, {Eisenstein}, {Elsworth}, {Emsellem},
  {Eracleous}, {Escoffier}, {Fan}, {Farr}, {Feng}, {Fern{\'a}ndez-Trincado},
  {Feuillet}, {Filipp}, {Fillingham}, {Frinchaboy}, {Fromenteau}, {Galbany},
  {Garc{\'\i}a}, {Garc{\'\i}a-Hern{\'a}ndez}, {Ge}, {Geisler}, {Gelfand},
  {G{\'e}ron}, {Gibson}, {Goddy}, {Godoy-Rivera}, {Grabowski}, {Green},
  {Greener}, {Grier}, {Griffith}, {Guo}, {Guy}, {Hadjara}, {Harding},
  {Hasselquist}, {Hayes}, {Hearty}, {Hern{\'a}ndez}, {Hill}, {Hogg},
  {Holtzman}, {Horta}, {Hsieh}, {Hsu}, {Hsu}, {Huber}, {Huertas-Company},
  {Hutchinson}, {Hwang}, {Ibarra-Medel}, {Chitham}, {Ilha}, {Imig}, {Jaekle},
  {Jayasinghe}, {Ji}, {Johnson}, {Jones}, {J{\"o}nsson}, {Katkov}, {Khalatyan},
  {Kinemuchi}, {Kisku}, {Knapen}, {Kneib}, {Kollmeier}, {Kong}, {Kounkel},
  {Kreckel}, {Krishnarao}, {Lacerna}, {Lane}, {Langgin}, {Lavender}, {Law},
  {Lazarz}, {Leung}, {Leung}, {Lewis}, {Li}, {Li}, {Lian}, {Liang}, {Lin},
  {Lin}, {Lin}, {Lintott}, {Long}, {Longa-Pe{\~n}a}, {L{\'o}pez-Cob{\'a}},
  {Lu}, {Lundgren}, {Luo}, {Mackereth}, {de la Macorra}, {Mahadevan},
  {Majewski}, {Manchado}, {Mandeville}, {Maraston}, {Margalef-Bentabol},
  {Masseron}, {Masters}, {Mathur}, {McDermid}, {Mckay}, {Merloni},
  {Merrifield}, {Meszaros}, {Miglio}, {Di Mille}, {Minniti}, {Minsley},
  {Monachesi}, {Moon}, {Mosser}, {Mulchaey}, {Muna}, {Mu{\~n}oz}, {Myers},
  {Myers}, {Nadathur}, {Nair}, {Nandra}, {Neumann}, {Newman}, {Nidever},
  {Nikakhtar}, {Nitschelm}, {O'Connell}, {Garma-Oehmichen}, {Luan Souza de
  Oliveira}, {Olney}, {Oravetz}, {Ortigoza-Urdaneta}, {Osorio}, {Otter},
  {Pace}, {Padilla}, {Pan}, {Pan}, {Parikh}, {Parker}, {Peirani}, {Pe{\~n}a
  Ram{\'\i}rez}, {Penny}, {Percival}, {Perez-Fournon}, {Pinsonneault},
  {Poidevin}, {Poovelil}, {Price-Whelan}, {B{\'a}rbara de Andrade Queiroz},
  {Raddick}, {Ray}, {Rembold}, {Riddle}, {Riffel}, {Riffel}, {Rix}, {Robin},
  {Rodr{\'\i}guez-Puebla}, {Roman-Lopes}, {Rom{\'a}n-Z{\'u}{\~n}iga}, {Rose},
  {Ross}, {Rossi}, {Rubin}, {Salvato}, {S{\'a}nchez}, {S{\'a}nchez-Gallego},
  {Sanderson}, {Santana Rojas}, {Sarceno}, {Sarmiento}, {Sayres}, {Sazonova},
  {Schaefer}, {Schiavon}, {Schlegel}, {Schneider}, {Schultheis}, {Schwope},
  {Serenelli}, {Serna}, {Shao}, {Shapiro}, {Sharma}, {Shen}, {Shetrone}, {Shu},
  {Simon}, {Skrutskie}, {Smethurst}, {Smith}, {Sobeck}, {Spoo}, {Sprague},
  {Stark}, {Stassun}, {Steinmetz}, {Stello}, {Stone-Martinez},
  {Storchi-Bergmann}, {Stringfellow}, {Stutz}, {Su}, {Taghizadeh-Popp},
  {Talbot}, {Tayar}, {Telles}, {Teske}, {Thakar}, {Theissen}, {Tkachenko},
  {Thomas}, {Tojeiro}, {Hernandez Toledo}, {Troup}, {Trump}, {Trussler},
  {Turner}, {Tuttle}, {Unda-Sanzana}, {V{\'a}zquez-Mata}, {Valentini},
  {Valenzuela}, {Vargas-Gonz{\'a}lez}, {Vargas-Maga{\~n}a}, {Alfaro},
  {Villanova}, {Vincenzo}, {Wake}, {Warfield}, {Washington}, {Weaver},
  {Weijmans}, {Weinberg}, {Weiss}, {Westfall}, {Wild}, {Wilde}, {Wilson},
  {Wilson}, {Wilson}, {Wolf}, {Wood-Vasey}, {Yan}, {Zamora}, {Zasowski},
  {Zhang}, {Zhao}, {Zheng}, {Zheng}, \& {Zhu}}]{APOGEE17}
{Abdurro'uf}, {Accetta}, K., {Aerts}, C., {et~al.} 2022, \apjs, 259, 35

\bibitem[{{Ahumada} {et~al.}(2020){Ahumada}, {Prieto}, {Almeida}, {Anders},
  {Anderson}, {Andrews}, {Anguiano}, {Arcodia}, {Armengaud}, {Aubert}, {Avila},
  {Avila-Reese}, {Badenes}, {Balland}, {Barger}, {Barrera-Ballesteros}, {Basu},
  {Bautista}, {Beaton}, {Beers}, {Benavides}, {Bender}, {Bernardi}, {Bershady},
  {Beutler}, {Bidin}, {Bird}, {Bizyaev}, {Blanc}, {Blanton}, {Boquien},
  {Borissova}, {Bovy}, {Brandt}, {Brinkmann}, {Brownstein}, {Bundy}, {Bureau},
  {Burgasser}, {Burtin}, {Cano-D{\'\i}az}, {Capasso}, {Cappellari}, {Carrera},
  {Chabanier}, {Chaplin}, {Chapman}, {Cherinka}, {Chiappini}, {Doohyun Choi},
  {Chojnowski}, {Chung}, {Clerc}, {Coffey}, {Comerford}, {Comparat}, {da
  Costa}, {Cousinou}, {Covey}, {Crane}, {Cunha}, {Ilha}, {Dai}, {Damsted},
  {Darling}, {Davidson}, {Davies}, {Dawson}, {De}, {de la Macorra}, {De Lee},
  {Queiroz}, {Deconto Machado}, {de la Torre}, {Dell'Agli}, {du Mas des
  Bourboux}, {Diamond-Stanic}, {Dillon}, {Donor}, {Drory}, {Duckworth},
  {Dwelly}, {Ebelke}, {Eftekharzadeh}, {Davis Eigenbrot}, {Elsworth},
  {Eracleous}, {Erfanianfar}, {Escoffier}, {Fan}, {Farr},
  {Fern{\'a}ndez-Trincado}, {Feuillet}, {Finoguenov}, {Fofie},
  {Fraser-McKelvie}, {Frinchaboy}, {Fromenteau}, {Fu}, {Galbany}, {Garcia},
  {Garc{\'\i}a-Hern{\'a}ndez}, {Oehmichen}, {Ge}, {Maia}, {Geisler}, {Gelfand},
  {Goddy}, {Gonzalez-Perez}, {Grabowski}, {Green}, {Grier}, {Guo}, {Guy},
  {Harding}, {Hasselquist}, {Hawken}, {Hayes}, {Hearty}, {Hekker}, {Hogg},
  {Holtzman}, {Horta}, {Hou}, {Hsieh}, {Huber}, {Hunt}, {Chitham}, {Imig},
  {Jaber}, {Angel}, {Johnson}, {Jones}, {J{\"o}nsson}, {Jullo}, {Kim},
  {Kinemuchi}, {Kirkpatrick}, {Kite}, {Klaene}, {Kneib}, {Kollmeier}, {Kong},
  {Kounkel}, {Krishnarao}, {Lacerna}, {Lan}, {Lane}, {Law}, {Le Goff}, {Leung},
  {Lewis}, {Li}, {Lian}, {Lin}, {Long}, {Longa-Pe{\~n}a}, {Lundgren}, {Lyke},
  {Ted Mackereth}, {MacLeod}, {Majewski}, {Manchado}, {Maraston}, {Martini},
  {Masseron}, {Masters}, {Mathur}, {McDermid}, {Merloni}, {Merrifield},
  {M{\'e}sz{\'a}ros}, {Miglio}, {Minniti}, {Minsley}, {Miyaji}, {Mohammad},
  {Mosser}, {Mueller}, {Muna}, {Mu{\~n}oz-Guti{\'e}rrez}, {Myers}, {Nadathur},
  {Nair}, {Nandra}, {do Nascimento}, {Nevin}, {Newman}, {Nidever}, {Nitschelm},
  {Noterdaeme}, {O'Connell}, {Olmstead}, {Oravetz}, {Oravetz}, {Osorio},
  {Pace}, {Padilla}, {Palanque-Delabrouille}, {Palicio}, {Pan}, {Pan},
  {Parker}, {Paviot}, {Peirani}, {Ram{\'r}ez}, {Penny}, {Percival},
  {Perez-Fournon}, {P{\'e}rez-R{\`a}fols}, {Petitjean}, {Pieri},
  {Pinsonneault}, {Poovelil}, {Povick}, {Prakash}, {Price-Whelan}, {Raddick},
  {Raichoor}, {Ray}, {Rembold}, {Rezaie}, {Riffel}, {Riffel}, {Rix}, {Robin},
  {Roman-Lopes}, {Rom{\'a}n-Z{\'u}{\~n}iga}, {Rose}, {Ross}, {Rossi},
  {Rowlands}, {Rubin}, {Salvato}, {S{\'a}nchez}, {S{\'a}nchez-Menguiano},
  {S{\'a}nchez-Gallego}, {Sayres}, {Schaefer}, {Schiavon}, {Schimoia},
  {Schlafly}, {Schlegel}, {Schneider}, {Schultheis}, {Schwope}, {Seo},
  {Serenelli}, {Shafieloo}, {Shamsi}, {Shao}, {Shen}, {Shetrone}, {Shirley},
  {Aguirre}, {Simon}, {Skrutskie}, {Slosar}, {Smethurst}, {Sobeck}, {Sodi},
  {Souto}, {Stark}, {Stassun}, {Steinmetz}, {Stello}, {Stermer},
  {Storchi-Bergmann}, {Streblyanska}, {Stringfellow}, {Stutz}, {Su{\'a}rez},
  {Sun}, {Taghizadeh-Popp}, {Talbot}, {Tayar}, {Thakar}, {Theriault}, {Thomas},
  {Thomas}, {Tinker}, {Tojeiro}, {Toledo}, {Tremonti}, {Troup}, {Tuttle},
  {Unda-Sanzana}, {Valentini}, {Vargas-Gonz{\'a}lez}, {Vargas-Maga{\~n}a},
  {V{\'a}zquez-Mata}, {Vivek}, {Wake}, {Wang}, {Weaver}, {Weijmans}, {Wild},
  {Wilson}, {Wilson}, {Wolthuis}, {Wood-Vasey}, {Yan}, {Yang}, {Y{\`e}che},
  {Zamora}, {Zarrouk}, {Zasowski}, {Zhang}, {Zhao}, {Zhao}, {Zheng}, {Zheng},
  {Zhu}, \& {Zou}}]{Ahumada19}
{Ahumada}, R., {Prieto}, C.~A., {Almeida}, A., {et~al.} 2020, \apjs, 249, 3

\bibitem[{{Anders} {et~al.}(2017){Anders}, {Chiappini}, {Minchev}, {Miglio},
  {Montalb{\'a}n}, {Mosser}, {Rodrigues}, {Santiago}, {Baudin}, {Beers}, {da
  Costa}, {Garc{\'\i}a}, {Garc{\'\i}a-Hern{\'a}ndez}, {Holtzman}, {Maia},
  {Majewski}, {Mathur}, {Noels-Grotsch}, {Pan}, {Schneider}, {Schultheis},
  {Steinmetz}, {Valentini}, \& {Zamora}}]{Anders17}
{Anders}, F., {Chiappini}, C., {Minchev}, I., {et~al.} 2017, \aap, 600, A70

\bibitem[{{Baba} \& {Kawata}(2020)}]{Baba20}
{Baba}, J. \& {Kawata}, D. 2020, \mnras, 492, 4500

\bibitem[{{Balser} {et~al.}(2011){Balser}, {Rood}, {Bania}, \&
  {Anderson}}]{Balser11}
{Balser}, D.~S., {Rood}, R.~T., {Bania}, T.~M., \& {Anderson}, L.~D. 2011,
  \apj, 738, 27

\bibitem[{{Bernard}(2017)}]{Bernard17}
{Bernard}, E.~J. 2017, in SF2A-2017: Proceedings of the Annual meeting of the
  French Society of Astronomy and Astrophysics, ed. C.~{Reyl{\'e}}, P.~{Di
  Matteo}, F.~{Herpin}, E.~{Lagadec}, A.~{Lan{\c{c}}on}, Z.~{Meliani}, \&
  F.~{Royer}, Di

\bibitem[{{Bilitewski} \& {Sch{\"o}nrich}(2012)}]{Bilitewski12}
{Bilitewski}, T. \& {Sch{\"o}nrich}, R. 2012, \mnras, 426, 2266

\bibitem[{{Bovy}(2015)}]{Bovy15}
{Bovy}, J. 2015, \apjs, 216, 29

\bibitem[{{Bragaglia} {et~al.}(2022){Bragaglia}, {Alfaro}, {Flaccomio},
  {Blomme}, {Donati}, {Costado}, {Damiani}, {Franciosini}, {Prisinzano},
  {Randich}, {Friel}, {Hatztidimitriou}, {Vallenari}, {Spagna},
  {Balaguer-Nunez}, {Bonito}, {Cantat Gaudin}, {Casamiquela}, {Jeffries},
  {Jordi}, {Magrini}, {Drew}, {Jackson}, {Abbas}, {Caramazza}, {Hayes},
  {Jim{\'e}nez-Esteban}, {Re Fiorentin}, {Wright}, {Bayo}, {Bensby},
  {Bergemann}, {Gilmore}, {Gonneau}, {Heiter}, {Hourihane}, {Pancino}, {Sacco},
  {Smiljanic}, {Zaggia}, \& {Vink}}]{Bragaglia22}
{Bragaglia}, A., {Alfaro}, E.~J., {Flaccomio}, E., {et~al.} 2022, \aap, 659,
  A200

\bibitem[{{Bragan{\c{c}}a} {et~al.}(2019){Bragan{\c{c}}a}, {Daflon}, {Lanz},
  {Cunha}, {Bensby}, {McMillan}, {Garmany}, {Glaspey}, {Borges Fernandes},
  {Oey}, \& {Hubeny}}]{Braganca19}
{Bragan{\c{c}}a}, G.~A., {Daflon}, S., {Lanz}, T., {et~al.} 2019, \aap, 625,
  A120

\bibitem[{{Buck}(2020)}]{Buck20}
{Buck}, T. 2020, \mnras, 491, 5435

\bibitem[{{Buder} {et~al.}(2019){Buder}, {Lind}, {Ness}, {Asplund}, {Duong},
  {Lin}, {Kos}, {Casagrande}, {Casey}, {Bland-Hawthorn}, {de Silva}, {D'Orazi},
  {Freeman}, {Martell}, {Schlesinger}, {Sharma}, {Simpson}, {Zucker},
  {Zwitter}, {{\v{C}}otar}, {Dotter}, {Hayden}, {Hyde}, {Kafle}, {Lewis},
  {Nataf}, {Nordlander}, {Reid}, {Rix}, {Sk{\'u}lad{\'o}ttir}, {Stello},
  {Ting}, {Traven}, {Wyse}, \& {Galah Collaboration}}]{Buder19}
{Buder}, S., {Lind}, K., {Ness}, M.~K., {et~al.} 2019, \aap, 624, A19

\bibitem[{{Buder} {et~al.}(2021){Buder}, {Sharma}, {Kos}, {Amarsi},
  {Nordlander}, {Lind}, {Martell}, {Asplund}, {Bland-Hawthorn}, {Casey}, {de
  Silva}, {D'Orazi}, {Freeman}, {Hayden}, {Lewis}, {Lin}, {Schlesinger},
  {Simpson}, {Stello}, {Zucker}, {Zwitter}, {Beeson}, {Buck}, {Casagrande},
  {Clark}, {{\v{C}}otar}, {da Costa}, {de Grijs}, {Feuillet}, {Horner},
  {Kafle}, {Khanna}, {Kobayashi}, {Liu}, {Montet}, {Nandakumar}, {Nataf},
  {Ness}, {Spina}, {Tepper-Garc{\'\i}a}, {Ting}, {Traven},
  {Vogrin{\v{c}}i{\v{c}}}, {Wittenmyer}, {Wyse}, {{\v{Z}}erjal}, \& {Galah
  Collaboration}}]{Buder21}
{Buder}, S., {Sharma}, S., {Kos}, J., {et~al.} 2021, \mnras, 506, 150

\bibitem[{{Cantat-Gaudin} {et~al.}(2020){Cantat-Gaudin}, {Anders},
  {Castro-Ginard}, {Jordi}, {Romero-G{\'o}mez}, {Soubiran}, {Casamiquela},
  {Tarricq}, {Moitinho}, {Vallenari}, {Bragaglia}, {Krone-Martins}, \&
  {Kounkel}}]{Cantat2020A&A...640A...1C}
{Cantat-Gaudin}, T., {Anders}, F., {Castro-Ginard}, A., {et~al.} 2020, \aap,
  640, A1

\bibitem[{{Carbajo-Hijarrubia} {et~al.}(2024){Carbajo-Hijarrubia},
  {Casamiquela}, {Carrera}, {Balaguer-N{\'u}{\~n}ez}, {Jordi}, {Anders},
  {Gallart}, {Pancino}, {Drazdauskas}, {Stonkute}, {Tautvai{\v{s}}iene},
  {Carrasco}, {Masana}, {Cantat-Gaudin}, \& {Blanco-Cuaraesma}}]{Carbajo24}
{Carbajo-Hijarrubia}, J., {Casamiquela}, L., {Carrera}, R., {et~al.} 2024,
  arXiv e-prints, arXiv:2405.00110

\bibitem[{{Carraro} {et~al.}(2007){Carraro}, {Geisler}, {Villanova},
  {Frinchaboy}, \& {Majewski}}]{Carraro07}
{Carraro}, G., {Geisler}, D., {Villanova}, S., {Frinchaboy}, P.~M., \&
  {Majewski}, S.~R. 2007, \aap, 476, 217

\bibitem[{{Casali} {et~al.}(2020{\natexlab{a}}){Casali}, {Magrini}, {Frasca},
  {Bragaglia}, {Catanzaro}, {D'Orazi}, {Sordo}, {Carretta}, {Origlia},
  {Andreuzzi}, {Fu}, \& {Vallenari}}]{Casali2020A&A...643A..12C}
{Casali}, G., {Magrini}, L., {Frasca}, A., {et~al.} 2020{\natexlab{a}}, \aap,
  643, A12

\bibitem[{{Casali} {et~al.}(2019){Casali}, {Magrini}, {Tognelli}, {Jackson},
  {Jeffries}, {Lagarde}, {Tautvai{\v{s}}ien{\.{e}}}, {Masseron},
  {Degl'Innocenti}, {Prada Moroni}, {Kordopatis}, {Pancino}, {Randich},
  {Feltzing}, {Sahlholdt}, {Spina}, {Friel}, {Roccatagliata}, {Sanna},
  {Bragaglia}, {Drazdauskas}, {Mikolaitis}, {Minkevi{\v{c}}i{\={u}}t{\.{e}}},
  {Stonkut{\.{e}}}, {Chorniy}, {Bagdonas}, {Jimenez-Esteban}, {Martell}, {Van
  der Swaelmen}, {Gilmore}, {Vallenari}, {Bensby}, {Koposov}, {Korn}, {Worley},
  {Smiljanic}, {Bergemann}, {Carraro}, {Damiani}, {Prisinzano}, {Bonito},
  {Franciosini}, {Gonneau}, {Hourihane}, {Jofre}, {Lewis}, {Morbidelli},
  {Sacco}, {Sousa}, {Zaggia}, {Lanzafame}, {Heiter}, {Frasca}, \&
  {Bayo}}]{Casali2019A&A...629A..62C}
{Casali}, G., {Magrini}, L., {Tognelli}, E., {et~al.} 2019, \aap, 629, A62

\bibitem[{{Casali} {et~al.}(2020{\natexlab{b}}){Casali}, {Spina}, {Magrini},
  {Karakas}, {Kobayashi}, {Casey}, {Feltzing}, {Van der Swaelmen}, {Tsantaki},
  {Jofr{\'e}}, {Bragaglia}, {Feuillet}, {Bensby}, {Biazzo}, {Gonneau},
  {Tautvai{\v{s}}ien{\.{e}}}, {Baratella}, {Roccatagliata}, {Pancino}, {Sousa},
  {Adibekyan}, {Martell}, {Bayo}, {Jackson}, {Jeffries}, {Gilmore}, {Randich},
  {Alfaro}, {Koposov}, {Korn}, {Recio-Blanco}, {Smiljanic}, {Franciosini},
  {Hourihane}, {Monaco}, {Morbidelli}, {Sacco}, {Worley}, \&
  {Zaggia}}]{Casali2020A&A...639A.127C}
{Casali}, G., {Spina}, L., {Magrini}, L., {et~al.} 2020{\natexlab{b}}, \aap,
  639, A127

\bibitem[{{Cavichia} {et~al.}(2014){Cavichia}, {Moll{\'a}}, {Costa}, \&
  {Maciel}}]{Cavichia14}
{Cavichia}, O., {Moll{\'a}}, M., {Costa}, R.~D.~D., \& {Maciel}, W.~J. 2014,
  \mnras, 437, 3688

\bibitem[{{Chiappini} {et~al.}(1997){Chiappini}, {Matteucci}, \&
  {Gratton}}]{Chiappini97}
{Chiappini}, C., {Matteucci}, F., \& {Gratton}, R. 1997, \apj, 477, 765

\bibitem[{{Chiappini} {et~al.}(2001){Chiappini}, {Matteucci}, \&
  {Romano}}]{Chiappini01}
{Chiappini}, C., {Matteucci}, F., \& {Romano}, D. 2001, \apj, 554, 1044

\bibitem[{{Colavitti} {et~al.}(2009){Colavitti}, {Cescutti}, {Matteucci}, \&
  {Murante}}]{Colavitti09}
{Colavitti}, E., {Cescutti}, G., {Matteucci}, F., \& {Murante}, G. 2009, \aap,
  496, 429

\bibitem[{{da Silva} {et~al.}(2022){da Silva}, {Crestani}, {Bono}, {Braga},
  {D'Orazi}, {Lemasle}, {Bergemann}, {Dall'Ora}, {Fiorentino},
  {Fran{\c{c}}ois}, {Groenewegen}, {Inno}, {Kovtyukh}, {Kudritzki},
  {Matsunaga}, {Monelli}, {Pietrinferni}, {Porcelli}, {Storm}, {Tantalo}, \&
  {Th{\'e}v{\'e}nin}}]{DaSilva22}
{da Silva}, R., {Crestani}, J., {Bono}, G., {et~al.} 2022, \aap, 661, A104

\bibitem[{{da Silva} {et~al.}(2023){da Silva}, {D'Orazi}, {Palla}, {Bono},
  {Braga}, {Fabrizio}, {Lemasle}, {Spitoni}, {Matteucci}, {J{\"o}nsson},
  {Kovtyukh}, {Magrini}, {Bergemann}, {Dall'Ora}, {Ferraro}, {Fiorentino},
  {Fran{\c{c}}ois}, {Iannicola}, {Inno}, {Kudritzki}, {Matsunaga}, {Monelli},
  {Nonino}, {Sneden}, {Storm}, {Th{\'e}v{\'e}nin}, {Tsujimoto}, \&
  {Zocchi}}]{DaSilva23}
{da Silva}, R., {D'Orazi}, V., {Palla}, M., {et~al.} 2023, \aap, 678, A195

\bibitem[{{Daflon} \& {Cunha}(2004)}]{Daflon04}
{Daflon}, S. \& {Cunha}, K. 2004, \apj, 617, 1115

\bibitem[{{Dame}(1993)}]{Dame93}
{Dame}, T.~M. 1993, in American Institute of Physics Conference Series, Vol.
  278, Back to the Galaxy, ed. S.~S. {Holt} \& F.~{Verter}, 267--278

\bibitem[{{de Jong} {et~al.}(2012){de Jong}, {Bellido-Tirado}, {Chiappini},
  {Depagne}, {Haynes}, {Johl}, {Schnurr}, {Schwope}, {Walcher}, {Dionies},
  {Haynes}, {Kelz}, {Kitaura}, {Lamer}, {Minchev}, {M{\"u}ller}, {Nuza},
  {Olaya}, {Piffl}, {Popow}, {Steinmetz}, {Ural}, {Williams}, {Winkler},
  {Wisotzki}, {Ansorge}, {Banerji}, {Gonzalez Solares}, {Irwin}, {Kennicutt},
  {King}, {McMahon}, {Koposov}, {Parry}, {Sun}, {Walton}, {Finger}, {Iwert},
  {Krumpe}, {Lizon}, {Vincenzo}, {Amans}, {Bonifacio}, {Cohen}, {Francois},
  {Jagourel}, {Mignot}, {Royer}, {Sartoretti}, {Bender}, {Grupp}, {Hess},
  {Lang-Bardl}, {Muschielok}, {B{\"o}hringer}, {Boller}, {Bongiorno}, {Brusa},
  {Dwelly}, {Merloni}, {Nandra}, {Salvato}, {Pragt}, {Navarro}, {Gerlofsma},
  {Roelfsema}, {Dalton}, {Middleton}, {Tosh}, {Boeche}, {Caffau}, {Christlieb},
  {Grebel}, {Hansen}, {Koch}, {Ludwig}, {Quirrenbach}, {Sbordone}, {Seifert},
  {Thimm}, {Trifonov}, {Helmi}, {Trager}, {Feltzing}, {Korn}, \&
  {Boland}}]{4MOSTpaper}
{de Jong}, R.~S., {Bellido-Tirado}, O., {Chiappini}, C., {et~al.} 2012, in
  Society of Photo-Optical Instrumentation Engineers (SPIE) Conference Series,
  Vol. 8446, Ground-based and Airborne Instrumentation for Astronomy IV, ed.
  I.~S. {McLean}, S.~K. {Ramsay}, \& H.~{Takami}, 84460T

\bibitem[{{Di Teodoro} \& {Peek}(2021)}]{DiTeodoro21}
{Di Teodoro}, E.~M. \& {Peek}, J.~E.~G. 2021, \apj, 923, 220

\bibitem[{{Donor} {et~al.}(2020){Donor}, {Frinchaboy}, {Cunha}, {O'Connell},
  {Allende Prieto}, {Almeida}, {Anders}, {Beaton}, {Bizyaev}, {Brownstein},
  {Carrera}, {Chiappini}, {Cohen}, {Garc{\'\i}a-Hern{\'a}ndez}, {Geisler},
  {Hasselquist}, {J{\"o}nsson}, {Lane}, {Majewski}, {Minniti}, {Bidin}, {Pan},
  {Roman-Lopes}, {Sobeck}, \& {Zasowski}}]{Donor20}
{Donor}, J., {Frinchaboy}, P.~M., {Cunha}, K., {et~al.} 2020, \aj, 159, 199

\bibitem[{{Esteban} {et~al.}(2017){Esteban}, {Fang}, {Garc{\'\i}a-Rojas}, \&
  {Toribio San Cipriano}}]{Esteban17}
{Esteban}, C., {Fang}, X., {Garc{\'\i}a-Rojas}, J., \& {Toribio San Cipriano},
  L. 2017, \mnras, 471, 987

\bibitem[{{Frankel} {et~al.}(2018){Frankel}, {Rix}, {Ting}, {Ness}, \&
  {Hogg}}]{Frankel18}
{Frankel}, N., {Rix}, H.-W., {Ting}, Y.-S., {Ness}, M., \& {Hogg}, D.~W. 2018,
  \apj, 865, 96

\bibitem[{{Frankel} {et~al.}(2020){Frankel}, {Sanders}, {Ting}, \&
  {Rix}}]{Frankel20}
{Frankel}, N., {Sanders}, J., {Ting}, Y.-S., \& {Rix}, H.-W. 2020, \apj, 896,
  15

\bibitem[{{Gaia Collaboration} {et~al.}(2018){Gaia Collaboration}, {Brown},
  {Vallenari}, {Prusti}, {de Bruijne}, {Babusiaux}, {Bailer-Jones}, {Biermann},
  {Evans}, {Eyer}, {Jansen}, {Jordi}, {Klioner}, {Lammers}, {Lindegren},
  {Luri}, {Mignard}, {Panem}, {Pourbaix}, {Randich}, {Sartoretti}, {Siddiqui},
  {Soubiran}, {van Leeuwen}, {Walton}, {Arenou}, {Bastian}, {Cropper},
  {Drimmel}, {Katz}, {Lattanzi}, {Bakker}, {Cacciari}, {Casta{\~n}eda},
  {Chaoul}, {Cheek}, {De Angeli}, {Fabricius}, {Guerra}, {Holl}, {Masana},
  {Messineo}, {Mowlavi}, {Nienartowicz}, {Panuzzo}, {Portell}, {Riello},
  {Seabroke}, {Tanga}, {Th{\'e}venin}, {Gracia-Abril}, {Comoretto},
  {Garcia-Reinaldos}, {Teyssier}, {Altmann}, {Andrae}, {Audard},
  {Bellas-Velidis}, {Benson}, {Berthier}, {Blomme}, {Burgess}, {Busso},
  {Carry}, {Cellino}, {Clementini}, {Clotet}, {Creevey}, {Davidson}, {De
  Ridder}, {Delchambre}, {Dell'Oro}, {Ducourant},
  {Fern{\'a}ndez-Hern{\'a}ndez}, {Fouesneau}, {Fr{\'e}mat}, {Galluccio},
  {Garc{\'\i}a-Torres}, {Gonz{\'a}lez-N{\'u}{\~n}ez}, {Gonz{\'a}lez-Vidal},
  {Gosset}, {Guy}, {Halbwachs}, {Hambly}, {Harrison}, {Hern{\'a}ndez},
  {Hestroffer}, {Hodgkin}, {Hutton}, {Jasniewicz}, {Jean-Antoine-Piccolo},
  {Jordan}, {Korn}, {Krone-Martins}, {Lanzafame}, {Lebzelter}, {L{\"o}ffler},
  {Manteiga}, {Marrese}, {Mart{\'\i}n-Fleitas}, {Moitinho}, {Mora}, {Muinonen},
  {Osinde}, {Pancino}, {Pauwels}, {Petit}, {Recio-Blanco}, {Richards},
  {Rimoldini}, {Robin}, {Sarro}, {Siopis}, {Smith}, {Sozzetti}, {S{\"u}veges},
  {Torra}, {van Reeven}, {Abbas}, {Abreu Aramburu}, {Accart}, {Aerts},
  {Altavilla}, {{\'A}lvarez}, {Alvarez}, {Alves}, {Anderson}, {Andrei},
  {Anglada Varela}, {Antiche}, {Antoja}, {Arcay}, {Astraatmadja}, {Bach},
  {Baker}, {Balaguer-N{\'u}{\~n}ez}, {Balm}, {Barache}, {Barata}, {Barbato},
  {Barblan}, {Barklem}, {Barrado}, {Barros}, {Barstow}, {Bartholom{\'e}
  Mu{\~n}oz}, {Bassilana}, {Becciani}, {Bellazzini}, {Berihuete}, {Bertone},
  {Bianchi}, {Bienaym{\'e}}, {Blanco-Cuaresma}, {Boch}, {Boeche}, {Bombrun},
  {Borrachero}, {Bossini}, {Bouquillon}, {Bourda}, {Bragaglia}, {Bramante},
  {Breddels}, {Bressan}, {Brouillet}, {Br{\"u}semeister}, {Brugaletta},
  {Bucciarelli}, {Burlacu}, {Busonero}, {Butkevich}, {Buzzi}, {Caffau},
  {Cancelliere}, {Cannizzaro}, {Cantat-Gaudin}, {Carballo}, {Carlucci},
  {Carrasco}, {Casamiquela}, {Castellani}, {Castro-Ginard}, {Charlot},
  {Chemin}, {Chiavassa}, {Cocozza}, {Costigan}, {Cowell}, {Crifo}, {Crosta},
  {Crowley}, {Cuypers}, {Dafonte}, {Damerdji}, {Dapergolas}, {David}, {David},
  {de Laverny}, {De Luise}, {De March}, {de Martino}, {de Souza}, {de Torres},
  {Debosscher}, {del Pozo}, {Delbo}, {Delgado}, {Delgado}, {Di Matteo},
  {Diakite}, {Diener}, {Distefano}, {Dolding}, {Drazinos}, {Dur{\'a}n},
  {Edvardsson}, {Enke}, {Eriksson}, {Esquej}, {Eynard Bontemps}, {Fabre},
  {Fabrizio}, {Faigler}, {Falc{\~a}o}, {Farr{\`a}s Casas}, {Federici},
  {Fedorets}, {Fernique}, {Figueras}, {Filippi}, {Findeisen}, {Fonti},
  {Fraile}, {Fraser}, {Fr{\'e}zouls}, {Gai}, {Galleti}, {Garabato},
  {Garc{\'\i}a-Sedano}, {Garofalo}, {Garralda}, {Gavel}, {Gavras}, {Gerssen},
  {Geyer}, {Giacobbe}, {Gilmore}, {Girona}, {Giuffrida}, {Glass}, {Gomes},
  {Granvik}, {Gueguen}, {Guerrier}, {Guiraud}, {Guti{\'e}rrez-S{\'a}nchez},
  {Haigron}, {Hatzidimitriou}, {Hauser}, {Haywood}, {Heiter}, {Helmi}, {Heu},
  {Hilger}, {Hobbs}, {Hofmann}, {Holland}, {Huckle}, {Hypki}, {Icardi},
  {Jan{\ss}en}, {Jevardat de Fombelle}, {Jonker}, {Juh{\'a}sz}, {Julbe},
  {Karampelas}, {Kewley}, {Klar}, {Kochoska}, {Kohley}, {Kolenberg},
  {Kontizas}, {Kontizas}, {Koposov}, {Kordopatis}, {Kostrzewa-Rutkowska},
  {Koubsky}, {Lambert}, {Lanza}, {Lasne}, {Lavigne}, {Le Fustec}, {Le
  Poncin-Lafitte}, {Lebreton}, {Leccia}, {Leclerc}, {Lecoeur-Taibi},
  {Lenhardt}, {Leroux}, {Liao}, {Licata}, {Lindstr{\o}m}, {Lister}, {Livanou},
  {Lobel}, {L{\'o}pez}, {Managau}, {Mann}, {Mantelet}, {Marchal}, {Marchant},
  {Marconi}, {Marinoni}, {Marschalk{\'o}}, {Marshall}, {Martino}, {Marton},
  {Mary}, {Massari}, {Matijevi{\v{c}}}, {Mazeh}, {McMillan}, {Messina},
  {Michalik}, {Millar}, {Molina}, {Molinaro}, {Moln{\'a}r}, {Montegriffo},
  {Mor}, {Morbidelli}, {Morel}, {Morris}, {Mulone}, {Muraveva}, {Musella},
  {Nelemans}, {Nicastro}, {Noval}, {O'Mullane}, {Ord{\'e}novic},
  {Ord{\'o}{\~n}ez-Blanco}, {Osborne}, {Pagani}, {Pagano}, {Pailler},
  {Palacin}, {Palaversa}, {Panahi}, {Pawlak}, {Piersimoni}, {Pineau}, {Plachy},
  {Plum}, {Poggio}, {Poujoulet}, {Pr{\v{s}}a}, {Pulone}, {Racero}, {Ragaini},
  {Rambaux}, {Ramos-Lerate}, {Regibo}, {Reyl{\'e}}, {Riclet}, {Ripepi}, {Riva},
  {Rivard}, {Rixon}, {Roegiers}, {Roelens}, {Romero-G{\'o}mez}, {Rowell},
  {Royer}, {Ruiz-Dern}, {Sadowski}, {Sagrist{\`a} Sell{\'e}s}, {Sahlmann},
  {Salgado}, {Salguero}, {Sanna}, {Santana-Ros}, {Sarasso}, {Savietto},
  {Schultheis}, {Sciacca}, {Segol}, {Segovia}, {S{\'e}gransan}, {Shih},
  {Siltala}, {Silva}, {Smart}, {Smith}, {Solano}, {Solitro}, {Sordo}, {Soria
  Nieto}, {Souchay}, {Spagna}, {Spoto}, {Stampa}, {Steele},
  {Steidelm{\"u}ller}, {Stephenson}, {Stoev}, {Suess}, {Surdej}, {Szabados},
  {Szegedi-Elek}, {Tapiador}, {Taris}, {Tauran}, {Taylor}, {Teixeira},
  {Terrett}, {Teyssand ier}, {Thuillot}, {Titarenko}, {Torra Clotet}, {Turon},
  {Ulla}, {Utrilla}, {Uzzi}, {Vaillant}, {Valentini}, {Valette}, {van Elteren},
  {Van Hemelryck}, {van Leeuwen}, {Vaschetto}, {Vecchiato}, {Veljanoski},
  {Viala}, {Vicente}, {Vogt}, {von Essen}, {Voss}, {Votruba}, {Voutsinas},
  {Walmsley}, {Weiler}, {Wertz}, {Wevers}, {Wyrzykowski}, {Yoldas},
  {{\v{Z}}erjal}, {Ziaeepour}, {Zorec}, {Zschocke}, {Zucker}, {Zurbach}, \&
  {Zwitter}}]{Gaia18}
{Gaia Collaboration}, {Brown}, A.~G.~A., {Vallenari}, A., {et~al.} 2018, \aap,
  616, A1

\bibitem[{{Gaia Collaboration} {et~al.}(2021){Gaia Collaboration}, {Brown},
  {Vallenari}, {Prusti}, {de Bruijne}, {Babusiaux}, {Biermann}, {Creevey},
  {Evans}, {Eyer}, {Hutton}, {Jansen}, {Jordi}, {Klioner}, {Lammers},
  {Lindegren}, {Luri}, {Mignard}, {Panem}, {Pourbaix}, {Randich}, {Sartoretti},
  {Soubiran}, {Walton}, {Arenou}, {Bailer-Jones}, {Bastian}, {Cropper},
  {Drimmel}, {Katz}, {Lattanzi}, {van Leeuwen}, {Bakker}, {Cacciari},
  {Casta{\~n}eda}, {De Angeli}, {Ducourant}, {Fabricius}, {Fouesneau},
  {Fr{\'e}mat}, {Guerra}, {Guerrier}, {Guiraud}, {Jean-Antoine Piccolo},
  {Masana}, {Messineo}, {Mowlavi}, {Nicolas}, {Nienartowicz}, {Pailler},
  {Panuzzo}, {Riclet}, {Roux}, {Seabroke}, {Sordo}, {Tanga}, {Th{\'e}venin},
  {Gracia-Abril}, {Portell}, {Teyssier}, {Altmann}, {Andrae}, {Bellas-Velidis},
  {Benson}, {Berthier}, {Blomme}, {Brugaletta}, {Burgess}, {Busso}, {Carry},
  {Cellino}, {Cheek}, {Clementini}, {Damerdji}, {Davidson}, {Delchambre},
  {Dell'Oro}, {Fern{\'a}ndez-Hern{\'a}ndez}, {Galluccio}, {Garc{\'\i}a-Lario},
  {Garcia-Reinaldos}, {Gonz{\'a}lez-N{\'u}{\~n}ez}, {Gosset}, {Haigron},
  {Halbwachs}, {Hambly}, {Harrison}, {Hatzidimitriou}, {Heiter},
  {Hern{\'a}ndez}, {Hestroffer}, {Hodgkin}, {Holl}, {Jan{\ss}en}, {Jevardat de
  Fombelle}, {Jordan}, {Krone-Martins}, {Lanzafame}, {L{\"o}ffler}, {Lorca},
  {Manteiga}, {Marchal}, {Marrese}, {Moitinho}, {Mora}, {Muinonen}, {Osborne},
  {Pancino}, {Pauwels}, {Petit}, {Recio-Blanco}, {Richards}, {Riello},
  {Rimoldini}, {Robin}, {Roegiers}, {Rybizki}, {Sarro}, {Siopis}, {Smith},
  {Sozzetti}, {Ulla}, {Utrilla}, {van Leeuwen}, {van Reeven}, {Abbas}, {Abreu
  Aramburu}, {Accart}, {Aerts}, {Aguado}, {Ajaj}, {Altavilla}, {{\'A}lvarez},
  {{\'A}lvarez Cid-Fuentes}, {Alves}, {Anderson}, {Anglada Varela}, {Antoja},
  {Audard}, {Baines}, {Baker}, {Balaguer-N{\'u}{\~n}ez}, {Balbinot}, {Balog},
  {Barache}, {Barbato}, {Barros}, {Barstow}, {Bartolom{\'e}}, {Bassilana},
  {Bauchet}, {Baudesson-Stella}, {Becciani}, {Bellazzini}, {Bernet}, {Bertone},
  {Bianchi}, {Blanco-Cuaresma}, {Boch}, {Bombrun}, {Bossini}, {Bouquillon},
  {Bragaglia}, {Bramante}, {Breedt}, {Bressan}, {Brouillet}, {Bucciarelli},
  {Burlacu}, {Busonero}, {Butkevich}, {Buzzi}, {Caffau}, {Cancelliere},
  {C{\'a}novas}, {Cantat-Gaudin}, {Carballo}, {Carlucci}, {Carnerero},
  {Carrasco}, {Casamiquela}, {Castellani}, {Castro-Ginard}, {Castro Sampol},
  {Chaoul}, {Charlot}, {Chemin}, {Chiavassa}, {Cioni}, {Comoretto}, {Cooper},
  {Cornez}, {Cowell}, {Crifo}, {Crosta}, {Crowley}, {Dafonte}, {Dapergolas},
  {David}, {David}, {de Laverny}, {De Luise}, {De March}, {De Ridder}, {de
  Souza}, {de Teodoro}, {de Torres}, {del Peloso}, {del Pozo}, {Delbo},
  {Delgado}, {Delgado}, {Delisle}, {Di Matteo}, {Diakite}, {Diener},
  {Distefano}, {Dolding}, {Eappachen}, {Edvardsson}, {Enke}, {Esquej}, {Fabre},
  {Fabrizio}, {Faigler}, {Fedorets}, {Fernique}, {Fienga}, {Figueras},
  {Fouron}, {Fragkoudi}, {Fraile}, {Franke}, {Gai}, {Garabato},
  {Garcia-Gutierrez}, {Garc{\'\i}a-Torres}, {Garofalo}, {Gavras}, {Gerlach},
  {Geyer}, {Giacobbe}, {Gilmore}, {Girona}, {Giuffrida}, {Gomel}, {Gomez},
  {Gonzalez-Santamaria}, {Gonz{\'a}lez-Vidal}, {Granvik},
  {Guti{\'e}rrez-S{\'a}nchez}, {Guy}, {Hauser}, {Haywood}, {Helmi}, {Hidalgo},
  {Hilger}, {H{\l}adczuk}, {Hobbs}, {Holland}, {Huckle}, {Jasniewicz},
  {Jonker}, {Juaristi Campillo}, {Julbe}, {Karbevska}, {Kervella}, {Khanna},
  {Kochoska}, {Kontizas}, {Kordopatis}, {Korn}, {Kostrzewa-Rutkowska},
  {Kruszy{\'n}ska}, {Lambert}, {Lanza}, {Lasne}, {Le Campion}, {Le Fustec},
  {Lebreton}, {Lebzelter}, {Leccia}, {Leclerc}, {Lecoeur-Taibi}, {Liao},
  {Licata}, {Lindstr{\o}m}, {Lister}, {Livanou}, {Lobel}, {Madrero Pardo},
  {Managau}, {Mann}, {Marchant}, {Marconi}, {Marcos Santos}, {Marinoni},
  {Marocco}, {Marshall}, {Martin Polo}, {Mart{\'\i}n-Fleitas}, {Masip},
  {Massari}, {Mastrobuono-Battisti}, {Mazeh}, {McMillan}, {Messina},
  {Michalik}, {Millar}, {Mints}, {Molina}, {Molinaro}, {Moln{\'a}r},
  {Montegriffo}, {Mor}, {Morbidelli}, {Morel}, {Morris}, {Mulone}, {Munoz},
  {Muraveva}, {Murphy}, {Musella}, {Noval}, {Ord{\'e}novic}, {Orr{\`u}},
  {Osinde}, {Pagani}, {Pagano}, {Palaversa}, {Palicio}, {Panahi}, {Pawlak},
  {Pe{\~n}alosa Esteller}, {Penttil{\"a}}, {Piersimoni}, {Pineau}, {Plachy},
  {Plum}, {Poggio}, {Poretti}, {Poujoulet}, {Pr{\v{s}}a}, {Pulone}, {Racero},
  {Ragaini}, {Rainer}, {Raiteri}, {Rambaux}, {Ramos}, {Ramos-Lerate}, {Re
  Fiorentin}, {Regibo}, {Reyl{\'e}}, {Ripepi}, {Riva}, {Rixon}, {Robichon},
  {Robin}, {Roelens}, {Rohrbasser}, {Romero-G{\'o}mez}, {Rowell}, {Royer},
  {Rybicki}, {Sadowski}, {Sagrist{\`a} Sell{\'e}s}, {Sahlmann}, {Salgado},
  {Salguero}, {Samaras}, {Sanchez Gimenez}, {Sanna}, {Santove{\~n}a},
  {Sarasso}, {Schultheis}, {Sciacca}, {Segol}, {Segovia}, {S{\'e}gransan},
  {Semeux}, {Shahaf}, {Siddiqui}, {Siebert}, {Siltala}, {Slezak}, {Smart},
  {Solano}, {Solitro}, {Souami}, {Souchay}, {Spagna}, {Spoto}, {Steele},
  {Steidelm{\"u}ller}, {Stephenson}, {S{\"u}veges}, {Szabados}, {Szegedi-Elek},
  {Taris}, {Tauran}, {Taylor}, {Teixeira}, {Thuillot}, {Tonello}, {Torra},
  {Torra}, {Turon}, {Unger}, {Vaillant}, {van Dillen}, {Vanel}, {Vecchiato},
  {Viala}, {Vicente}, {Voutsinas}, {Weiler}, {Wevers}, {Wyrzykowski}, {Yoldas},
  {Yvard}, {Zhao}, {Zorec}, {Zucker}, {Zurbach}, \& {Zwitter}}]{Gaia21}
{Gaia Collaboration}, {Brown}, A.~G.~A., {Vallenari}, A., {et~al.} 2021, \aap,
  649, A1

\bibitem[{{Gaia Collaboration} {et~al.}(2016){Gaia Collaboration}, {Brown},
  {Vallenari}, {Prusti}, {de Bruijne}, {Mignard}, {Drimmel}, {Babusiaux},
  {Bailer-Jones}, {Bastian}, {Biermann}, {Evans}, {Eyer}, {Jansen}, {Jordi},
  {Katz}, {Klioner}, {Lammers}, {Lindegren}, {Luri}, {O'Mullane}, {Panem},
  {Pourbaix}, {Randich}, {Sartoretti}, {Siddiqui}, {Soubiran}, {Valette}, {van
  Leeuwen}, {Walton}, {Aerts}, {Arenou}, {Cropper}, {H{\o}g}, {Lattanzi},
  {Grebel}, {Holland}, {Huc}, {Passot}, {Perryman}, {Bramante}, {Cacciari},
  {Casta{\~n}eda}, {Chaoul}, {Cheek}, {De Angeli}, {Fabricius}, {Guerra},
  {Hern{\'a}ndez}, {Jean-Antoine-Piccolo}, {Masana}, {Messineo}, {Mowlavi},
  {Nienartowicz}, {Ord{\'o}{\~n}ez-Blanco}, {Panuzzo}, {Portell}, {Richards},
  {Riello}, {Seabroke}, {Tanga}, {Th{\'e}venin}, {Torra}, {Els},
  {Gracia-Abril}, {Comoretto}, {Garcia-Reinaldos}, {Lock}, {Mercier},
  {Altmann}, {Andrae}, {Astraatmadja}, {Bellas-Velidis}, {Benson}, {Berthier},
  {Blomme}, {Busso}, {Carry}, {Cellino}, {Clementini}, {Cowell}, {Creevey},
  {Cuypers}, {Davidson}, {De Ridder}, {de Torres}, {Delchambre}, {Dell'Oro},
  {Ducourant}, {Fr{\'e}mat}, {Garc{\'\i}a-Torres}, {Gosset}, {Halbwachs},
  {Hambly}, {Harrison}, {Hauser}, {Hestroffer}, {Hodgkin}, {Huckle}, {Hutton},
  {Jasniewicz}, {Jordan}, {Kontizas}, {Korn}, {Lanzafame}, {Manteiga},
  {Moitinho}, {Muinonen}, {Osinde}, {Pancino}, {Pauwels}, {Petit},
  {Recio-Blanco}, {Robin}, {Sarro}, {Siopis}, {Smith}, {Smith}, {Sozzetti},
  {Thuillot}, {van Reeven}, {Viala}, {Abbas}, {Abreu Aramburu}, {Accart},
  {Aguado}, {Allan}, {Allasia}, {Altavilla}, {{\'A}lvarez}, {Alves},
  {Anderson}, {Andrei}, {Anglada Varela}, {Antiche}, {Antoja}, {Ant{\'o}n},
  {Arcay}, {Bach}, {Baker}, {Balaguer-N{\'u}{\~n}ez}, {Barache}, {Barata},
  {Barbier}, {Barblan}, {Barrado y Navascu{\'e}s}, {Barros}, {Barstow},
  {Becciani}, {Bellazzini}, {Bello Garc{\'\i}a}, {Belokurov}, {Bendjoya},
  {Berihuete}, {Bianchi}, {Bienaym{\'e}}, {Billebaud}, {Blagorodnova},
  {Blanco-Cuaresma}, {Boch}, {Bombrun}, {Borrachero}, {Bouquillon}, {Bourda},
  {Bouy}, {Bragaglia}, {Breddels}, {Brouillet}, {Br{\"u}semeister},
  {Bucciarelli}, {Burgess}, {Burgon}, {Burlacu}, {Busonero}, {Buzzi}, {Caffau},
  {Cambras}, {Campbell}, {Cancelliere}, {Cantat-Gaudin}, {Carlucci},
  {Carrasco}, {Castellani}, {Charlot}, {Charnas}, {Chiavassa}, {Clotet},
  {Cocozza}, {Collins}, {Costigan}, {Crifo}, {Cross}, {Crosta}, {Crowley},
  {Dafonte}, {Damerdji}, {Dapergolas}, {David}, {David}, {De Cat}, {de Felice},
  {de Laverny}, {De Luise}, {De March}, {de Martino}, {de Souza}, {Debosscher},
  {del Pozo}, {Delbo}, {Delgado}, {Delgado}, {Di Matteo}, {Diakite},
  {Distefano}, {Dolding}, {Dos Anjos}, {Drazinos}, {Duran}, {Dzigan},
  {Edvardsson}, {Enke}, {Evans}, {Eynard Bontemps}, {Fabre}, {Fabrizio},
  {Faigler}, {Falc{\~a}o}, {Farr{\`a}s Casas}, {Federici}, {Fedorets},
  {Fern{\'a}ndez-Hern{\'a}ndez}, {Fernique}, {Fienga}, {Figueras}, {Filippi},
  {Findeisen}, {Fonti}, {Fouesneau}, {Fraile}, {Fraser}, {Fuchs}, {Gai},
  {Galleti}, {Galluccio}, {Garabato}, {Garc{\'\i}a-Sedano}, {Garofalo},
  {Garralda}, {Gavras}, {Gerssen}, {Geyer}, {Gilmore}, {Girona}, {Giuffrida},
  {Gomes}, {Gonz{\'a}lez-Marcos}, {Gonz{\'a}lez-N{\'u}{\~n}ez},
  {Gonz{\'a}lez-Vidal}, {Granvik}, {Guerrier}, {Guillout}, {Guiraud},
  {G{\'u}rpide}, {Guti{\'e}rrez-S{\'a}nchez}, {Guy}, {Haigron},
  {Hatzidimitriou}, {Haywood}, {Heiter}, {Helmi}, {Hobbs}, {Hofmann}, {Holl},
  {Holland}, {Hunt}, {Hypki}, {Icardi}, {Irwin}, {Jevardat de Fombelle},
  {Jofr{\'e}}, {Jonker}, {Jorissen}, {Julbe}, {Karampelas}, {Kochoska},
  {Kohley}, {Kolenberg}, {Kontizas}, {Koposov}, {Kordopatis}, {Koubsky},
  {Krone-Martins}, {Kudryashova}, {Kull}, {Bachchan}, {Lacoste-Seris}, {Lanza},
  {Lavigne}, {Le Poncin-Lafitte}, {Lebreton}, {Lebzelter}, {Leccia}, {Leclerc},
  {Lecoeur-Taibi}, {Lemaitre}, {Lenhardt}, {Leroux}, {Liao}, {Licata},
  {Lindstr{\o}m}, {Lister}, {Livanou}, {Lobel}, {L{\"o}ffler}, {L{\'o}pez},
  {Lorenz}, {MacDonald}, {Magalh{\~a}es Fernandes}, {Managau}, {Mann},
  {Mantelet}, {Marchal}, {Marchant}, {Marconi}, {Marinoni}, {Marrese},
  {Marschalk{\'o}}, {Marshall}, {Mart{\'\i}n-Fleitas}, {Martino}, {Mary},
  {Matijevi{\v{c}}}, {Mazeh}, {McMillan}, {Messina}, {Michalik}, {Millar},
  {Miranda}, {Molina}, {Molinaro}, {Molinaro}, {Moln{\'a}r}, {Moniez},
  {Montegriffo}, {Mor}, {Mora}, {Morbidelli}, {Morel}, {Morgenthaler},
  {Morris}, {Mulone}, {Muraveva}, {Musella}, {Narbonne}, {Nelemans},
  {Nicastro}, {Noval}, {Ord{\'e}novic}, {Ordieres-Mer{\'e}}, {Osborne},
  {Pagani}, {Pagano}, {Pailler}, {Palacin}, {Palaversa}, {Parsons}, {Pecoraro},
  {Pedrosa}, {Pentik{\"a}inen}, {Pichon}, {Piersimoni}, {Pineau}, {Plachy},
  {Plum}, {Poujoulet}, {Pr{\v{s}}a}, {Pulone}, {Ragaini}, {Rago}, {Rambaux},
  {Ramos-Lerate}, {Ranalli}, {Rauw}, {Read}, {Regibo}, {Reyl{\'e}}, {Ribeiro},
  {Rimoldini}, {Ripepi}, {Riva}, {Rixon}, {Roelens}, {Romero-G{\'o}mez},
  {Rowell}, {Royer}, {Ruiz-Dern}, {Sadowski}, {Sagrist{\`a} Sell{\'e}s},
  {Sahlmann}, {Salgado}, {Salguero}, {Sarasso}, {Savietto}, {Schultheis},
  {Sciacca}, {Segol}, {Segovia}, {Segransan}, {Shih}, {Smareglia}, {Smart},
  {Solano}, {Solitro}, {Sordo}, {Soria Nieto}, {Souchay}, {Spagna}, {Spoto},
  {Stampa}, {Steele}, {Steidelm{\"u}ller}, {Stephenson}, {Stoev}, {Suess},
  {S{\"u}veges}, {Surdej}, {Szabados}, {Szegedi-Elek}, {Tapiador}, {Taris},
  {Tauran}, {Taylor}, {Teixeira}, {Terrett}, {Tingley}, {Trager}, {Turon},
  {Ulla}, {Utrilla}, {Valentini}, {van Elteren}, {Van Hemelryck}, {van
  Leeuwen}, {Varadi}, {Vecchiato}, {Veljanoski}, {Via}, {Vicente}, {Vogt},
  {Voss}, {Votruba}, {Voutsinas}, {Walmsley}, {Weiler}, {Weingrill}, {Wevers},
  {Wyrzykowski}, {Yoldas}, {{\v{Z}}erjal}, {Zucker}, {Zurbach}, {Zwitter},
  {Alecu}, {Allen}, {Allende Prieto}, {Amorim}, {Anglada-Escud{\'e}},
  {Arsenijevic}, {Azaz}, {Balm}, {Beck}, {Bernstein}, {Bigot}, {Bijaoui},
  {Blasco}, {Bonfigli}, {Bono}, {Boudreault}, {Bressan}, {Brown}, {Brunet},
  {Bunclark}, {Buonanno}, {Butkevich}, {Carret}, {Carrion}, {Chemin},
  {Ch{\'e}reau}, {Corcione}, {Darmigny}, {de Boer}, {de Teodoro}, {de Zeeuw},
  {Delle Luche}, {Domingues}, {Dubath}, {Fodor}, {Fr{\'e}zouls}, {Fries},
  {Fustes}, {Fyfe}, {Gallardo}, {Gallegos}, {Gardiol}, {Gebran}, {Gomboc},
  {G{\'o}mez}, {Grux}, {Gueguen}, {Heyrovsky}, {Hoar}, {Iannicola}, {Isasi
  Parache}, {Janotto}, {Joliet}, {Jonckheere}, {Keil}, {Kim}, {Klagyivik},
  {Klar}, {Knude}, {Kochukhov}, {Kolka}, {Kos}, {Kutka}, {Lainey}, {LeBouquin},
  {Liu}, {Loreggia}, {Makarov}, {Marseille}, {Martayan}, {Martinez-Rubi},
  {Massart}, {Meynadier}, {Mignot}, {Munari}, {Nguyen}, {Nordlander}, {Ocvirk},
  {O'Flaherty}, {Olias Sanz}, {Ortiz}, {Osorio}, {Oszkiewicz}, {Ouzounis},
  {Palmer}, {Park}, {Pasquato}, {Peltzer}, {Peralta}, {P{\'e}turaud},
  {Pieniluoma}, {Pigozzi}, {Poels}, {Prat}, {Prod'homme}, {Raison}, {Rebordao},
  {Risquez}, {Rocca-Volmerange}, {Rosen}, {Ruiz-Fuertes}, {Russo}, {Sembay},
  {Serraller Vizcaino}, {Short}, {Siebert}, {Silva}, {Sinachopoulos}, {Slezak},
  {Soffel}, {Sosnowska}, {Strai{\v{z}}ys}, {ter Linden}, {Terrell}, {Theil},
  {Tiede}, {Troisi}, {Tsalmantza}, {Tur}, {Vaccari}, {Vachier}, {Valles}, {Van
  Hamme}, {Veltz}, {Virtanen}, {Wallut}, {Wichmann}, {Wilkinson}, {Ziaeepour},
  \& {Zschocke}}]{Gaia16}
{Gaia Collaboration}, {Brown}, A.~G.~A., {Vallenari}, A., {et~al.} 2016, \aap,
  595, A2

\bibitem[{{Gaia Collaboration} {et~al.}(2023{\natexlab{a}}){Gaia
  Collaboration}, {Recio-Blanco}, {Kordopatis}, {de Laverny}, {Palicio},
  {Spagna}, {Spina}, {Katz}, {Re Fiorentin}, {Poggio}, {McMillan}, {Vallenari},
  {Lattanzi}, {Seabroke}, {Casamiquela}, {Bragaglia}, {Antoja}, {Bailer-Jones},
  {Schultheis}, {Andrae}, {Fouesneau}, {Cropper}, {Cantat-Gaudin}, {Bijaoui},
  {Heiter}, {Brown}, {Prusti}, {de Bruijne}, {Arenou}, {Babusiaux}, {Biermann},
  {Creevey}, {Ducourant}, {Evans}, {Eyer}, {Guerra}, {Hutton}, {Jordi},
  {Klioner}, {Lammers}, {Lindegren}, {Luri}, {Mignard}, {Panem}, {Pourbaix},
  {Randich}, {Sartoretti}, {Soubiran}, {Tanga}, {Walton}, {Bastian}, {Drimmel},
  {Jansen}, {van Leeuwen}, {Bakker}, {Cacciari}, {Casta{\~n}eda}, {De Angeli},
  {Fabricius}, {Fr{\'e}mat}, {Galluccio}, {Guerrier}, {Masana}, {Messineo},
  {Mowlavi}, {Nicolas}, {Nienartowicz}, {Pailler}, {Panuzzo}, {Riclet}, {Roux},
  {Sordo}, {Th{\'e}venin}, {Gracia-Abril}, {Portell}, {Teyssier}, {Altmann},
  {Audard}, {Bellas-Velidis}, {Benson}, {Berthier}, {Blomme}, {Burgess},
  {Busonero}, {Busso}, {C{\'a}novas}, {Carry}, {Cellino}, {Cheek},
  {Clementini}, {Damerdji}, {Davidson}, {de Teodoro}, {Nu{\~n}ez Campos},
  {Delchambre}, {Dell'Oro}, {Esquej}, {Fern{\'a}ndez-Hern{\'a}ndez}, {Fraile},
  {Garabato}, {Garc{\'\i}a-Lario}, {Gosset}, {Haigron}, {Halbwachs}, {Hambly},
  {Harrison}, {Hern{\'a}ndez}, {Hestroffer}, {Hodgkin}, {Holl}, {Jan{\ss}en},
  {Jevardat de Fombelle}, {Jordan}, {Krone-Martins}, {Lanzafame},
  {L{\"o}ffler}, {Marchal}, {Marrese}, {Moitinho}, {Muinonen}, {Osborne},
  {Pancino}, {Pauwels}, {Reyl{\'e}}, {Riello}, {Rimoldini}, {Roegiers},
  {Rybizki}, {Sarro}, {Siopis}, {Smith}, {Sozzetti}, {Utrilla}, {van Leeuwen},
  {Abbas}, {{\'A}brah{\'a}m}, {Abreu Aramburu}, {Aerts}, {Aguado}, {Ajaj},
  {Aldea-Montero}, {Altavilla}, {{\'A}lvarez}, {Alves}, {Anders}, {Anderson},
  {Anglada Varela}, {Baines}, {Baker}, {Balaguer-N{\'u}{\~n}ez}, {Balbinot},
  {Balog}, {Barache}, {Barbato}, {Barros}, {Barstow}, {Bartolom{\'e}},
  {Bassilana}, {Bauchet}, {Becciani}, {Bellazzini}, {Berihuete}, {Bernet},
  {Bertone}, {Bianchi}, {Binnenfeld}, {Blanco-Cuaresma}, {Boch}, {Bombrun},
  {Bossini}, {Bouquillon}, {Bramante}, {Breedt}, {Bressan}, {Brouillet},
  {Brugaletta}, {Bucciarelli}, {Burlacu}, {Butkevich}, {Buzzi}, {Caffau},
  {Cancelliere}, {Carballo}, {Carlucci}, {Carnerero}, {Carrasco}, {Castellani},
  {Castro-Ginard}, {Chaoul}, {Charlot}, {Chemin}, {Chiaramida}, {Chiavassa},
  {Chornay}, {Comoretto}, {Contursi}, {Cooper}, {Cornez}, {Cowell}, {Crifo},
  {Crosta}, {Crowley}, {Dafonte}, {Dapergolas}, {David}, {De Luise}, {De
  March}, {De Ridder}, {de Souza}, {de Torres}, {del Peloso}, {del Pozo},
  {Delbo}, {Delgado}, {Delisle}, {Demouchy}, {Dharmawardena}, {Di Matteo},
  {Diakite}, {Diener}, {Distefano}, {Dolding}, {Edvardsson}, {Enke}, {Fabre},
  {Fabrizio}, {Faigler}, {Fedorets}, {Fernique}, {Figueras}, {Fournier},
  {Fouron}, {Fragkoudi}, {Gai}, {Garcia-Gutierrez}, {Garcia-Reinaldos},
  {Garc{\'\i}a-Torres}, {Garofalo}, {Gavel}, {Gavras}, {Gerlach}, {Geyer},
  {Giacobbe}, {Gilmore}, {Girona}, {Giuffrida}, {Gomel}, {Gomez},
  {Gonz{\'a}lez-N{\'u}{\~n}ez}, {Gonz{\'a}lez-Santamar{\'\i}a},
  {Gonz{\'a}lez-Vidal}, {Granvik}, {Guillout}, {Guiraud},
  {Guti{\'e}rrez-S{\'a}nchez}, {Guy}, {Hatzidimitriou}, {Hauser}, {Haywood},
  {Helmer}, {Helmi}, {Sarmiento}, {Hidalgo}, {H{\l}adczuk}, {Hobbs}, {Holland},
  {Huckle}, {Jardine}, {Jasniewicz}, {Jean-Antoine Piccolo},
  {Jim{\'e}nez-Arranz}, {Juaristi Campillo}, {Julbe}, {Karbevska}, {Kervella},
  {Khanna}, {Korn}, {K{\'o}sp{\'a}l}, {Kostrzewa-Rutkowska}, {Kruszy{\'n}ska},
  {Kun}, {Laizeau}, {Lambert}, {Lanza}, {Lasne}, {Le Campion}, {Lebreton},
  {Lebzelter}, {Leccia}, {Leclerc}, {Lecoeur-Taibi}, {Liao}, {Licata},
  {Lindstr{\o}m}, {Lister}, {Livanou}, {Lobel}, {Lorca}, {Loup}, {Madrero
  Pardo}, {Magdaleno Romeo}, {Managau}, {Mann}, {Manteiga}, {Marchant},
  {Marconi}, {Marcos}, {Marcos Santos}, {Mar{\'\i}n Pina}, {Marinoni},
  {Marocco}, {Marshall}, {Martin Polo}, {Mart{\'\i}n-Fleitas}, {Marton},
  {Mary}, {Masip}, {Massari}, {Mastrobuono-Battisti}, {Mazeh}, {Messina},
  {Michalik}, {Millar}, {Mints}, {Molina}, {Molinaro}, {Moln{\'a}r}, {Monari},
  {Mongui{\'o}}, {Montegriffo}, {Montero}, {Mor}, {Mora}, {Morbidelli},
  {Morel}, {Morris}, {Muraveva}, {Murphy}, {Musella}, {Nagy}, {Noval},
  {Oca{\~n}a}, {Ogden}, {Ordenovic}, {Osinde}, {Pagani}, {Pagano}, {Palaversa},
  {Pallas-Quintela}, {Panahi}, {Payne-Wardenaar}, {Pe{\~n}alosa Esteller},
  {Penttil{\"a}}, {Pichon}, {Piersimoni}, {Pineau}, {Plachy}, {Plum},
  {Pr{\v{s}}a}, {Pulone}, {Racero}, {Ragaini}, {Rainer}, {Raiteri}, {Ramos},
  {Ramos-Lerate}, {Regibo}, {Richards}, {Rios Diaz}, {Ripepi}, {Riva}, {Rix},
  {Rixon}, {Robichon}, {Robin}, {Robin}, {Roelens}, {Rogues}, {Rohrbasser},
  {Romero-G{\'o}mez}, {Rowell}, {Royer}, {Ruz Mieres}, {Rybicki}, {Sadowski},
  {S{\'a}ez N{\'u}{\~n}ez}, {Sagrist{\`a} Sell{\'e}s}, {Sahlmann}, {Salguero},
  {Samaras}, {Sanchez Gimenez}, {Sanna}, {Santove{\~n}a}, {Sarasso}, {Sciacca},
  {Segol}, {Segovia}, {S{\'e}gransan}, {Semeux}, {Shahaf}, {Siddiqui},
  {Siebert}, {Siltala}, {Silvelo}, {Slezak}, {Slezak}, {Smart}, {Snaith},
  {Solano}, {Solitro}, {Souami}, {Souchay}, {Spoto}, {Steele},
  {Steidelm{\"u}ller}, {Stephenson}, {S{\"u}veges}, {Surdej}, {Szabados},
  {Szegedi-Elek}, {Taris}, {Taylor}, {Teixeira}, {Tolomei}, {Tonello}, {Torra},
  {Torra}, {Torralba Elipe}, {Trabucchi}, {Tsounis}, {Turon}, {Ulla}, {Unger},
  {Vaillant}, {van Dillen}, {van Reeven}, {Vanel}, {Vecchiato}, {Viala},
  {Vicente}, {Voutsinas}, {Weiler}, {Wevers}, {Wyrzykowski}, {Yoldas}, {Yvard},
  {Zhao}, {Zorec}, {Zucker}, \& {Zwitter}}]{Gaia23}
{Gaia Collaboration}, {Recio-Blanco}, A., {Kordopatis}, G., {et~al.}
  2023{\natexlab{a}}, \aap, 674, A38

\bibitem[{{Gaia Collaboration} {et~al.}(2023{\natexlab{b}}){Gaia
  Collaboration}, {Vallenari}, {Brown}, {Prusti}, {de Bruijne}, {Arenou},
  {Babusiaux}, {Biermann}, {Creevey}, {Ducourant}, {Evans}, {Eyer}, {Guerra},
  {Hutton}, {Jordi}, {Klioner}, {Lammers}, {Lindegren}, {Luri}, {Mignard},
  {Panem}, {Pourbaix}, {Randich}, {Sartoretti}, {Soubiran}, {Tanga}, {Walton},
  {Bailer-Jones}, {Bastian}, {Drimmel}, {Jansen}, {Katz}, {Lattanzi}, {van
  Leeuwen}, {Bakker}, {Cacciari}, {Casta{\~n}eda}, {De Angeli}, {Fabricius},
  {Fouesneau}, {Fr{\'e}mat}, {Galluccio}, {Guerrier}, {Heiter}, {Masana},
  {Messineo}, {Mowlavi}, {Nicolas}, {Nienartowicz}, {Pailler}, {Panuzzo},
  {Riclet}, {Roux}, {Seabroke}, {Sordo}, {Th{\'e}venin}, {Gracia-Abril},
  {Portell}, {Teyssier}, {Altmann}, {Andrae}, {Audard}, {Bellas-Velidis},
  {Benson}, {Berthier}, {Blomme}, {Burgess}, {Busonero}, {Busso},
  {C{\'a}novas}, {Carry}, {Cellino}, {Cheek}, {Clementini}, {Damerdji},
  {Davidson}, {de Teodoro}, {Nu{\~n}ez Campos}, {Delchambre}, {Dell'Oro},
  {Esquej}, {Fern{\'a}ndez-Hern{\'a}ndez}, {Fraile}, {Garabato},
  {Garc{\'\i}a-Lario}, {Gosset}, {Haigron}, {Halbwachs}, {Hambly}, {Harrison},
  {Hern{\'a}ndez}, {Hestroffer}, {Hodgkin}, {Holl}, {Jan{\ss}en}, {Jevardat de
  Fombelle}, {Jordan}, {Krone-Martins}, {Lanzafame}, {L{\"o}ffler}, {Marchal},
  {Marrese}, {Moitinho}, {Muinonen}, {Osborne}, {Pancino}, {Pauwels},
  {Recio-Blanco}, {Reyl{\'e}}, {Riello}, {Rimoldini}, {Roegiers}, {Rybizki},
  {Sarro}, {Siopis}, {Smith}, {Sozzetti}, {Utrilla}, {van Leeuwen}, {Abbas},
  {{\'A}brah{\'a}m}, {Abreu Aramburu}, {Aerts}, {Aguado}, {Ajaj},
  {Aldea-Montero}, {Altavilla}, {{\'A}lvarez}, {Alves}, {Anders}, {Anderson},
  {Anglada Varela}, {Antoja}, {Baines}, {Baker}, {Balaguer-N{\'u}{\~n}ez},
  {Balbinot}, {Balog}, {Barache}, {Barbato}, {Barros}, {Barstow},
  {Bartolom{\'e}}, {Bassilana}, {Bauchet}, {Becciani}, {Bellazzini},
  {Berihuete}, {Bernet}, {Bertone}, {Bianchi}, {Binnenfeld}, {Blanco-Cuaresma},
  {Blazere}, {Boch}, {Bombrun}, {Bossini}, {Bouquillon}, {Bragaglia},
  {Bramante}, {Breedt}, {Bressan}, {Brouillet}, {Brugaletta}, {Bucciarelli},
  {Burlacu}, {Butkevich}, {Buzzi}, {Caffau}, {Cancelliere}, {Cantat-Gaudin},
  {Carballo}, {Carlucci}, {Carnerero}, {Carrasco}, {Casamiquela}, {Castellani},
  {Castro-Ginard}, {Chaoul}, {Charlot}, {Chemin}, {Chiaramida}, {Chiavassa},
  {Chornay}, {Comoretto}, {Contursi}, {Cooper}, {Cornez}, {Cowell}, {Crifo},
  {Cropper}, {Crosta}, {Crowley}, {Dafonte}, {Dapergolas}, {David}, {David},
  {de Laverny}, {De Luise}, {De March}, {De Ridder}, {de Souza}, {de Torres},
  {del Peloso}, {del Pozo}, {Delbo}, {Delgado}, {Delisle}, {Demouchy},
  {Dharmawardena}, {Di Matteo}, {Diakite}, {Diener}, {Distefano}, {Dolding},
  {Edvardsson}, {Enke}, {Fabre}, {Fabrizio}, {Faigler}, {Fedorets}, {Fernique},
  {Fienga}, {Figueras}, {Fournier}, {Fouron}, {Fragkoudi}, {Gai},
  {Garcia-Gutierrez}, {Garcia-Reinaldos}, {Garc{\'\i}a-Torres}, {Garofalo},
  {Gavel}, {Gavras}, {Gerlach}, {Geyer}, {Giacobbe}, {Gilmore}, {Girona},
  {Giuffrida}, {Gomel}, {Gomez}, {Gonz{\'a}lez-N{\'u}{\~n}ez},
  {Gonz{\'a}lez-Santamar{\'\i}a}, {Gonz{\'a}lez-Vidal}, {Granvik}, {Guillout},
  {Guiraud}, {Guti{\'e}rrez-S{\'a}nchez}, {Guy}, {Hatzidimitriou}, {Hauser},
  {Haywood}, {Helmer}, {Helmi}, {Sarmiento}, {Hidalgo}, {Hilger},
  {H{\l}adczuk}, {Hobbs}, {Holland}, {Huckle}, {Jardine}, {Jasniewicz},
  {Jean-Antoine Piccolo}, {Jim{\'e}nez-Arranz}, {Jorissen}, {Juaristi
  Campillo}, {Julbe}, {Karbevska}, {Kervella}, {Khanna}, {Kontizas},
  {Kordopatis}, {Korn}, {K{\'o}sp{\'a}l}, {Kostrzewa-Rutkowska},
  {Kruszy{\'n}ska}, {Kun}, {Laizeau}, {Lambert}, {Lanza}, {Lasne}, {Le
  Campion}, {Lebreton}, {Lebzelter}, {Leccia}, {Leclerc}, {Lecoeur-Taibi},
  {Liao}, {Licata}, {Lindstr{\o}m}, {Lister}, {Livanou}, {Lobel}, {Lorca},
  {Loup}, {Madrero Pardo}, {Magdaleno Romeo}, {Managau}, {Mann}, {Manteiga},
  {Marchant}, {Marconi}, {Marcos}, {Marcos Santos}, {Mar{\'\i}n Pina},
  {Marinoni}, {Marocco}, {Marshall}, {Martin Polo}, {Mart{\'\i}n-Fleitas},
  {Marton}, {Mary}, {Masip}, {Massari}, {Mastrobuono-Battisti}, {Mazeh},
  {McMillan}, {Messina}, {Michalik}, {Millar}, {Mints}, {Molina}, {Molinaro},
  {Moln{\'a}r}, {Monari}, {Mongui{\'o}}, {Montegriffo}, {Montero}, {Mor},
  {Mora}, {Morbidelli}, {Morel}, {Morris}, {Muraveva}, {Murphy}, {Musella},
  {Nagy}, {Noval}, {Oca{\~n}a}, {Ogden}, {Ordenovic}, {Osinde}, {Pagani},
  {Pagano}, {Palaversa}, {Palicio}, {Pallas-Quintela}, {Panahi},
  {Payne-Wardenaar}, {Pe{\~n}alosa Esteller}, {Penttil{\"a}}, {Pichon},
  {Piersimoni}, {Pineau}, {Plachy}, {Plum}, {Poggio}, {Pr{\v{s}}a}, {Pulone},
  {Racero}, {Ragaini}, {Rainer}, {Raiteri}, {Rambaux}, {Ramos}, {Ramos-Lerate},
  {Re Fiorentin}, {Regibo}, {Richards}, {Rios Diaz}, {Ripepi}, {Riva}, {Rix},
  {Rixon}, {Robichon}, {Robin}, {Robin}, {Roelens}, {Rogues}, {Rohrbasser},
  {Romero-G{\'o}mez}, {Rowell}, {Royer}, {Ruz Mieres}, {Rybicki}, {Sadowski},
  {S{\'a}ez N{\'u}{\~n}ez}, {Sagrist{\`a} Sell{\'e}s}, {Sahlmann}, {Salguero},
  {Samaras}, {Sanchez Gimenez}, {Sanna}, {Santove{\~n}a}, {Sarasso},
  {Schultheis}, {Sciacca}, {Segol}, {Segovia}, {S{\'e}gransan}, {Semeux},
  {Shahaf}, {Siddiqui}, {Siebert}, {Siltala}, {Silvelo}, {Slezak}, {Slezak},
  {Smart}, {Snaith}, {Solano}, {Solitro}, {Souami}, {Souchay}, {Spagna},
  {Spina}, {Spoto}, {Steele}, {Steidelm{\"u}ller}, {Stephenson}, {S{\"u}veges},
  {Surdej}, {Szabados}, {Szegedi-Elek}, {Taris}, {Taylor}, {Teixeira},
  {Tolomei}, {Tonello}, {Torra}, {Torra}, {Torralba Elipe}, {Trabucchi},
  {Tsounis}, {Turon}, {Ulla}, {Unger}, {Vaillant}, {van Dillen}, {van Reeven},
  {Vanel}, {Vecchiato}, {Viala}, {Vicente}, {Voutsinas}, {Weiler}, {Wevers},
  {Wyrzykowski}, {Yoldas}, {Yvard}, {Zhao}, {Zorec}, {Zucker}, \&
  {Zwitter}}]{Gaia23GENERAL}
{Gaia Collaboration}, {Vallenari}, A., {Brown}, A.~G.~A., {et~al.}
  2023{\natexlab{b}}, \aap, 674, A1

\bibitem[{{Genovali} {et~al.}(2015){Genovali}, {Lemasle}, {da Silva}, {Bono},
  {Fabrizio}, {Bergemann}, {Buonanno}, {Ferraro}, {Fran{\c{c}}ois},
  {Iannicola}, {Inno}, {Laney}, {Kudritzki}, {Matsunaga}, {Nonino}, {Primas},
  {Romaniello}, {Urbaneja}, \& {Th{\'e}venin}}]{Genovali15}
{Genovali}, K., {Lemasle}, B., {da Silva}, R., {et~al.} 2015, \aap, 580, A17

\bibitem[{{Gilmore} {et~al.}(2012){Gilmore}, {Randich}, {Asplund}, {Binney},
  {Bonifacio}, {Drew}, {Feltzing}, {Ferguson}, {Jeffries}, {Micela},
  {Negueruela}, {Prusti}, {Rix}, {Vallenari}, {Alfaro}, {Allende-Prieto},
  {Babusiaux}, {Bensby}, {Blomme}, {Bragaglia}, {Flaccomio}, {Fran{\c{c}}ois},
  {Irwin}, {Koposov}, {Korn}, {Lanzafame}, {Pancino}, {Paunzen},
  {Recio-Blanco}, {Sacco}, {Smiljanic}, {Van Eck}, {Walton}, {Aden}, {Aerts},
  {Affer}, {Alcala}, {Altavilla}, {Alves}, {Antoja}, {Arenou}, {Argiroffi},
  {Asensio Ramos}, {Bailer-Jones}, {Balaguer-Nunez}, {Bayo}, {Barbuy},
  {Barisevicius}, {Barrado y Navascues}, {Battistini}, {Bellas Velidis},
  {Bellazzini}, {Belokurov}, {Bergemann}, {Bertelli}, {Biazzo}, {Bienayme},
  {Bland-Hawthorn}, {Boeche}, {Bonito}, {Boudreault}, {Bouvier}, {Brandao},
  {Brown}, {de Bruijne}, {Burleigh}, {Caballero}, {Caffau}, {Calura},
  {Capuzzo-Dolcetta}, {Caramazza}, {Carraro}, {Casagrande}, {Casewell},
  {Chapman}, {Chiappini}, {Chorniy}, {Christlieb}, {Cignoni}, {Cocozza},
  {Colless}, {Collet}, {Collins}, {Correnti}, {Covino}, {Crnojevic}, {Cropper},
  {Cunha}, {Damiani}, {David}, {Delgado}, {Duffau}, {Edvardsson}, {Eldridge},
  {Enke}, {Eriksson}, {Evans}, {Eyer}, {Famaey}, {Fellhauer}, {Ferreras},
  {Figueras}, {Fiorentino}, {Flynn}, {Folha}, {Franciosini}, {Frasca},
  {Freeman}, {Fremat}, {Friel}, {Gaensicke}, {Gameiro}, {Garzon}, {Geier},
  {Geisler}, {Gerhard}, {Gibson}, {Gomboc}, {Gomez}, {Gonzalez-Fernandez},
  {Gonzalez Hernandez}, {Gosset}, {Grebel}, {Greimel}, {Groenewegen},
  {Grundahl}, {Guarcello}, {Gustafsson}, {Hadrava}, {Hatzidimitriou}, {Hambly},
  {Hammersley}, {Hansen}, {Haywood}, {Heber}, {Heiter}, {Held}, {Helmi},
  {Hensler}, {Herrero}, {Hill}, {Hodgkin}, {Huelamo}, {Huxor}, {Ibata},
  {Jackson}, {de Jong}, {Jonker}, {Jordan}, {Jordi}, {Jorissen}, {Katz},
  {Kawata}, {Keller}, {Kharchenko}, {Klement}, {Klutsch}, {Knude}, {Koch},
  {Kochukhov}, {Kontizas}, {Koubsky}, {Lallement}, {de Laverny}, {van Leeuwen},
  {Lemasle}, {Lewis}, {Lind}, {Lindstrom}, {Lobel}, {Lopez Santiago}, {Lucas},
  {Ludwig}, {Lueftinger}, {Magrini}, {Maiz Apellaniz}, {Maldonado}, {Marconi},
  {Marino}, {Martayan}, {Martinez-Valpuesta}, {Matijevic}, {McMahon},
  {Messina}, {Meyer}, {Miglio}, {Mikolaitis}, {Minchev}, {Minniti}, {Moitinho},
  {Momany}, {Monaco}, {Montalto}, {Monteiro}, {Monier}, {Montes}, {Mora},
  {Moraux}, {Morel}, {Mowlavi}, {Mucciarelli}, {Munari}, {Napiwotzki},
  {Nardetto}, {Naylor}, {Naze}, {Nelemans}, {Okamoto}, {Ortolani}, {Pace},
  {Palla}, {Palous}, {Parker}, {Penarrubia}, {Pillitteri}, {Piotto}, {Posbic},
  {Prisinzano}, {Puzeras}, {Quirrenbach}, {Ragaini}, {Read}, {Read}, {Reyle},
  {De Ridder}, {Robichon}, {Robin}, {Roeser}, {Romano}, {Royer}, {Ruchti},
  {Ruzicka}, {Ryan}, {Ryde}, {Santos}, {Sanz Forcada}, {Sarro Baro},
  {Sbordone}, {Schilbach}, {Schmeja}, {Schnurr}, {Schoenrich}, {Scholz},
  {Seabroke}, {Sharma}, {De Silva}, {Smith}, {Solano}, {Sordo}, {Soubiran},
  {Sousa}, {Spagna}, {Steffen}, {Steinmetz}, {Stelzer}, {Stempels},
  {Tabernero}, {Tautvaisiene}, {Thevenin}, {Torra}, {Tosi}, {Tolstoy}, {Turon},
  {Walker}, {Wambsganss}, {Worley}, {Venn}, {Vink}, {Wyse}, {Zaggia},
  {Zeilinger}, {Zoccali}, {Zorec}, {Zucker}, {Zwitter}, \& {Gaia-ESO Survey
  Team}}]{Gilmore12}
{Gilmore}, G., {Randich}, S., {Asplund}, M., {et~al.} 2012, The Messenger, 147,
  25

\bibitem[{{Gilmore} {et~al.}(2022){Gilmore}, {Randich}, {Worley}, {Hourihane},
  {Gonneau}, {Sacco}, {Lewis}, {Magrini}, {Fran{\c{c}}ois}, {Jeffries},
  {Koposov}, {Bragaglia}, {Alfaro}, {Allende Prieto}, {Blomme}, {Korn},
  {Lanzafame}, {Pancino}, {Recio-Blanco}, {Smiljanic}, {Van Eck}, {Zwitter},
  {Bensby}, {Flaccomio}, {Irwin}, {Franciosini}, {Morbidelli}, {Damiani},
  {Bonito}, {Friel}, {Vink}, {Prisinzano}, {Abbas}, {Hatzidimitriou}, {Held},
  {Jordi}, {Paunzen}, {Spagna}, {Jackson}, {Ma{\'\i}z Apell{\'a}niz},
  {Asplund}, {Bonifacio}, {Feltzing}, {Binney}, {Drew}, {Ferguson}, {Micela},
  {Negueruela}, {Prusti}, {Rix}, {Vallenari}, {Bergemann}, {Casey}, {de
  Laverny}, {Frasca}, {Hill}, {Lind}, {Sbordone}, {Sousa}, {Adibekyan},
  {Caffau}, {Daflon}, {Feuillet}, {Gebran}, {Gonzalez Hernandez}, {Guiglion},
  {Herrero}, {Lobel}, {Merle}, {Mikolaitis}, {Montes}, {Morel}, {Ruchti},
  {Soubiran}, {Tabernero}, {Tautvai{\v{s}}ien{\.{e}}}, {Traven}, {Valentini},
  {Van der Swaelmen}, {Villanova}, {Viscasillas V{\'a}zquez}, {Bayo}, {Biazzo},
  {Carraro}, {Edvardsson}, {Heiter}, {Jofr{\'e}}, {Marconi}, {Martayan},
  {Masseron}, {Monaco}, {Walton}, {Zaggia}, {Aguirre B{\o}rsen-Koch}, {Alves},
  {Balaguer-Nunez}, {Barklem}, {Barrado}, {Bellazzini}, {Berlanas}, {Binks},
  {Bressan}, {Capuzzo-Dolcetta}, {Casagrande}, {Casamiquela}, {Collins},
  {D'Orazi}, {Dantas}, {Debattista}, {Delgado-Mena}, {Di Marcantonio},
  {Drazdauskas}, {Evans}, {Famaey}, {Franchini}, {Fr{\'e}mat}, {Fu}, {Geisler},
  {Gerhard}, {Gonz{\'a}lez Solares}, {Grebel}, {Guti{\'e}rrez Albarr{\'a}n},
  {Jim{\'e}nez-Esteban}, {J{\"o}nsson}, {Khachaturyants}, {Kordopatis}, {Kos},
  {Lagarde}, {Ludwig}, {Mahy}, {Mapelli}, {Marfil}, {Martell}, {Messina},
  {Miglio}, {Minchev}, {Moitinho}, {Montalban}, {Monteiro}, {Morossi},
  {Mowlavi}, {Mucciarelli}, {Murphy}, {Nardetto}, {Ortolani}, {Paletou},
  {Palou{\v{s}}}, {Pickering}, {Quirrenbach}, {Re Fiorentin}, {Read}, {Romano},
  {Ryde}, {Sanna}, {Santos}, {Seabroke}, {Spina}, {Steinmetz}, {Stonkut{\'e}},
  {Sutorius}, {Th{\'e}venin}, {Tosi}, {Tsantaki}, {Wright}, {Wyse}, {Zoccali},
  {Zorec}, \& {Zucker}}]{Gilmore22}
{Gilmore}, G., {Randich}, S., {Worley}, C.~C., {et~al.} 2022, \aap, 666, A120

\bibitem[{{Grand} {et~al.}(2018){Grand}, {Bustamante}, {G{\'o}mez}, {Kawata},
  {Marinacci}, {Pakmor}, {Rix}, {Simpson}, {Sparre}, \& {Springel}}]{Grand18}
{Grand}, R. J.~J., {Bustamante}, S., {G{\'o}mez}, F.~A., {et~al.} 2018, \mnras,
  474, 3629

\bibitem[{{Green}(2014)}]{Green14}
{Green}, D.~A. 2014, in Supernova Environmental Impacts, ed. A.~{Ray} \& R.~A.
  {McCray}, Vol. 296, 188--196

\bibitem[{{Grisoni} {et~al.}(2018){Grisoni}, {Spitoni}, \&
  {Matteucci}}]{Grisoni18}
{Grisoni}, V., {Spitoni}, E., \& {Matteucci}, F. 2018, \mnras, 481, 2570

\bibitem[{{Halle} {et~al.}(2015){Halle}, {Di Matteo}, {Haywood}, \&
  {Combes}}]{Halle15}
{Halle}, A., {Di Matteo}, P., {Haywood}, M., \& {Combes}, F. 2015, \aap, 578,
  A58

\bibitem[{{Halle} {et~al.}(2018){Halle}, {Di Matteo}, {Haywood}, \&
  {Combes}}]{Halle18}
{Halle}, A., {Di Matteo}, P., {Haywood}, M., \& {Combes}, F. 2018, \aap, 616,
  A86

\bibitem[{{Hayden} {et~al.}(2015){Hayden}, {Bovy}, {Holtzman}, {Nidever},
  {Bird}, {Weinberg}, {Andrews}, {Majewski}, {Allende Prieto}, {Anders},
  {Beers}, {Bizyaev}, {Chiappini}, {Cunha}, {Frinchaboy},
  {Garc{\'\i}a-Her{\'n}andez}, {Garc{\'\i}a P{\'e}rez}, {Girardi}, {Harding},
  {Hearty}, {Johnson}, {M{\'e}sz{\'a}ros}, {Minchev}, {O'Connell}, {Pan},
  {Robin}, {Schiavon}, {Schneider}, {Schultheis}, {Shetrone}, {Skrutskie},
  {Steinmetz}, {Smith}, {Wilson}, {Zamora}, \& {Zasowski}}]{Hayden15}
{Hayden}, M.~R., {Bovy}, J., {Holtzman}, J.~A., {et~al.} 2015, \apj, 808, 132

\bibitem[{{Henry} {et~al.}(2010){Henry}, {Kwitter}, {Jaskot}, {Balick},
  {Morrison}, \& {Milingo}}]{Henry10}
{Henry}, R.~B.~C., {Kwitter}, K.~B., {Jaskot}, A.~E., {et~al.} 2010, \apj, 724,
  748

\bibitem[{{Hopkins} {et~al.}(2023){Hopkins}, {Gurvich}, {Shen}, {Hafen},
  {Grudi{\'c}}, {Kurinchi-Vendhan}, {Hayward}, {Jiang}, {Orr}, {Wetzel},
  {Kere{\v{s}}}, {Stern}, {Faucher-Gigu{\`e}re}, {Bullock}, {Wheeler},
  {El-Badry}, {Loebman}, {Moreno}, {Boylan-Kolchin}, \& {Quataert}}]{Hopkins23}
{Hopkins}, P.~F., {Gurvich}, A.~B., {Shen}, X., {et~al.} 2023, \mnras, 525,
  2241

\bibitem[{{Hourihane} {et~al.}(2023){Hourihane}, {Fran{\c{c}}ois}, {Worley},
  {Magrini}, {Gonneau}, {Casey}, {Gilmore}, {Randich}, {Sacco}, {Recio-Blanco},
  {Korn}, {Allende Prieto}, {Smiljanic}, {Blomme}, {Bragaglia}, {Walton}, {Van
  Eck}, {Bensby}, {Lanzafame}, {Frasca}, {Franciosini}, {Damiani}, {Lind},
  {Bergemann}, {Bonifacio}, {Hill}, {Lobel}, {Montes}, {Feuillet},
  {Tautvai{\v{s}}ien{\.{e}}}, {Guiglion}, {Tabernero}, {Gonz{\'a}lez
  Hern{\'a}ndez}, {Gebran}, {Van der Swaelmen}, {Mikolaitis}, {Daflon},
  {Merle}, {Morel}, {Lewis}, {Gonz{\'a}lez Solares}, {Murphy}, {Jeffries},
  {Jackson}, {Feltzing}, {Prusti}, {Carraro}, {Biazzo}, {Prisinzano},
  {Jofr{\'e}}, {Zaggia}, {Drazdauskas}, {Stonkut{\'e}}, {Marfil},
  {Jim{\'e}nez-Esteban}, {Mahy}, {Guti{\'e}rrez Albarr{\'a}n}, {Berlanas},
  {Santos}, {Morbidelli}, {Spina}, \&
  {Minkevi{\v{c}}i{\={u}}t{\.{e}}}}]{Hourihane2023A&A...676A.129H}
{Hourihane}, A., {Fran{\c{c}}ois}, P., {Worley}, C.~C., {et~al.} 2023, \aap,
  676, A129

\bibitem[{{Isern}(2019)}]{Isern19}
{Isern}, J. 2019, \apjl, 878, L11

\bibitem[{{Iwamoto} {et~al.}(1999){Iwamoto}, {Brachwitz}, {Nomoto},
  {Kishimoto}, {Umeda}, {Hix}, \& {Thielemann}}]{Iwa99}
{Iwamoto}, K., {Brachwitz}, F., {Nomoto}, K., {et~al.} 1999, \apjs, 125, 439

\bibitem[{{Jackson} {et~al.}(2022){Jackson}, {Jeffries}, {Wright}, {Randich},
  {Sacco}, {Bragaglia}, {Hourihane}, {Tognelli}, {Degl'Innocenti}, {Prada
  Moroni}, {Gilmore}, {Bensby}, {Pancino}, {Smiljanic}, {Bergemann}, {Carraro},
  {Franciosini}, {Gonneau}, {Jofr{\'e}}, {Lewis}, {Magrini}, {Morbidelli},
  {Prisinzano}, {Worley}, {Zaggia}, {Tautvai{\v{s}}iene}, {Guti{\'e}rrez
  Albarr{\'a}n}, {Montes}, \&
  {Jim{\'e}nez-Esteban}}]{Jackson2022MNRAS.509.1664J}
{Jackson}, R.~J., {Jeffries}, R.~D., {Wright}, N.~J., {et~al.} 2022, \mnras,
  509, 1664

\bibitem[{{Jacobson} {et~al.}(2016){Jacobson}, {Friel}, {J{\'\i}lkov{\'a}},
  {Magrini}, {Bragaglia}, {Vallenari}, {Tosi}, {Randich}, {Donati},
  {Cantat-Gaudin}, {Sordo}, {Smiljanic}, {Overbeek}, {Carraro},
  {Tautvai{\v{s}}ien{\.{e}}}, {San Roman}, {Villanova}, {Geisler}, {Mu{\~n}oz},
  {Jim{\'e}nez-Esteban}, {Tang}, {Gilmore}, {Alfaro}, {Bensby}, {Flaccomio},
  {Koposov}, {Korn}, {Pancino}, {Recio-Blanco}, {Casey}, {Costado},
  {Franciosini}, {Heiter}, {Hill}, {Hourihane}, {Lardo}, {de Laverny}, {Lewis},
  {Monaco}, {Morbidelli}, {Sacco}, {Sousa}, {Worley}, \&
  {Zaggia}}]{Jacobson2016A&A...591A..37J}
{Jacobson}, H.~R., {Friel}, E.~D., {J{\'\i}lkov{\'a}}, L., {et~al.} 2016, \aap,
  591, A37

\bibitem[{{Jeffries} {et~al.}(2023){Jeffries}, {Jackson}, {Wright}, {Weaver},
  {Gilmore}, {Randich}, {Bragaglia}, {Korn}, {Smiljanic}, {Biazzo}, {Casey},
  {Frasca}, {Gonneau}, {Guiglion}, {Morbidelli}, {Prisinzano}, {Sacco},
  {Tautvai{\v{s}}ien{\.{e}}}, {Worley}, \& {Zaggia}}]{Jeffries23}
{Jeffries}, R.~D., {Jackson}, R.~J., {Wright}, N.~J., {et~al.} 2023, \mnras,
  523, 802

\bibitem[{{Jin} {et~al.}(2024){Jin}, {Trager}, {Dalton}, {Aguerri}, {Drew},
  {Falc{\'o}n-Barroso}, {G{\"a}nsicke}, {Hill}, {Iovino}, {Pieri}, {Poggianti},
  {Smith}, {Vallenari}, {Abrams}, {Aguado}, {Antoja}, {Arag{\'o}n-Salamanca},
  {Ascasibar}, {Babusiaux}, {Balcells}, {Barrena}, {Battaglia}, {Belokurov},
  {Bensby}, {Bonifacio}, {Bragaglia}, {Carrasco}, {Carrera}, {Cornwell},
  {Dom{\'\i}nguez-Palmero}, {Duncan}, {Famaey}, {Fari{\~n}a}, {Gonzalez},
  {Guest}, {Hatch}, {Hess}, {Hoskin}, {Irwin}, {Knapen}, {Koposov}, {Kuchner},
  {Laigle}, {Lewis}, {Longhetti}, {Lucatello}, {M{\'e}ndez-Abreu}, {Mercurio},
  {Molaeinezhad}, {Mongui{\'o}}, {Morrison}, {Murphy}, {Peralta de Arriba},
  {P{\'e}rez}, {P{\'e}rez-R{\`a}fols}, {Pic{\'o}}, {Raddi}, {Romero-G{\'o}mez},
  {Royer}, {Siebert}, {Seabroke}, {Som}, {Terrett}, {Thomas}, {Wesson},
  {Worley}, {Alfaro}, {Allende Prieto}, {Alonso-Santiago}, {Amos}, {Ashley},
  {Balaguer-N{\'u}{\~n}ez}, {Balbinot}, {Bellazzini}, {Benn}, {Berlanas},
  {Bernard}, {Best}, {Bettoni}, {Bianco}, {Bishop}, {Blomqvist}, {Boeche},
  {Bolzonella}, {Bonoli}, {Bosma}, {Britavskiy}, {Busarello}, {Caffau},
  {Cantat-Gaudin}, {Castro-Ginard}, {Couto}, {Carbajo-Hijarrubia}, {Carter},
  {Casamiquela}, {Conrado}, {Corcho-Caballero}, {Costantin}, {Deason}, {de
  Burgos}, {De Grandi}, {Di Matteo}, {Dom{\'\i}nguez-G{\'o}mez}, {Dorda},
  {Drake}, {Dutta}, {Erkal}, {Feltzing}, {Ferr{\'e}-Mateu}, {Feuillet},
  {Figueras}, {Fossati}, {Franciosini}, {Frasca}, {Fumagalli}, {Gallazzi},
  {Garc{\'\i}a-Benito}, {Gentile Fusillo}, {Gebran}, {Gilbert}, {Gledhill},
  {Gonz{\'a}lez Delgado}, {Greimel}, {Guarcello}, {Guerra}, {Gullieuszik},
  {Haines}, {Hardcastle}, {Harris}, {Haywood}, {Helmi}, {Hernandez}, {Herrero},
  {Hughes}, {Ir{\v{s}}i{\v{c}}}, {Jablonka}, {Jarvis}, {Jordi}, {Kondapally},
  {Kordopatis}, {Krogager}, {La Barbera}, {Lam}, {Larsen}, {Lemasle}, {Lewis},
  {Lhom{\'e}}, {Lind}, {Lodi}, {Longobardi}, {Lonoce}, {Magrini}, {Ma{\'\i}z
  Apell{\'a}niz}, {Marchal}, {Marco}, {Martin}, {Matsuno}, {Maurogordato},
  {Merluzzi}, {Miralda-Escud{\'e}}, {Molinari}, {Monari}, {Morelli}, {Mottram},
  {Naylor}, {Negueruela}, {O{\~n}orbe}, {Pancino}, {Peirani}, {Peletier},
  {Pozzetti}, {Rainer}, {Ramos}, {Read}, {Rossi}, {R{\"o}ttgering},
  {Rubi{\~n}o-Mart{\'\i}n}, {Sabater}, {San Juan}, {Sanna}, {Schallig},
  {Schiavon}, {Schultheis}, {Serra}, {Shimwell}, {Sim{\'o}n-D{\'\i}az},
  {Smith}, {Sordo}, {Sorini}, {Soubiran}, {Starkenburg}, {Steele}, {Stott},
  {Stuik}, {Tolstoy}, {Tortora}, {Tsantaki}, {Van der Swaelmen}, {van Weeren},
  {Vergani}, {Verheijen}, {Verro}, {Vink}, {Vioque}, {Walcher}, {Walton},
  {Wegg}, {Weijmans}, {Williams}, {Wilson}, {Wright}, {Xylakis-Dornbusch},
  {Youakim}, {Zibetti}, \& {Zurita}}]{WEAVEpaper}
{Jin}, S., {Trager}, S.~C., {Dalton}, G.~B., {et~al.} 2024, \mnras, 530, 2688

\bibitem[{{Kennicutt}(1998)}]{Kennicutt98}
{Kennicutt}, Robert~C., J. 1998, \apj, 498, 541

\bibitem[{{Kobayashi} {et~al.}(2011){Kobayashi}, {Karakas}, \&
  {Umeda}}]{Koba11}
{Kobayashi}, C., {Karakas}, A.~I., \& {Umeda}, H. 2011, \mnras, 414, 3231

\bibitem[{{Kobayashi} {et~al.}(2020){Kobayashi}, {Leung}, \& {Nomoto}}]{Koba20}
{Kobayashi}, C., {Leung}, S.-C., \& {Nomoto}, K. 2020, \apj, 895, 138

\bibitem[{{Kobayashi} {et~al.}(2006){Kobayashi}, {Umeda}, {Nomoto}, {Tominaga},
  \& {Ohkubo}}]{Koba06}
{Kobayashi}, C., {Umeda}, H., {Nomoto}, K., {Tominaga}, N., \& {Ohkubo}, T.
  2006, \apj, 653, 1145

\bibitem[{{Kordopatis} {et~al.}(2015){Kordopatis}, {Binney}, {Gilmore}, {Wyse},
  {Belokurov}, {McMillan}, {Hatfield}, {Grebel}, {Steinmetz}, {Navarro},
  {Seabroke}, {Minchev}, {Chiappini}, {Bienaym{\'e}}, {Bland-Hawthorn},
  {Freeman}, {Gibson}, {Helmi}, {Munari}, {Parker}, {Reid}, {Siebert},
  {Siviero}, \& {Zwitter}}]{Kordopatis15}
{Kordopatis}, G., {Binney}, J., {Gilmore}, G., {et~al.} 2015, \mnras, 447, 3526

\bibitem[{{Kovtyukh} {et~al.}(2022){Kovtyukh}, {Lemasle}, {Bono}, {Usenko}, {da
  Silva}, {Kniazev}, {Grebel}, {Andronov}, {Shakun}, \&
  {Chinarova}}]{Kovtyukh22}
{Kovtyukh}, V., {Lemasle}, B., {Bono}, G., {et~al.} 2022, \mnras, 510, 1894

\bibitem[{{Kreckel} {et~al.}(2019){Kreckel}, {Ho}, {Blanc}, {Groves},
  {Santoro}, {Schinnerer}, {Bigiel}, {Chevance}, {Congiu}, {Emsellem}, {Faesi},
  {Glover}, {Grasha}, {Kruijssen}, {Lang}, {Leroy}, {Meidt}, {McElroy}, {Pety},
  {Rosolowsky}, {Saito}, {Sandstrom}, {Sanchez-Blazquez}, \&
  {Schruba}}]{Kreckel19}
{Kreckel}, K., {Ho}, I.~T., {Blanc}, G.~A., {et~al.} 2019, \apj, 887, 80

\bibitem[{{Kroupa} {et~al.}(1993){Kroupa}, {Tout}, \& {Gilmore}}]{Kroupa93}
{Kroupa}, P., {Tout}, C.~A., \& {Gilmore}, G. 1993, \mnras, 262, 545

\bibitem[{{Lemasle} {et~al.}(2007){Lemasle}, {Fran{\c{c}}ois}, {Bono},
  {Mottini}, {Primas}, \& {Romaniello}}]{Lemasle07}
{Lemasle}, B., {Fran{\c{c}}ois}, P., {Bono}, G., {et~al.} 2007, \aap, 467, 283

\bibitem[{{Lemasle} {et~al.}(2008){Lemasle}, {Fran{\c{c}}ois}, {Piersimoni},
  {Pedicelli}, {Bono}, {Laney}, {Primas}, \& {Romaniello}}]{Lemasle08}
{Lemasle}, B., {Fran{\c{c}}ois}, P., {Piersimoni}, A., {et~al.} 2008, \aap,
  490, 613

\bibitem[{{Leung} \& {Nomoto}(2018)}]{Leung18}
{Leung}, S.-C. \& {Nomoto}, K. 2018, \apj, 861, 143

\bibitem[{{Leung} \& {Nomoto}(2020)}]{Leung20}
{Leung}, S.-C. \& {Nomoto}, K. 2020, \apj, 888, 80

\bibitem[{{Limongi} \& {Chieffi}(2018)}]{Limongi18}
{Limongi}, M. \& {Chieffi}, A. 2018, \apjs, 237, 13

\bibitem[{{Lin} {et~al.}(2020){Lin}, {Li}, {Du}, {Wang}, {Xiao}, {Bureau},
  {Fraser-McKelvie}, {Masters}, {Lin}, {Wake}, \& {Hao}}]{Lin20}
{Lin}, L., {Li}, C., {Du}, C., {et~al.} 2020, \mnras, 499, 1406

\bibitem[{{Luck} \& {Lambert}(2011)}]{Luck11}
{Luck}, R.~E. \& {Lambert}, D.~L. 2011, \aj, 142, 136

\bibitem[{{Maciel} {et~al.}(2003){Maciel}, {Costa}, \& {Uchida}}]{Maciel04}
{Maciel}, W.~J., {Costa}, R.~D.~D., \& {Uchida}, M.~M.~M. 2003, \aap, 397, 667

\bibitem[{{Mackereth} {et~al.}(2018){Mackereth}, {Crain}, {Schiavon}, {Schaye},
  {Theuns}, \& {Schaller}}]{Mackereth18}
{Mackereth}, J.~T., {Crain}, R.~A., {Schiavon}, R.~P., {et~al.} 2018, \mnras,
  477, 5072

\bibitem[{{Magrini} {et~al.}(2017){Magrini}, {Randich}, {Kordopatis},
  {Prantzos}, {Romano}, {Chieffi}, {Limongi}, {Fran{\c{c}}ois}, {Pancino},
  {Friel}, {Bragaglia}, {Tautvai{\v{s}}ien{\.{e}}}, {Spina}, {Overbeek},
  {Cantat-Gaudin}, {Donati}, {Vallenari}, {Sordo}, {Jim{\'e}nez-Esteban},
  {Tang}, {Drazdauskas}, {Sousa}, {Duffau}, {Jofr{\'e}}, {Gilmore}, {Feltzing},
  {Alfaro}, {Bensby}, {Flaccomio}, {Koposov}, {Lanzafame}, {Smiljanic}, {Bayo},
  {Carraro}, {Casey}, {Costado}, {Damiani}, {Franciosini}, {Hourihane},
  {Lardo}, {Lewis}, {Monaco}, {Morbidelli}, {Sacco}, {Sbordone}, {Worley}, \&
  {Zaggia}}]{Magrini17}
{Magrini}, L., {Randich}, S., {Kordopatis}, G., {et~al.} 2017, \aap, 603, A2

\bibitem[{{Magrini} {et~al.}(2010){Magrini}, {Randich}, {Zoccali}, {Jilkova},
  {Carraro}, {Galli}, {Maiorca}, \& {Busso}}]{Magrini10}
{Magrini}, L., {Randich}, S., {Zoccali}, M., {et~al.} 2010, \aap, 523, A11

\bibitem[{{Magrini} {et~al.}(2021{\natexlab{a}}){Magrini}, {Smiljanic},
  {Franciosini}, {Pasquini}, {Randich}, {Casali}, {Viscasillas V{\'a}zquez},
  {Bragaglia}, {Spina}, {Biazzo}, {Tautvai{\v{s}}ien{\.{e}}}, {Masseron}, {Van
  der Swaelmen}, {Pancino}, {Jim{\'e}nez-Esteban}, {Guiglion}, {Martell},
  {Bensby}, {D'Orazi}, {Baratella}, {Korn}, {Jofre}, {Gilmore}, {Worley},
  {Hourihane}, {Gonneau}, {Sacco}, \&
  {Morbidelli}}]{Magrini2021A&A...655A..23M}
{Magrini}, L., {Smiljanic}, R., {Franciosini}, E., {et~al.} 2021{\natexlab{a}},
  \aap, 655, A23

\bibitem[{{Magrini} {et~al.}(2018){Magrini}, {Spina}, {Randich}, {Friel},
  {Kordopatis}, {Worley}, {Pancino}, {Bragaglia}, {Donati},
  {Tautvai{\v{s}}ien{\.{e}}}, {Bagdonas}, {Delgado-Mena}, {Adibekyan}, {Sousa},
  {Jim{\'e}nez-Esteban}, {Sanna}, {Roccatagliata}, {Bonito}, {Sbordone},
  {Duffau}, {Gilmore}, {Feltzing}, {Jeffries}, {Vallenari}, {Alfaro}, {Bensby},
  {Francois}, {Koposov}, {Korn}, {Recio-Blanco}, {Smiljanic}, {Bayo},
  {Carraro}, {Casey}, {Costado}, {Damiani}, {Franciosini}, {Frasca},
  {Hourihane}, {Jofr{\'e}}, {de Laverny}, {Lewis}, {Masseron}, {Monaco},
  {Morbidelli}, {Prisinzano}, {Sacco}, \&
  {Zaggia}}]{Magrini2018A&A...617A.106M}
{Magrini}, L., {Spina}, L., {Randich}, S., {et~al.} 2018, \aap, 617, A106

\bibitem[{{Magrini} {et~al.}(2022){Magrini}, {V{\'a}zquez}, {Casali},
  {Baratella}, {D'Orazi}, {Spina}, {Randich}, {Cristallo}, \&
  {Vescovi}}]{Magrini2022Univ....8...64M}
{Magrini}, L., {V{\'a}zquez}, C.~V., {Casali}, G., {et~al.} 2022, Universe, 8,
  64

\bibitem[{{Magrini} {et~al.}(2021{\natexlab{b}}){Magrini}, {Vescovi}, {Casali},
  {Cristallo}, {Viscasillas V{\'a}zquez}, {Cescutti}, {Spina}, {Van Der
  Swaelmen}, \& {Randich}}]{Magrini2021A&A...646L...2M}
{Magrini}, L., {Vescovi}, D., {Casali}, G., {et~al.} 2021{\natexlab{b}}, \aap,
  646, L2

\bibitem[{{Magrini} {et~al.}(2023){Magrini}, {Viscasillas V{\'a}zquez},
  {Spina}, {Randich}, {Romano}, {Franciosini}, {Recio-Blanco}, {Nordlander},
  {D'Orazi}, {Baratella}, {Smiljanic}, {Dantas}, {Pasquini}, {Spitoni},
  {Casali}, {Van der Swaelmen}, {Bensby}, {Stonkute}, {Feltzing}, {Sacco},
  {Bragaglia}, {Pancino}, {Heiter}, {Biazzo}, {Gilmore}, {Bergemann},
  {Tautvai{\v{s}}ien{\.{e}}}, {Worley}, {Hourihane}, {Gonneau}, \&
  {Morbidelli}}]{Magrini23}
{Magrini}, L., {Viscasillas V{\'a}zquez}, C., {Spina}, L., {et~al.} 2023, \aap,
  669, A119

\bibitem[{{Mainieri} {et~al.}(2024){Mainieri}, {Anderson}, {Brinchmann},
  {Cimatti}, {Ellis}, {Hill}, {Kneib}, {McLeod}, {Opitom}, {Roth},
  {Sanchez-Saez}, {Smiljanic}, {Tolstoy}, {Bacon}, {Randich}, {Adamo},
  {Annibali}, {Arevalo}, {Audard}, {Barsanti}, {Battaglia}, {Bayo Aran},
  {Belfiore}, {Bellazzini}, {Bellini}, {Beltran}, {Berni}, {Bianchi}, {Biazzo},
  {Bisero}, {Bisogni}, {Bland-Hawthorn}, {Blondin}, {Bodensteiner}, {Boffin},
  {Bonito}, {Bono}, {Bouche}, {Bowman}, {Braga}, {Bragaglia}, {Branchesi},
  {Brucalassi}, {Bryant}, {Bryson}, {Busa}, {Camera}, {Carbone}, {Casali},
  {Casali}, {Casasola}, {Castro}, {Catelan}, {Cavallo}, {Chiappini}, {Cioni},
  {Colless}, {Colzi}, {Contarini}, {Couch}, {D'Ammando}, {d'Assignies D.},
  {D'Orazi}, {da Silva}, {Dainotti}, {Damiani}, {Danielski}, {De Cia}, {de
  Jong}, {Dhawan}, {Dierickx}, {Driver}, {Dupletsa}, {Escoffier}, {Escorza},
  {Fabrizio}, {Fiorentino}, {Fontana}, {Fontani}, {Forero Sanchez}, {Franois},
  {Galindo-Guil}, {Gallazzi}, {Galli}, {Garcia}, {Garcia-Rojas}, {Garilli},
  {Grand}, {Guarcello}, {Hazra}, {Helmi}, {Herrero}, {Iglesias}, {Ilic},
  {Irsic}, {Ivanov}, {Izzo}, {Jablonka}, {Joachimi}, {Kakkad}, {Kamann},
  {Koposov}, {Kordopatis}, {Kovacevic}, {Kraljic}, {Kuncarayakti}, {Kwon}, {La
  Forgia}, {Lahav}, {Laigle}, {Lazzarin}, {Leaman}, {Leclercq}, {Lee}, {Lee},
  {Lehnert}, {Lira}, {Loffredo}, {Lucatello}, {Magrini}, {Maguire}, {Mahler},
  {Zahra Majidi}, {Malavasi}, {Mannucci}, {Marconi}, {Martin}, {Marulli},
  {Massari}, {Matsuno}, {Mattheee}, {McGee}, {Merc}, {Merle}, {Miglio},
  {Migliorini}, {Minchev}, {Minniti}, {Miret-Roig}, {Monreal Ibero}, {Montano},
  {Montet}, {Moresco}, {Moretti}, {Moscardini}, {Moya}, {Mueller},
  {Nanayakkara}, {Nicholl}, {Nordlander}, {Onori}, {Padovani}, {Pala}, {Panda},
  {Pandey-Pommier}, {Pasquini}, {Pawlak}, {Pessi}, {Pisani}, {Popovic},
  {Prisinzano}, {Raddi}, {Rainer}, {Rebassa-Mansergas}, {Richard}, {Rigault},
  {Rocher}, {Romano}, {Rosati}, {Sacco}, {Sanchez-Janssen}, {Sander},
  {Sanders}, {Sargent}, {Sarpa}, {Schimd}, {Schipani}, {Sefusatti}, {Smith},
  {Spina}, {Steinmetz}, {Tacchella}, {Tautvaisiene}, {Theissen}, {Thomas},
  {Ting}, {Travouillon}, {Tresse}, {Trivedi}, {Tsantaki}, {Tsedrik}, {Urrutia},
  {Valenti}, {Van der Swaelmen}, {Van Eck}, {Verdiani}, {Verdier}, {Vergani},
  {Verhamme}, {Vernet}, {Verza}, {Viel}, {Vielzeuf}, {Vietri}, {Vink},
  {Viscasillas Vazquez}, {Wang}, {Weilbacher}, {Wendt}, {Wright}, {Ye},
  {Yeche}, {Yu}, {Zafar}, {Zibetti}, {Ziegler}, \& {Zinchenko}}]{WSTpaper}
{Mainieri}, V., {Anderson}, R.~I., {Brinchmann}, J., {et~al.} 2024, arXiv
  e-prints, arXiv:2403.05398

\bibitem[{{Matteucci} \& {Francois}(1989)}]{Francois89}
{Matteucci}, F. \& {Francois}, P. 1989, \mnras, 239, 885

\bibitem[{{Matteucci} \& {Recchi}(2001)}]{MatteucciRecchi01}
{Matteucci}, F. \& {Recchi}, S. 2001, \apj, 558, 351

\bibitem[{{McKee} {et~al.}(2015){McKee}, {Parravano}, \&
  {Hollenbach}}]{McKee15}
{McKee}, C.~F., {Parravano}, A., \& {Hollenbach}, D.~J. 2015, \apj, 814, 13

\bibitem[{{Melioli} {et~al.}(2008){Melioli}, {Brighenti}, {D'Ercole}, \& {de
  Gouveia Dal Pino}}]{Meioli08}
{Melioli}, C., {Brighenti}, F., {D'Ercole}, A., \& {de Gouveia Dal Pino}, E.~M.
  2008, \mnras, 388, 573

\bibitem[{{Melioli} {et~al.}(2009){Melioli}, {Brighenti}, {D'Ercole}, \& {de
  Gouveia Dal Pino}}]{Meioli09}
{Melioli}, C., {Brighenti}, F., {D'Ercole}, A., \& {de Gouveia Dal Pino}, E.~M.
  2009, \mnras, 399, 1089

\bibitem[{{M{\'e}ndez-Delgado} {et~al.}(2022){M{\'e}ndez-Delgado}, {Amayo},
  {Arellano-C{\'o}rdova}, {Esteban}, {Garc{\'\i}a-Rojas}, {Carigi}, \&
  {Delgado-Inglada}}]{Mendez22}
{M{\'e}ndez-Delgado}, J.~E., {Amayo}, A., {Arellano-C{\'o}rdova}, K.~Z.,
  {et~al.} 2022, \mnras, 510, 4436

\bibitem[{{Miglio} {et~al.}(2021){Miglio}, {Chiappini}, {Mackereth}, {Davies},
  {Brogaard}, {Casagrande}, {Chaplin}, {Girardi}, {Kawata}, {Khan}, {Izzard},
  {Montalb{\'a}n}, {Mosser}, {Vincenzo}, {Bossini}, {Noels}, {Rodrigues},
  {Valentini}, \& {Mandel}}]{Miglio21}
{Miglio}, A., {Chiappini}, C., {Mackereth}, J.~T., {et~al.} 2021, \aap, 645,
  A85

\bibitem[{{Minchev} {et~al.}(2018){Minchev}, {Anders}, {Recio-Blanco},
  {Chiappini}, {de Laverny}, {Queiroz}, {Steinmetz}, {Adibekyan}, {Carrillo},
  {Cescutti}, {Guiglion}, {Hayden}, {de Jong}, {Kordopatis}, {Majewski},
  {Martig}, \& {Santiago}}]{Minchev18}
{Minchev}, I., {Anders}, F., {Recio-Blanco}, A., {et~al.} 2018, \mnras, 481,
  1645

\bibitem[{{Minchev} {et~al.}(2011){Minchev}, {Famaey}, {Combes}, {Di Matteo},
  {Mouhcine}, \& {Wozniak}}]{Minchev11}
{Minchev}, I., {Famaey}, B., {Combes}, F., {et~al.} 2011, \aap, 527, A147

\bibitem[{{Molero} {et~al.}(2023){Molero}, {Magrini}, {Matteucci}, {Romano},
  {Palla}, {Cescutti}, {Viscasillas V{\'a}zquez}, \&
  {Spitoni}}]{Molero2023MNRAS.523.2974M}
{Molero}, M., {Magrini}, L., {Matteucci}, F., {et~al.} 2023, \mnras, 523, 2974

\bibitem[{{Mor} {et~al.}(2019){Mor}, {Robin}, {Figueras}, {Roca-F{\`a}brega},
  \& {Luri}}]{Mor19}
{Mor}, R., {Robin}, A.~C., {Figueras}, F., {Roca-F{\`a}brega}, S., \& {Luri},
  X. 2019, \aap, 624, L1

\bibitem[{{Mott} {et~al.}(2013){Mott}, {Spitoni}, \& {Matteucci}}]{Mott13}
{Mott}, A., {Spitoni}, E., \& {Matteucci}, F. 2013, \mnras, 435, 2918

\bibitem[{{Myers} {et~al.}(2022){Myers}, {Donor}, {Spoo}, {Frinchaboy},
  {Cunha}, {Price-Whelan}, {Majewski}, {Beaton}, {Zasowski}, {O'Connell},
  {Ray}, {Bizyaev}, {Chiappini}, {Garc{\'\i}a-Hern{\'a}ndez}, {Geisler},
  {J{\"o}nsson}, {Lane}, {Longa-Pe{\~n}a}, {Minchev}, {Minniti}, {Nitschelm},
  \& {Roman-Lopes}}]{Myers22}
{Myers}, N., {Donor}, J., {Spoo}, T., {et~al.} 2022, \aj, 164, 85

\bibitem[{{Nakanishi} \& {Sofue}(2003)}]{Nakanishi03}
{Nakanishi}, H. \& {Sofue}, Y. 2003, \pasj, 55, 191

\bibitem[{{Nakanishi} \& {Sofue}(2006)}]{Nakanishi06}
{Nakanishi}, H. \& {Sofue}, Y. 2006, \pasj, 58, 847

\bibitem[{{Nepal} {et~al.}(2024){Nepal}, {Chiappini}, {Guiglion}, {Steinmetz},
  {P{\'e}rez-Villegas}, {Queiroz}, {Miglio}, {Dohme}, \& {Khalatyan}}]{Nepal24}
{Nepal}, S., {Chiappini}, C., {Guiglion}, G., {et~al.} 2024, \aap, 681, L8

\bibitem[{{Nissen} {et~al.}(2020){Nissen}, {Christensen-Dalsgaard},
  {Mosumgaard}, {Silva Aguirre}, {Spitoni}, \& {Verma}}]{Nissen20}
{Nissen}, P.~E., {Christensen-Dalsgaard}, J., {Mosumgaard}, J.~R., {et~al.}
  2020, \aap, 640, A81

\bibitem[{{Noguchi}(2018)}]{Noguchi18}
{Noguchi}, M. 2018, \nat, 559, 585

\bibitem[{{Overbeek} {et~al.}(2017){Overbeek}, {Friel}, {Donati}, {Smiljanic},
  {Jacobson}, {Hatzidimitriou}, {Held}, {Magrini}, {Bragaglia}, {Randich},
  {Vallenari}, {Cantat-Gaudin}, {Tautvai{\v{s}}ien{\.{e}}},
  {Jim{\'e}nez-Esteban}, {Frasca}, {Geisler}, {Villanova}, {Tang}, {Mu{\~n}oz},
  {Marconi}, {Carraro}, {San Roman}, {Drazdauskas}, {{\v{Z}}enovien{\.{e}}},
  {Gilmore}, {Jeffries}, {Flaccomio}, {Pancino}, {Bayo}, {Costado}, {Damiani},
  {Jofr{\'e}}, {Monaco}, {Prisinzano}, {Sousa}, \&
  {Zaggia}}]{Overbeek2017A&A...598A..68O}
{Overbeek}, J.~C., {Friel}, E.~D., {Donati}, P., {et~al.} 2017, \aap, 598, A68

\bibitem[{{Palla}(2021)}]{Palla21}
{Palla}, M. 2021, \mnras, 503, 3216

\bibitem[{{Palla} {et~al.}(2020{\natexlab{a}}){Palla}, {Matteucci}, {Calura},
  \& {Longo}}]{Palla20b}
{Palla}, M., {Matteucci}, F., {Calura}, F., \& {Longo}, F. 2020{\natexlab{a}},
  \apj, 889, 4

\bibitem[{{Palla} {et~al.}(2020{\natexlab{b}}){Palla}, {Matteucci}, {Spitoni},
  {Vincenzo}, \& {Grisoni}}]{Palla20}
{Palla}, M., {Matteucci}, F., {Spitoni}, E., {Vincenzo}, F., \& {Grisoni}, V.
  2020{\natexlab{b}}, \mnras, 498, 1710

\bibitem[{{Palla} {et~al.}(2022){Palla}, {Santos-Peral}, {Recio-Blanco}, \&
  {Matteucci}}]{Palla22}
{Palla}, M., {Santos-Peral}, P., {Recio-Blanco}, A., \& {Matteucci}, F. 2022,
  \aap, 663, A125

\bibitem[{{Pinsonneault} {et~al.}(2014){Pinsonneault}, {Elsworth}, {Epstein},
  {Hekker}, {M{\'e}sz{\'a}ros}, {Chaplin}, {Johnson}, {Garc{\'\i}a},
  {Holtzman}, {Mathur}, {Garc{\'\i}a P{\'e}rez}, {Silva Aguirre}, {Girardi},
  {Basu}, {Shetrone}, {Stello}, {Allende Prieto}, {An}, {Beck}, {Beers},
  {Bizyaev}, {Bloemen}, {Bovy}, {Cunha}, {De Ridder}, {Frinchaboy},
  {Garc{\'\i}a-Hern{\'a}ndez}, {Gilliland}, {Harding}, {Hearty}, {Huber},
  {Ivans}, {Kallinger}, {Majewski}, {Metcalfe}, {Miglio}, {Mosser}, {Muna},
  {Nidever}, {Schneider}, {Serenelli}, {Smith}, {Tayar}, {Zamora}, \&
  {Zasowski}}]{Pinsonneault14}
{Pinsonneault}, M.~H., {Elsworth}, Y., {Epstein}, C., {et~al.} 2014, \apjs,
  215, 19

\bibitem[{{Pinsonneault} {et~al.}(2018){Pinsonneault}, {Elsworth}, {Tayar},
  {Serenelli}, {Stello}, {Zinn}, {Mathur}, {Garc{\'\i}a}, {Johnson}, {Hekker},
  {Huber}, {Kallinger}, {M{\'e}sz{\'a}ros}, {Mosser}, {Stassun}, {Girardi},
  {Rodrigues}, {Silva Aguirre}, {An}, {Basu}, {Chaplin}, {Corsaro}, {Cunha},
  {Garc{\'\i}a-Hern{\'a}ndez}, {Holtzman}, {J{\"o}nsson}, {Shetrone}, {Smith},
  {Sobeck}, {Stringfellow}, {Zamora}, {Beers}, {Fern{\'a}ndez-Trincado},
  {Frinchaboy}, {Hearty}, \& {Nitschelm}}]{Pinsonneault18}
{Pinsonneault}, M.~H., {Elsworth}, Y.~P., {Tayar}, J., {et~al.} 2018, \apjs,
  239, 32

\bibitem[{{Portinari} \& {Chiosi}(2000)}]{Portinari00}
{Portinari}, L. \& {Chiosi}, C. 2000, \aap, 355, 929

\bibitem[{{Prantzos} {et~al.}(2018){Prantzos}, {Abia}, {Limongi}, {Chieffi}, \&
  {Cristallo}}]{Prantzos18}
{Prantzos}, N., {Abia}, C., {Limongi}, M., {Chieffi}, A., \& {Cristallo}, S.
  2018, \mnras, 476, 3432

\bibitem[{{Queiroz} {et~al.}(2020){Queiroz}, {Anders}, {Chiappini},
  {Khalatyan}, {Santiago}, {Steinmetz}, {Valentini}, {Miglio}, {Bossini},
  {Barbuy}, {Minchev}, {Minniti}, {Garc{\'\i}a Hern{\'a}ndez}, {Schultheis},
  {Beaton}, {Beers}, {Bizyaev}, {Brownstein}, {Cunha},
  {Fern{\'a}ndez-Trincado}, {Frinchaboy}, {Lane}, {Majewski}, {Nataf},
  {Nitschelm}, {Pan}, {Roman-Lopes}, {Sobeck}, {Stringfellow}, \&
  {Zamora}}]{Queiroz20}
{Queiroz}, A.~B.~A., {Anders}, F., {Chiappini}, C., {et~al.} 2020, \aap, 638,
  A76

\bibitem[{{Rana}(1991)}]{Rana91}
{Rana}, N.~C. 1991, \araa, 29, 129

\bibitem[{{Randich} {et~al.}(2013){Randich}, {Gilmore}, \& {Gaia-ESO
  Consortium}}]{Randich13}
{Randich}, S., {Gilmore}, G., \& {Gaia-ESO Consortium}. 2013, The Messenger,
  154, 47

\bibitem[{{Randich} {et~al.}(2022){Randich}, {Gilmore}, {Magrini}, {Sacco},
  {Jackson}, {Jeffries}, {Worley}, {Hourihane}, {Gonneau}, {Viscasillas
  Vazquez}, {Franciosini}, {Lewis}, {Alfaro}, {Allende Prieto}, {Bensby},
  {Blomme}, {Bragaglia}, {Flaccomio}, {Fran{\c{c}}ois}, {Irwin}, {Koposov},
  {Korn}, {Lanzafame}, {Pancino}, {Recio-Blanco}, {Smiljanic}, {Van Eck},
  {Zwitter}, {Asplund}, {Bonifacio}, {Feltzing}, {Binney}, {Drew}, {Ferguson},
  {Micela}, {Negueruela}, {Prusti}, {Rix}, {Vallenari}, {Bayo}, {Bergemann},
  {Biazzo}, {Carraro}, {Casey}, {Damiani}, {Frasca}, {Heiter}, {Hill},
  {Jofr{\'e}}, {de Laverny}, {Lind}, {Marconi}, {Martayan}, {Masseron},
  {Monaco}, {Morbidelli}, {Prisinzano}, {Sbordone}, {Sousa}, {Zaggia},
  {Adibekyan}, {Bonito}, {Caffau}, {Daflon}, {Feuillet}, {Gebran}, {Gonzalez
  Hernandez}, {Guiglion}, {Herrero}, {Lobel}, {Maiz Apellaniz}, {Merle},
  {Mikolaitis}, {Montes}, {Morel}, {Soubiran}, {Spina}, {Tabernero},
  {Tautvai{\v{s}}iene}, {Traven}, {Valentini}, {Van der Swaelmen}, {Villanova},
  {Wright}, {Abbas}, {Aguirre B{\o}rsen-Koch}, {Alves}, {Balaguer-Nunez},
  {Barklem}, {Barrado}, {Berlanas}, {Binks}, {Bressan}, {Capuzzo-Dolcetta},
  {Casagrande}, {Casamiquela}, {Collins}, {D'Orazi}, {Dantas}, {Debattista},
  {Delgado-Mena}, {Di Marcantonio}, {Drazdauskas}, {Evans}, {Famaey},
  {Franchini}, {Fr{\'e}mat}, {Friel}, {Fu}, {Geisler}, {Gerhard}, {Gonzalez
  Solares}, {Grebel}, {Gutierrez Albarran}, {Hatzidimitriou}, {Held},
  {Jim{\'e}nez-Esteban}, {J{\"o}nsson}, {Jordi}, {Khachaturyants},
  {Kordopatis}, {Kos}, {Lagarde}, {Mahy}, {Mapelli}, {Marfil}, {Martell},
  {Messina}, {Miglio}, {Minchev}, {Moitinho}, {Montalban}, {Monteiro},
  {Morossi}, {Mowlavi}, {Mucciarelli}, {Murphy}, {Nardetto}, {Ortolani},
  {Paletou}, {Palou{\v{s}}}, {Paunzen}, {Pickering}, {Quirrenbach}, {Re
  Fiorentin}, {Read}, {Romano}, {Ryde}, {Sanna}, {Santos}, {Seabroke},
  {Spagna}, {Steinmetz}, {Stonkut{\'e}}, {Sutorius}, {Th{\'e}venin}, {Tosi},
  {Tsantaki}, {Vink}, {Wright}, {Wyse}, {Zoccali}, {Zorec}, {Zucker}, \&
  {Walton}}]{Randich22}
{Randich}, S., {Gilmore}, G., {Magrini}, L., {et~al.} 2022, \aap, 666, A121

\bibitem[{{Randich} {et~al.}(2020){Randich}, {Pasquini}, {Franciosini},
  {Magrini}, {Jackson}, {Jeffries}, {d'Orazi}, {Romano}, {Sanna},
  {Tautvai{\v{s}}ien{\.{e}}}, {Tsantaki}, {Wright}, {Gilmore}, {Bensby},
  {Bragaglia}, {Pancino}, {Smiljanic}, {Bayo}, {Carraro}, {Gonneau},
  {Hourihane}, {Morbidelli}, \& {Worley}}]{Randich2020A&A...640L...1R}
{Randich}, S., {Pasquini}, L., {Franciosini}, E., {et~al.} 2020, \aap, 640, L1

\bibitem[{{Randich} {et~al.}(2003){Randich}, {Sestito}, \&
  {Pallavicini}}]{Randich03}
{Randich}, S., {Sestito}, P., \& {Pallavicini}, R. 2003, \aap, 399, 133

\bibitem[{{Rauer} {et~al.}(2022){Rauer}, {Aerts}, {Deleuil}, {Gizon}, {Goupil},
  {Heras}, {Mas-Hesse}, {Pagano}, {Piotto}, {Pollacco}, {Ragazzoni}, {Ramsay},
  \& {Udry}}]{PLATOpaper}
{Rauer}, H., {Aerts}, C., {Deleuil}, M., {et~al.} 2022, in European Planetary
  Science Congress, EPSC2022--453

\bibitem[{{Recio-Blanco} {et~al.}(2014){Recio-Blanco}, {de Laverny},
  {Kordopatis}, {Helmi}, {Hill}, {Gilmore}, {Wyse}, {Adibekyan}, {Randich},
  {Asplund}, {Feltzing}, {Jeffries}, {Micela}, {Vallenari}, {Alfaro}, {Allende
  Prieto}, {Bensby}, {Bragaglia}, {Flaccomio}, {Koposov}, {Korn}, {Lanzafame},
  {Pancino}, {Smiljanic}, {Jackson}, {Lewis}, {Magrini}, {Morbidelli},
  {Prisinzano}, {Sacco}, {Worley}, {Hourihane}, {Bergemann}, {Costado},
  {Heiter}, {Joffre}, {Lardo}, {Lind}, \& {Maiorca}}]{Recio14}
{Recio-Blanco}, A., {de Laverny}, P., {Kordopatis}, G., {et~al.} 2014, \aap,
  567, A5

\bibitem[{{Recio-Blanco} {et~al.}(2023){Recio-Blanco}, {de Laverny}, {Palicio},
  {Kordopatis}, {{\'A}lvarez}, {Schultheis}, {Contursi}, {Zhao}, {Torralba
  Elipe}, {Ordenovic}, {Manteiga}, {Dafonte}, {Oreshina-Slezak}, {Bijaoui},
  {Fr{\'e}mat}, {Seabroke}, {Pailler}, {Spitoni}, {Poggio}, {Creevey}, {Abreu
  Aramburu}, {Accart}, {Andrae}, {Bailer-Jones}, {Bellas-Velidis}, {Brouillet},
  {Brugaletta}, {Burlacu}, {Carballo}, {Casamiquela}, {Chiavassa}, {Cooper},
  {Dapergolas}, {Delchambre}, {Dharmawardena}, {Drimmel}, {Edvardsson},
  {Fouesneau}, {Garabato}, {Garc{\'\i}a-Lario}, {Garc{\'\i}a-Torres}, {Gavel},
  {Gomez}, {Gonz{\'a}lez-Santamar{\'\i}a}, {Hatzidimitriou}, {Heiter},
  {Jean-Antoine Piccolo}, {Kontizas}, {Korn}, {Lanzafame}, {Lebreton}, {Le
  Fustec}, {Licata}, {Lindstr{\o}m}, {Livanou}, {Lobel}, {Lorca}, {Magdaleno
  Romeo}, {Marocco}, {Marshall}, {Mary}, {Nicolas}, {Pallas-Quintela}, {Panem},
  {Pichon}, {Riclet}, {Robin}, {Rybizki}, {Santove{\~n}a}, {Silvelo}, {Smart},
  {Sarro}, {Sordo}, {Soubiran}, {S{\"u}veges}, {Ulla}, {Vallenari}, {Zorec},
  {Utrilla}, \& {Bakker}}]{RecioBlanco23}
{Recio-Blanco}, A., {de Laverny}, P., {Palicio}, P.~A., {et~al.} 2023, \aap,
  674, A29

\bibitem[{{Ripepi} {et~al.}(2022){Ripepi}, {Chemin}, {Molinaro}, {Cioni},
  {Bekki}, {Clementini}, {de Grijs}, {De Somma}, {El Youssoufi}, {Girardi},
  {Groenewegen}, {Ivanov}, {Marconi}, {McMillan}, \& {van Loon}}]{Ripepi22}
{Ripepi}, V., {Chemin}, L., {Molinaro}, R., {et~al.} 2022, \mnras, 512, 563

\bibitem[{{Roca-F{\`a}brega} {et~al.}(2021){Roca-F{\`a}brega}, {Llorente de
  Andr{\'e}s}, {Chavero}, {Cifuentes}, \& {de la Reza}}]{RocaFabrega21}
{Roca-F{\`a}brega}, S., {Llorente de Andr{\'e}s}, F., {Chavero}, C.,
  {Cifuentes}, C., \& {de la Reza}, R. 2021, \aap, 656, A64

\bibitem[{{Roederer} {et~al.}(2024){Roederer}, {Alvarado-G{\'o}mez}, {Allende
  Prieto}, {Adibekyan}, {Aguado}, {Amado}, {Amazo-G{\'o}mez}, {Baratella},
  {Barnes}, {Bensby}, {Bigot}, {Chiavassa}, {Domiciano de Souza}, {Gonz{\'a}lez
  Hern{\'a}ndez}, {Hansen}, {J{\"a}rvinen}, {Korn}, {Lucatello}, {Magrini},
  {Maiolino}, {Di Marcantonio}, {Marconi}, {De Medeiros}, {Mucciarelli},
  {Nardetto}, {Origlia}, {Peroux}, {Poppenh{\"a}ger}, {Reiners},
  {Rodr{\'\i}guez-L{\'o}pez}, {Romano}, {Salvadori}, {Tisserand}, {Venn},
  {Wade}, \& {Zanutta}}]{Roederer2024ExA....57...17R}
{Roederer}, I.~U., {Alvarado-G{\'o}mez}, J.~D., {Allende Prieto}, C., {et~al.}
  2024, Experimental Astronomy, 57, 17

\bibitem[{{Rojas-Arriagada} {et~al.}(2016){Rojas-Arriagada}, {Recio-Blanco},
  {de Laverny}, {Schultheis}, {Guiglion}, {Mikolaitis}, {Kordopatis}, {Hill},
  {Gilmore}, {Randich}, {Alfaro}, {Bensby}, {Koposov}, {Costado},
  {Franciosini}, {Hourihane}, {Jofr{\'e}}, {Lardo}, {Lewis}, {Lind}, {Magrini},
  {Monaco}, {Morbidelli}, {Sacco}, {Worley}, {Zaggia}, \&
  {Chiappini}}]{Rojas16}
{Rojas-Arriagada}, A., {Recio-Blanco}, A., {de Laverny}, P., {et~al.} 2016,
  \aap, 586, A39

\bibitem[{{Romano} {et~al.}(2005){Romano}, {Chiappini}, {Matteucci}, \&
  {Tosi}}]{Romano05}
{Romano}, D., {Chiappini}, C., {Matteucci}, F., \& {Tosi}, M. 2005, \aap, 430,
  491

\bibitem[{{Romano} {et~al.}(2010){Romano}, {Karakas}, {Tosi}, \&
  {Matteucci}}]{Romano10}
{Romano}, D., {Karakas}, A.~I., {Tosi}, M., \& {Matteucci}, F. 2010, \aap, 522,
  A32

\bibitem[{{Romano} {et~al.}(2021){Romano}, {Magrini}, {Randich}, {Casali},
  {Bonifacio}, {Jeffries}, {Matteucci}, {Franciosini}, {Spina}, {Guiglion},
  {Chiappini}, {Mucciarelli}, {Ventura}, {Grisoni}, {Bellazzini}, {Bensby},
  {Bragaglia}, {de Laverny}, {Korn}, {Martell}, {Tautvai{\v{s}}ien{\.{e}}},
  {Carraro}, {Gonneau}, {Jofr{\'e}}, {Pancino}, {Smiljanic}, {Vallenari}, {Fu},
  {Guti{\'e}rrez Albarr{\'a}n}, {Jim{\'e}nez-Esteban}, {Montes}, {Damiani},
  {Bergemann}, \& {Worley}}]{Romano2021A&A...653A..72R}
{Romano}, D., {Magrini}, L., {Randich}, S., {et~al.} 2021, \aap, 653, A72

\bibitem[{{Romano} {et~al.}(2000){Romano}, {Matteucci}, {Salucci}, \&
  {Chiappini}}]{Romano00}
{Romano}, D., {Matteucci}, F., {Salucci}, P., \& {Chiappini}, C. 2000, \apj,
  539, 235

\bibitem[{{Romano} {et~al.}(2019){Romano}, {Matteucci}, {Zhang}, {Ivison}, \&
  {Ventura}}]{Romano19}
{Romano}, D., {Matteucci}, F., {Zhang}, Z.-Y., {Ivison}, R.~J., \& {Ventura},
  P. 2019, \mnras, 490, 2838

\bibitem[{{Ruiz-Lara} {et~al.}(2020){Ruiz-Lara}, {Gallart}, {Bernard}, \&
  {Cassisi}}]{Ruiz20}
{Ruiz-Lara}, T., {Gallart}, C., {Bernard}, E.~J., \& {Cassisi}, S. 2020, Nature
  Astronomy, 4, 965

\bibitem[{{Sanders} \& {Binney}(2015)}]{Sanders15}
{Sanders}, J.~L. \& {Binney}, J. 2015, \mnras, 449, 3479

\bibitem[{{Santos-Peral} {et~al.}(2021){Santos-Peral}, {Recio-Blanco},
  {Kordopatis}, {Fern{\'a}ndez-Alvar}, \& {de Laverny}}]{Santos21}
{Santos-Peral}, P., {Recio-Blanco}, A., {Kordopatis}, G.,
  {Fern{\'a}ndez-Alvar}, E., \& {de Laverny}, P. 2021, \aap, arXiv:2106.13677

\bibitem[{{Sch{\"o}nrich} \& {Binney}(2009)}]{Schonrich09}
{Sch{\"o}nrich}, R. \& {Binney}, J. 2009, \mnras, 399, 1145

\bibitem[{{Sch{\"o}nrich} \& {McMillan}(2017)}]{Schonrich17}
{Sch{\"o}nrich}, R. \& {McMillan}, P.~J. 2017, \mnras, 467, 1154

\bibitem[{{Spina} {et~al.}(2022){Spina}, {Magrini}, \& {Cunha}}]{Spina22}
{Spina}, L., {Magrini}, L., \& {Cunha}, K. 2022, Universe, 8, 87

\bibitem[{{Spina} {et~al.}(2017){Spina}, {Randich}, {Magrini}, {Jeffries},
  {Friel}, {Sacco}, {Pancino}, {Bonito}, {Bravi}, {Franciosini}, {Klutsch},
  {Montes}, {Gilmore}, {Vallenari}, {Bensby}, {Bragaglia}, {Flaccomio},
  {Koposov}, {Korn}, {Lanzafame}, {Smiljanic}, {Bayo}, {Carraro}, {Casey},
  {Costado}, {Damiani}, {Donati}, {Frasca}, {Hourihane}, {Jofr{\'e}}, {Lewis},
  {Lind}, {Monaco}, {Morbidelli}, {Prisinzano}, {Sousa}, {Worley}, \&
  {Zaggia}}]{Spina2017A&A...601A..70S}
{Spina}, L., {Randich}, S., {Magrini}, L., {et~al.} 2017, \aap, 601, A70

\bibitem[{{Spina} {et~al.}(2021){Spina}, {Ting}, {De Silva}, {Frankel},
  {Sharma}, {Cantat-Gaudin}, {Joyce}, {Stello}, {Karakas}, {Asplund},
  {Nordlander}, {Casagrande}, {D'Orazi}, {Casey}, {Cottrell},
  {Tepper-Garc{\'\i}a}, {Baratella}, {Kos}, {{\v{C}}otar}, {Bland-Hawthorn},
  {Buder}, {Freeman}, {Hayden}, {Lewis}, {Lin}, {Lind}, {Martell},
  {Schlesinger}, {Simpson}, {Zucker}, \& {Zwitter}}]{Spina21}
{Spina}, L., {Ting}, Y.~S., {De Silva}, G.~M., {et~al.} 2021, \mnras, 503, 3279

\bibitem[{{Spitoni} \& {Matteucci}(2011)}]{Spitoni11}
{Spitoni}, E. \& {Matteucci}, F. 2011, \aap, 531, A72

\bibitem[{{Spitoni} {et~al.}(2009){Spitoni}, {Matteucci}, {Recchi}, {Cescutti},
  \& {Pipino}}]{Spitoni09}
{Spitoni}, E., {Matteucci}, F., {Recchi}, S., {Cescutti}, G., \& {Pipino}, A.
  2009, \aap, 504, 87

\bibitem[{{Spitoni} {et~al.}(2023){Spitoni}, {Recio-Blanco}, {de Laverny},
  {Palicio}, {Kordopatis}, {Schultheis}, {Contursi}, {Poggio}, {Romano}, \&
  {Matteucci}}]{Spitoni23}
{Spitoni}, E., {Recio-Blanco}, A., {de Laverny}, P., {et~al.} 2023, \aap, 670,
  A109

\bibitem[{{Spitoni} {et~al.}(2019){Spitoni}, {Silva Aguirre}, {Matteucci},
  {Calura}, \& {Grisoni}}]{Spitoni19}
{Spitoni}, E., {Silva Aguirre}, V., {Matteucci}, F., {Calura}, F., \&
  {Grisoni}, V. 2019, \aap, 623, A60

\bibitem[{{Spitoni} {et~al.}(2021){Spitoni}, {Verma}, {Silva Aguirre},
  {Vincenzo}, {Matteucci}, {Vai{\v{c}}ekauskait{\.{e}}}, {Palla}, {Grisoni}, \&
  {Calura}}]{Spitoni21}
{Spitoni}, E., {Verma}, K., {Silva Aguirre}, V., {et~al.} 2021, \aap, 647, A73

\bibitem[{{Stahler} \& {Palla}(2005)}]{Staler05}
{Stahler}, S.~W. \& {Palla}, F. 2005, {The Formation of Stars}

\bibitem[{{Stanghellini} \& {Haywood}(2010)}]{Stanghellini10}
{Stanghellini}, L. \& {Haywood}, M. 2010, \apj, 714, 1096

\bibitem[{{Stanghellini} \& {Haywood}(2018)}]{Stanghellini18}
{Stanghellini}, L. \& {Haywood}, M. 2018, \apj, 862, 45

\bibitem[{{Stonkut{\.{e}}} {et~al.}(2016){Stonkut{\.{e}}}, {Koposov}, {Howes},
  {Feltzing}, {Worley}, {Gilmore}, {Ruchti}, {Kordopatis}, {Randich},
  {Zwitter}, {Bensby}, {Bragaglia}, {Smiljanic}, {Costado},
  {Tautvai{\v{s}}ien{\.{e}}}, {Casey}, {Korn}, {Lanzafame}, {Pancino},
  {Franciosini}, {Hourihane}, {Jofr{\'e}}, {Lardo}, {Lewis}, {Magrini},
  {Monaco}, {Morbidelli}, {Sacco}, \& {Sbordone}}]{Stonkut2016MNRAS.460.1131S}
{Stonkut{\.{e}}}, E., {Koposov}, S.~E., {Howes}, L.~M., {et~al.} 2016, \mnras,
  460, 1131

\bibitem[{{Trentin} {et~al.}(2024){Trentin}, {Catanzaro}, {Ripepi},
  {Alonso-Santiago}, {Molinaro}, {Storm}, {De Somma}, {Marconi}, {Bhardwaj},
  {Gatto}, {Musella}, \& {Testa}}]{Ripepi24}
{Trentin}, E., {Catanzaro}, G., {Ripepi}, V., {et~al.} 2024, arXiv e-prints,
  arXiv:2404.17299

\bibitem[{{Van der Swaelmen} {et~al.}(2023){Van der Swaelmen}, {Viscasillas
  V{\'a}zquez}, {Cescutti}, {Magrini}, {Cristallo}, {Vescovi}, {Randich},
  {Tautvai{\v{s}}ien{\.{e}}}, {Bagdonas}, {Bensby}, {Bergemann}, {Bragaglia},
  {Drazdauskas}, {Jim{\'e}nez-Esteban}, {Guiglion}, {Korn}, {Masseron},
  {Minkevii{\={u}}t{\.{e}}}, {Smiljanic}, {Spina}, {Stonkut{\.{e}}}, \&
  {Zaggia}}]{Van2023A&A...670A.129V}
{Van der Swaelmen}, M., {Viscasillas V{\'a}zquez}, C., {Cescutti}, G., {et~al.}
  2023, \aap, 670, A129

\bibitem[{{Ventura} {et~al.}(2020){Ventura}, {Dell'Agli}, {Lugaro}, {Romano},
  {Tailo}, \& {Yag{\"u}e}}]{Ventura20}
{Ventura}, P., {Dell'Agli}, F., {Lugaro}, M., {et~al.} 2020, \aap, 641, A103

\bibitem[{{Ventura} {et~al.}(2013){Ventura}, {Di Criscienzo}, {Carini}, \&
  {D'Antona}}]{Ventura13}
{Ventura}, P., {Di Criscienzo}, M., {Carini}, R., \& {D'Antona}, F. 2013,
  \mnras, 431, 3642

\bibitem[{{Ventura} {et~al.}(2018){Ventura}, {Karakas}, {Dell'Agli},
  {Garc{\'\i}a-Hern{\'a}ndez}, \& {Guzman-Ramirez}}]{Ventura18}
{Ventura}, P., {Karakas}, A., {Dell'Agli}, F., {Garc{\'\i}a-Hern{\'a}ndez},
  D.~A., \& {Guzman-Ramirez}, L. 2018, \mnras, 475, 2282

\bibitem[{{Vincenzo} \& {Kobayashi}(2020)}]{Vincenzo20}
{Vincenzo}, F. \& {Kobayashi}, C. 2020, \mnras, 496, 80

\bibitem[{{Viscasillas V{\'a}zquez} {et~al.}(2022){Viscasillas V{\'a}zquez},
  {Magrini}, {Casali}, {Tautvai{\v{s}}ien{\.{e}}}, {Spina}, {Van der Swaelmen},
  {Randich}, {Bensby}, {Bragaglia}, {Friel}, {Feltzing}, {Sacco}, {Turchi},
  {Jim{\'e}nez-Esteban}, {D'Orazi}, {Delgado-Mena}, {Mikolaitis},
  {Drazdauskas}, {Minkevi{\v{c}}i{\={u}}t{\.{e}}}, {Stonkut{\.{e}}},
  {Bagdonas}, {Montes}, {Guiglion}, {Baratella}, {Tabernero}, {Gilmore},
  {Alfaro}, {Francois}, {Korn}, {Smiljanic}, {Bergemann}, {Franciosini},
  {Gonneau}, {Hourihane}, {Worley}, \&
  {Zaggia}}]{Viscasillas2022A&A...660A.135V}
{Viscasillas V{\'a}zquez}, C., {Magrini}, L., {Casali}, G., {et~al.} 2022,
  \aap, 660, A135

\bibitem[{{Viscasillas V{\'a}zquez} {et~al.}(2023){Viscasillas V{\'a}zquez},
  {Magrini}, {Spina}, {Tautvai{\v{s}}ien{\.{e}}}, {Van der Swaelmen},
  {Randich}, \& {Sacco}}]{Viscasillas23}
{Viscasillas V{\'a}zquez}, C., {Magrini}, L., {Spina}, L., {et~al.} 2023, \aap,
  679, A122

\bibitem[{{Willett} {et~al.}(2023){Willett}, {Miglio}, {Mackereth},
  {Chiappini}, {Lyttle}, {Elsworth}, {Mosser}, {Khan}, {Anders}, {Casali}, \&
  {Grisoni}}]{Willett23}
{Willett}, E., {Miglio}, A., {Mackereth}, J.~T., {et~al.} 2023, \mnras, 526,
  2141

\bibitem[{{Wong} {et~al.}(2004){Wong}, {Blitz}, \& {Bosma}}]{Wong04}
{Wong}, T., {Blitz}, L., \& {Bosma}, A. 2004, \apj, 605, 183

\bibitem[{{Yong} {et~al.}(2012){Yong}, {Carney}, \& {Friel}}]{Yong12}
{Yong}, D., {Carney}, B.~W., \& {Friel}, E.~D. 2012, \aj, 144, 95

\bibitem[{{Zhang} {et~al.}(2021){Zhang}, {Xiang}, {Zhang}, {Ting}, {Rix}, {Wu},
  {Huang}, {Sun}, {Tian}, {Wang}, \& {Liu}}]{Zhang21}
{Zhang}, M., {Xiang}, M., {Zhang}, H.-W., {et~al.} 2021, \apj, 922, 145

\end{thebibliography}


\begin{appendix} 
\section{Probing present-day gradients overestimation with different yield sets}
\label{appendixA}

As mentioned in \ref{ss:nucleo} and \ref{ss:results_2inf}, we perform additional model runs testing different massive stars and Type Ia SN yields than the reference ones used in this paper (\citealt{Limongi18} and \citealt{Iwa99}, respectively).

\begin{figure*}
    \centering
    \includegraphics[width=0.95\columnwidth]{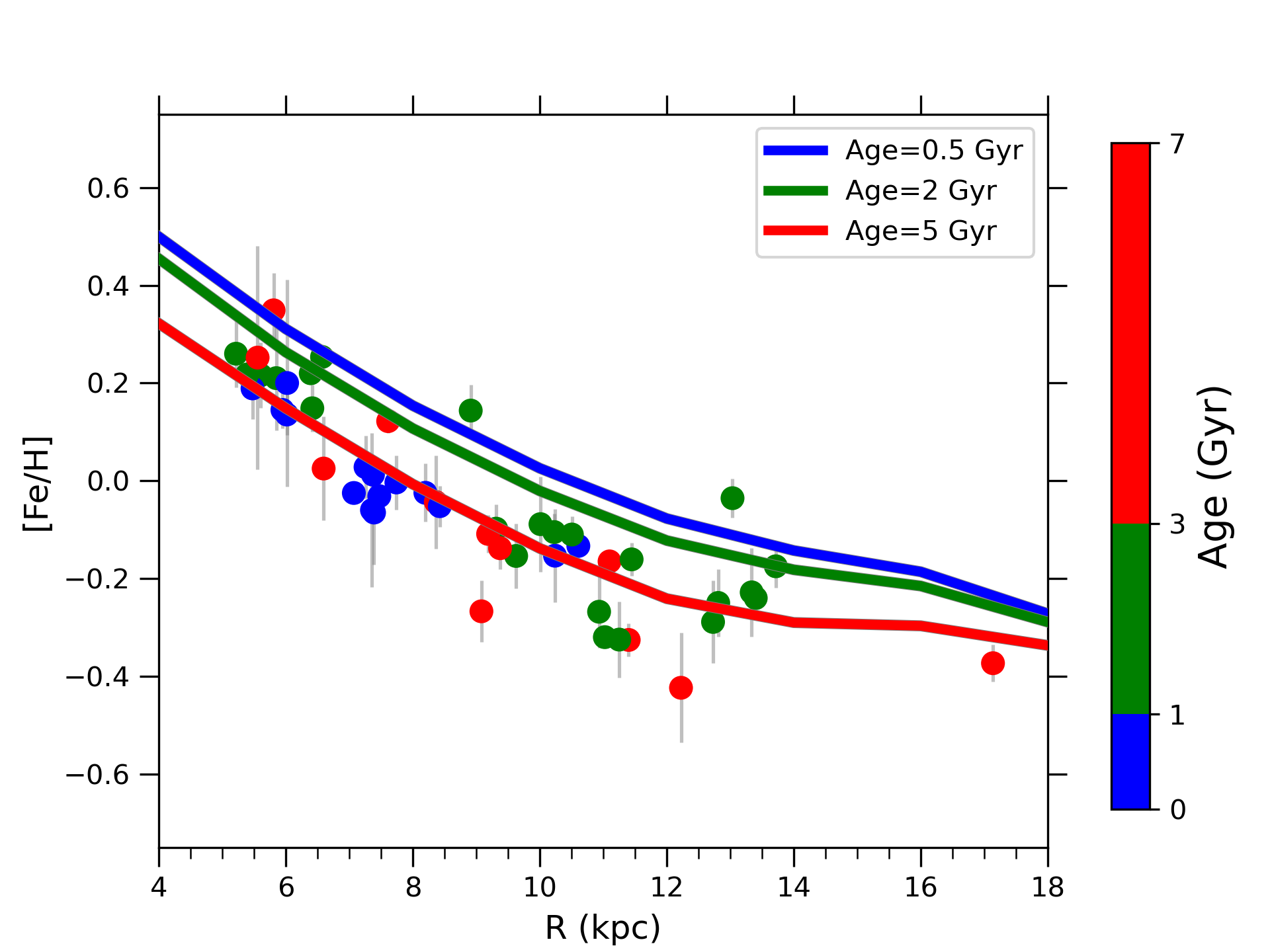}
    \includegraphics[width=0.95\columnwidth]{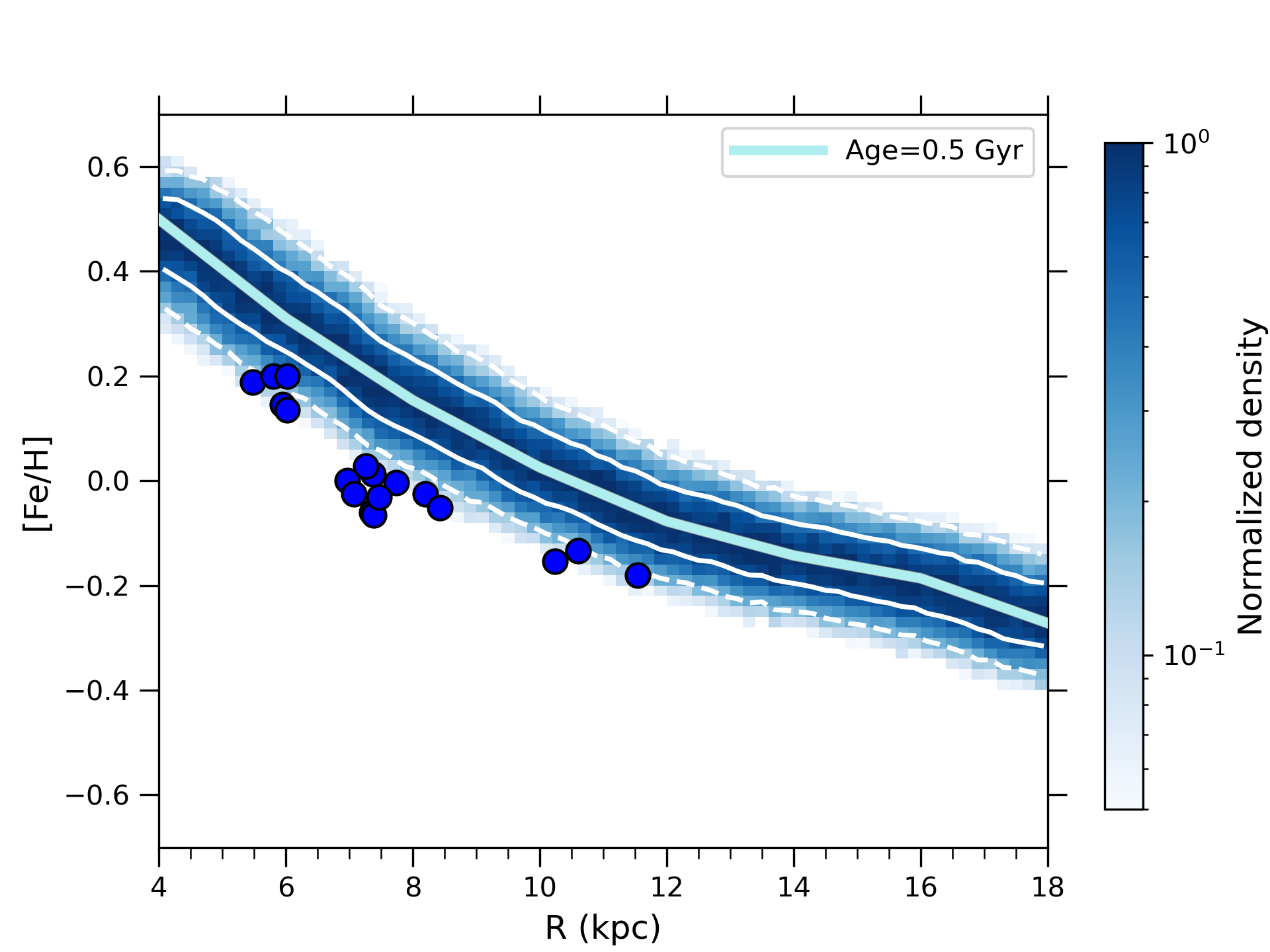}
    \caption{Predicted radial [Fe/H] gradient from the model 2INF using \citet{Koba06} instead of \citet{Limongi18} yields for massive stars.
    Left panel: same as Fig. \ref{fig:gradient_2inf}, but showing the data for the restricted OC sample.
    Right panel: same of Fig. \ref{fig:gradient_2inf_restricted}, but showing only the results for the young age bin ($< 1$ Gyr).}
    \label{fig:app_Koba}
\end{figure*}

\begin{figure*}
    \centering
    \includegraphics[width=0.95\columnwidth]{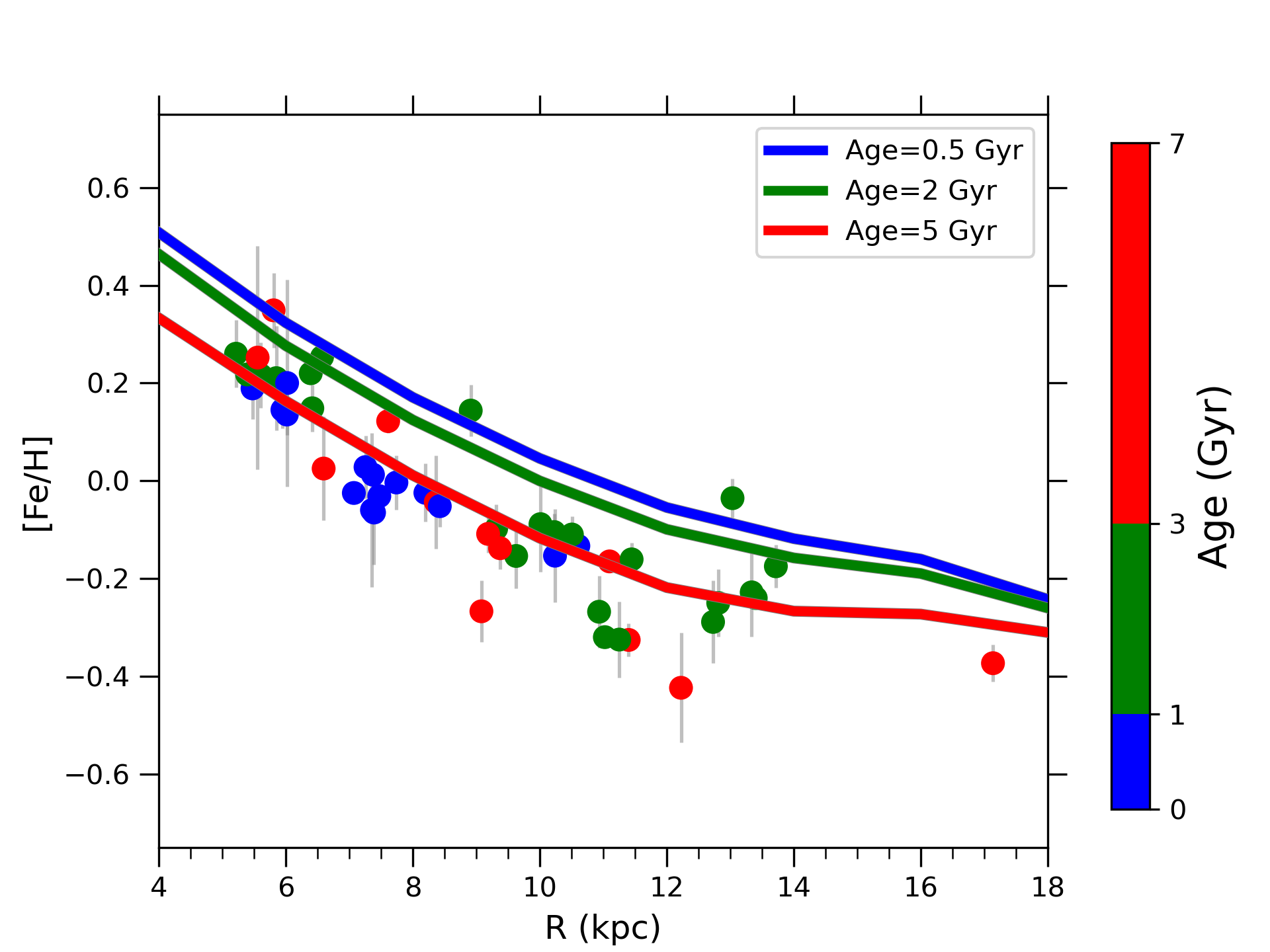}
    \includegraphics[width=0.95\columnwidth]{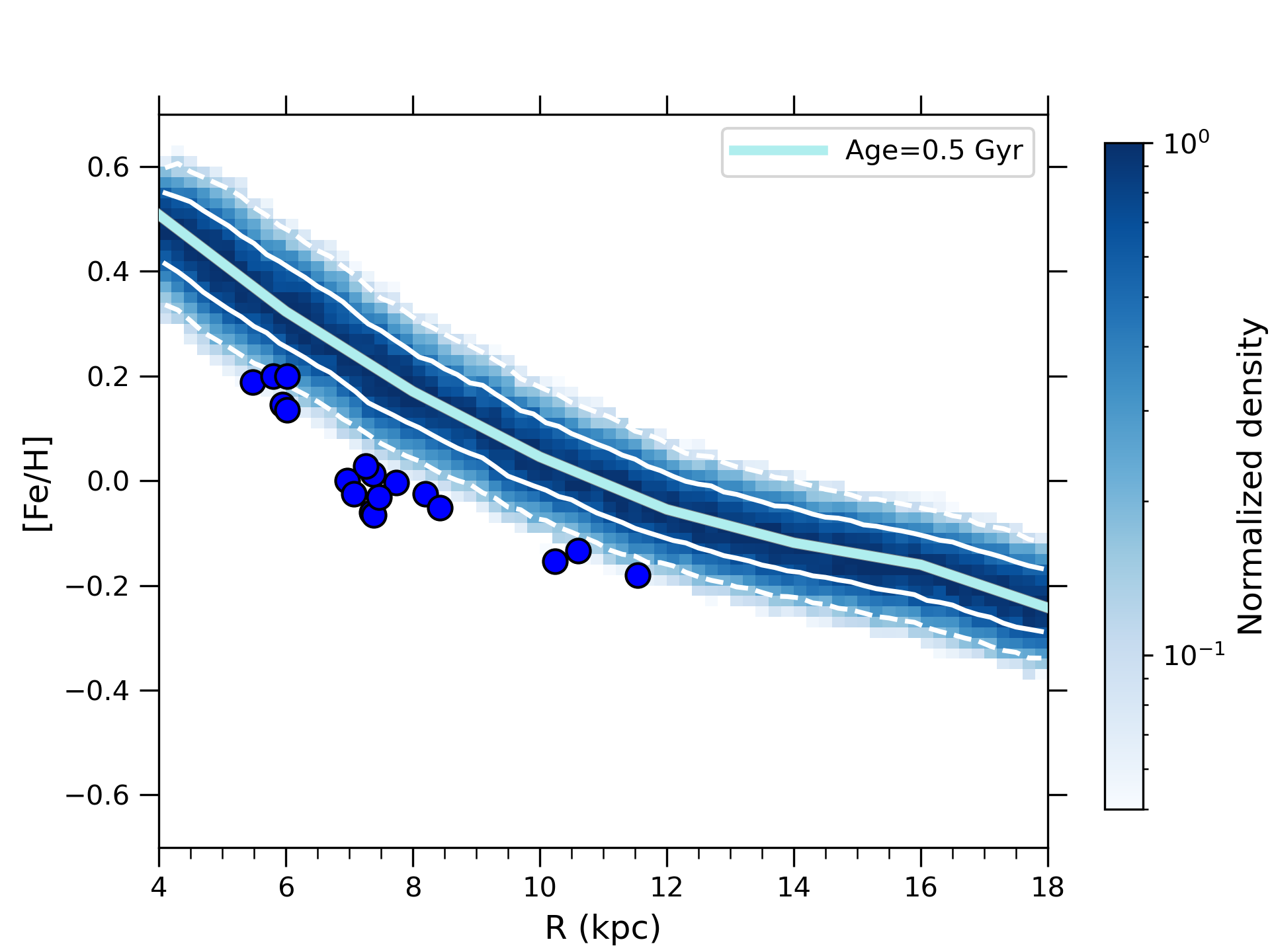}
    \caption{Predicted radial [Fe/H] gradient from the model 2INF using \citet{Leung18,Leung20} instead of \citet{Iwa99} yields for Type Ia SNe.
    Left panel: same as Fig. \ref{fig:gradient_2inf}, but showing the data for the restricted OC sample.
    Right panel: same of Fig. \ref{fig:gradient_2inf_restricted}, but showing only the results for the young age bin ($< 1$ Gyr).}
    \label{fig:app_Leung}
\end{figure*}

In the following Figures we show the results of this experiment, by reporting the gradient evolution (left panels) and the gradient in the young age bin considering the effects of stellar migration and abundance uncertainties in the model (right panels)
In particular, Fig. \ref{fig:app_Koba} shows the predicted radial [Fe/H] gradient obtained by adopting the massive star yields from \citet{Koba06} instead of those from \citet{Limongi18}.
Fig. \ref{fig:app_Leung} instead reports the predicted radial [Fe/H] gradient obtained by adopting the Type Ia SN yields from an equal mixture of \citet{Leung18} (near-$M_{Ch}$ progenitors) and \citet{Leung20} (sub-$M_{Ch}$ progenitors) instead of those \citet{Iwa99}. Regarding the latter Figure, it is worth mentioning that no significant changes are obtained by employing different mixtures in Type Ia yields.

\section{Gradients in physical quantities from the three-infall model}
\label{appendixB}

As mentioned in \ref{ss:discuss}, we checked whether our three-infall scenario reproduces the constraints in other gradients than those in chemical abundances, i.e. the present-day gradients in the physical quantities.

\begin{figure*}
    \centering
    \includegraphics[width=1\columnwidth]{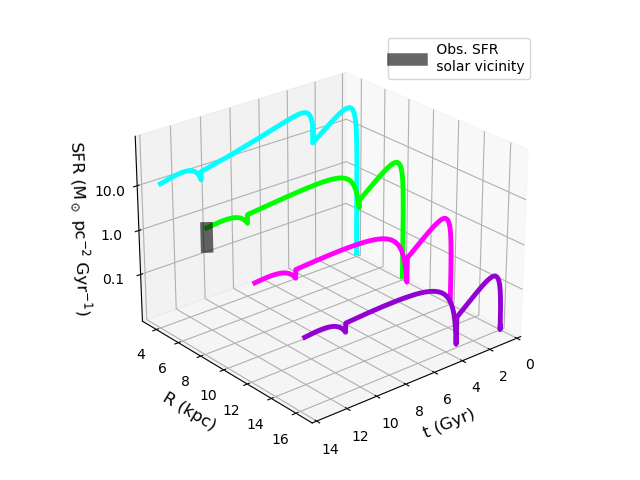}
    \includegraphics[width=0.95\columnwidth]{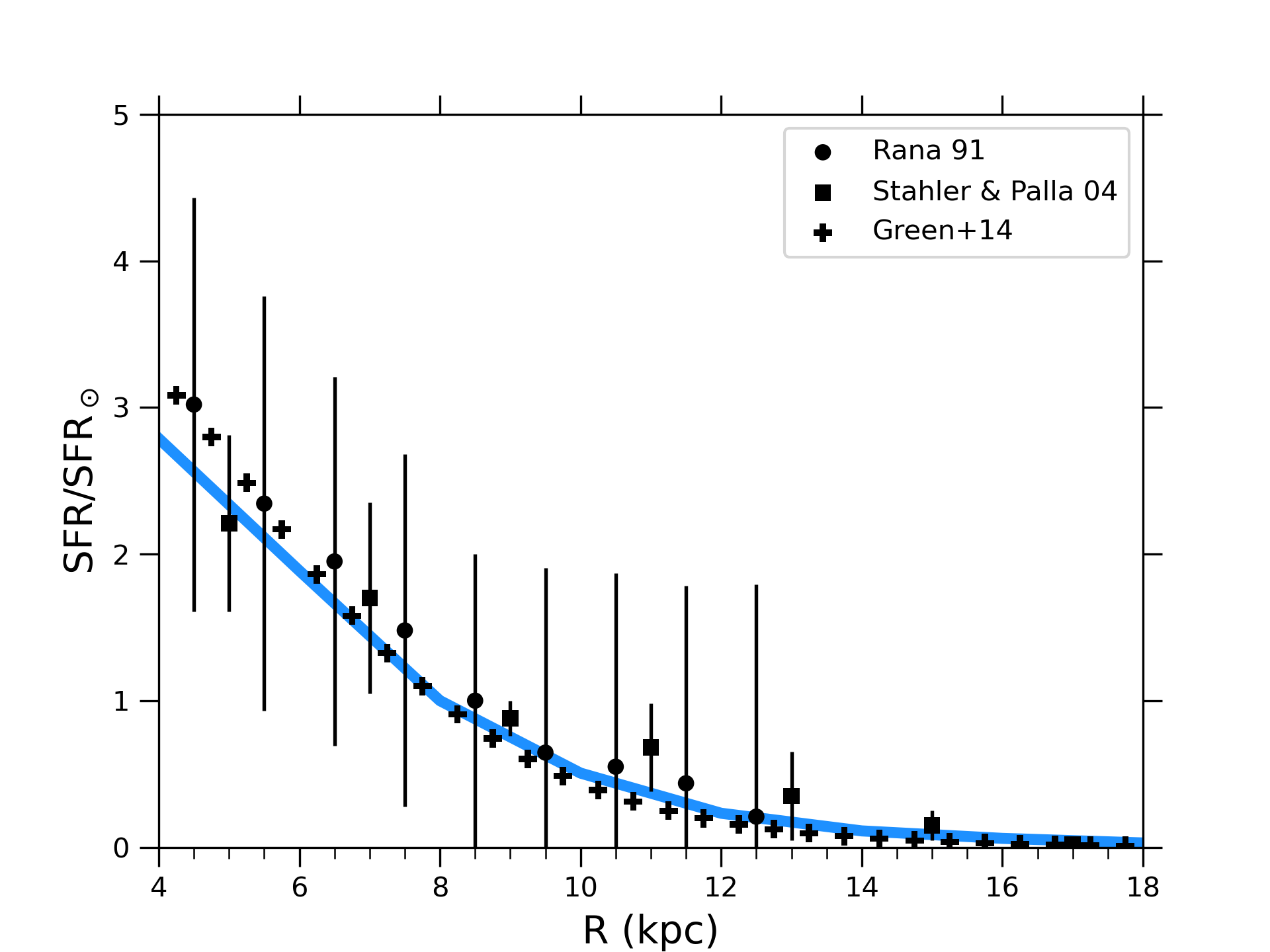}
    \caption{Left panel: SFR surface density time evolution at 4, 8, 12 and 16 kpc. The present-day observed value for the solar vicinity is taken from \citet{Prantzos18}.
    Right panel: present-day radial SFR surface density gradient. Data are from \citet{Rana91} (black points with errorbar), \citet{Staler05} (black squares with errorbar), \citet{Green14} (black crosses).}
    \label{fig:1}
\end{figure*}

\begin{figure*}
    \centering
    \includegraphics[width=1\columnwidth]{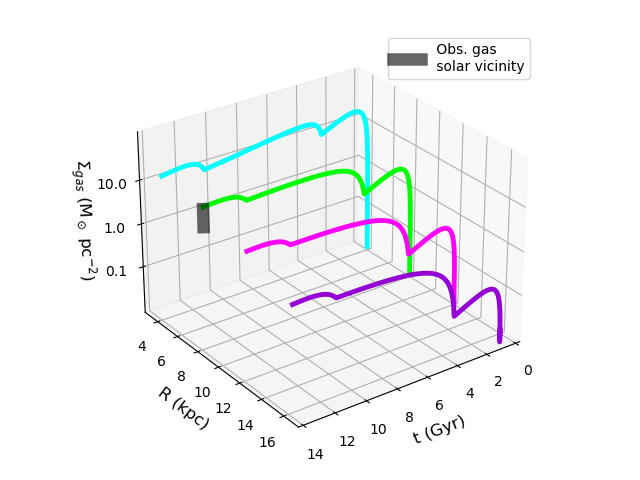}
    \includegraphics[width=0.95\columnwidth]{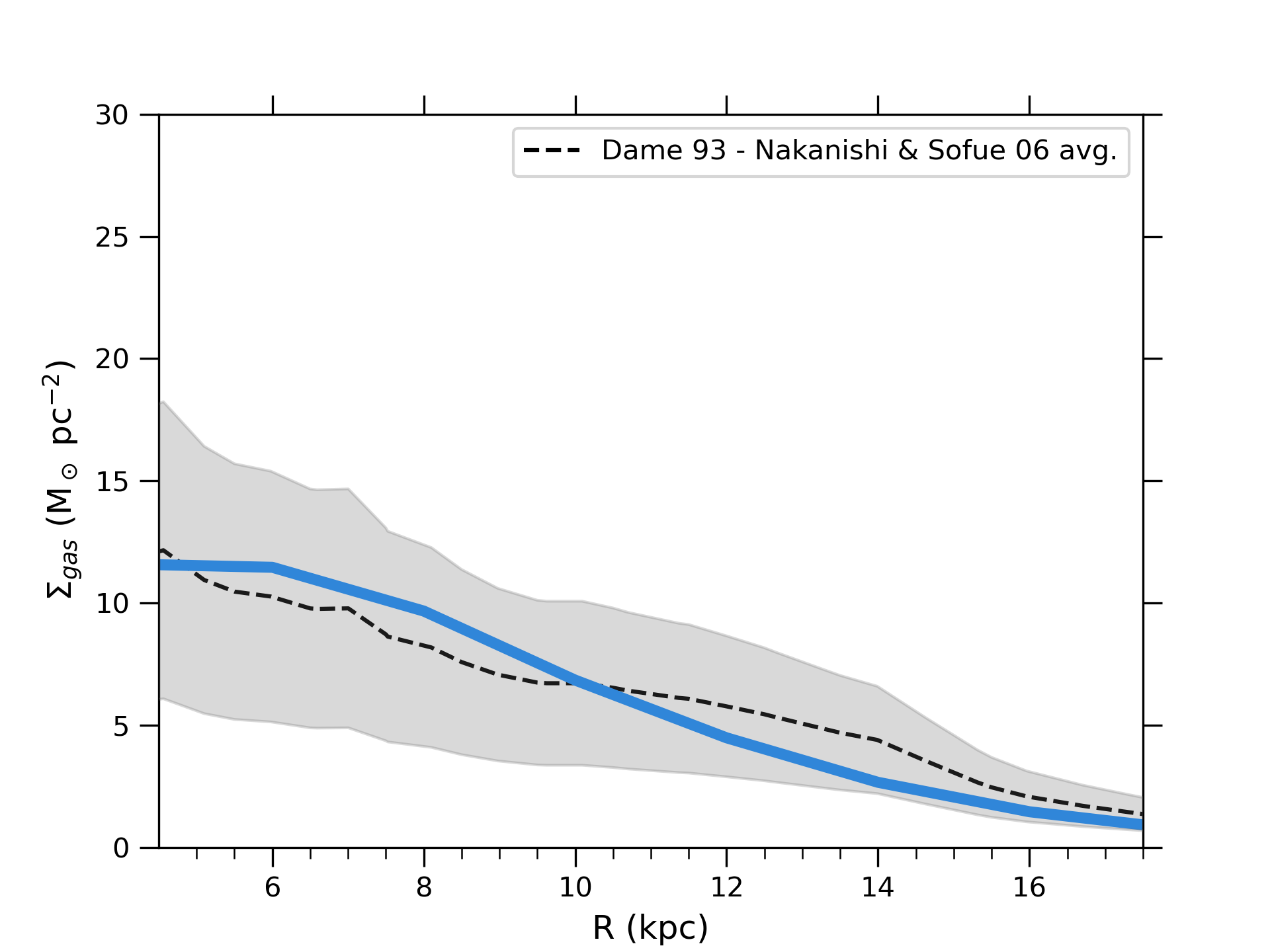}
    \caption{Left panel: gas surface density time evolution at 4, 8, 12 and 16 kpc. The present-day observed value for the solar vicinity is taken from \citet{Dame93,Nakanishi03,Nakanishi06}.
    Right panel: present-day radial gas surface density gradient. The dashed curve is the average between the \citet{Dame93} and \citet{Nakanishi03,Nakanishi06} data sets. The grey shaded region represents the typical uncertainty at each radius (see \citealt{Palla20} for more details).}
    \label{fig:2}
\end{figure*}

Results of the comparison between our best model 3INF-2 and literature constraints are shown in Fig. \ref{fig:1} and \ref{fig:2}.
In Fig. \ref{fig:1}, we show the prediction of the model for the SFR surface density. In the left panel, we show the time evolution of this quantity at different radii, with the present-day SFR observed in the solar vicinity (see \citealt{Prantzos18}) shown as reference. In the right panel, we show instead the comparison between the predicted present-day gradient and measurements from the literature (\citealt{Rana91,Staler05,Green14}).
In Fig. \ref{fig:2}, we display the same scheme as Fig. \ref{fig:1}, but for the gas surface density, where the reference value for the solar vicinity and the observed gradients are taken from \citet{Dame93,Nakanishi03,Nakanishi06}. 
For a detailed discussion on the adopted data sets, we refer to \citet{Palla20}.

\section{The restricted OCs sample}
\label{appendixC}

As described in \ref{ss:restricted_sample}, to avoid observational biases in the computation of the mean abundance of OCs, we build a restricted sample of member stars, where only stars with $\log g$  > 2.5 and $\xi$ < 1.8 km~$^{-1}$ are considered to compute the mean cluster abundances. 

In Tab. \ref{tab_clusters}, we provide the average abundances obtained with this membership selection, as well as all the obtained cluster parameters (see also \citealt{Viscasillas2022A&A...660A.135V}).

\begin{table*}
    \centering
    \caption{Excerpt from the list mean cluster abundances and cluster parameters from our adopted restricted sample (see \ref{ss:restricted_sample}).} 
    \begin{tabular}{c  c  c   c   c  c  c  c  c  c  c  c  c}
         \hline\\[-1.95ex]
         Cluster name &  [Fe/H] & A(O) & A(Mg) & A(Al) & A(Si) & ... & Age & R$_{GC}$ & $ecc$ & R$_{guide}$ & z$_{max}$ \\
          & (dex) &  (dex) & (dex) & (dex) & (dex)   & ... & (Gyr) & (kpc) &  & (kpc) & (kpc) \\[0.2cm]
         \hline\\[-1.95ex]
         Br81 & 0.22   &     -     &  7.69  & 6.58 & 7.71  & ... & 1.15 & 5.88  &  0.22 &  5.61 & 0.18  \\
         Rup134 & 0.26   &     8.87     &  7.83 & 6.67 & 7.79 & ... & 1.66 & 6.09  &  0.15 &   5.22 & 0.11\\
         ... & ...   &     ...     & ... & ... & ... & ... & ... &  ... &   ... & ... & ...\\
         Br20 & -0.32   &     -     & 7.28 & 6.38 & 7.23 & ... & 4.79 & 16.32 &  0.24 &  14.46 &  2.87 \\
         Br29  & -0.37   &     -     & 7.23 & 6.32 & 7.15 & ... & 3.09 & 20.58 &  0.16 &  17.14 & 1.88 \\[0.1cm]
         \hline\\
    \end{tabular}\\[0.1cm]
    {\bf Notes.} For the orbital parameters, we use the \textsc{Galpy} code, with the axis-symmetric potential \textsc{MWPotential2014} (\citealt{Bovy15}).\\ The complete Table will be available at the CDS.
    \label{tab_clusters}
\end{table*}

\end{appendix}

\end{document}